\documentclass[%
 reprint,
 nofootinbib,
 amsmath,amssymb,
 aps,
 prd,
]{revtex4-2}

\usepackage[shortlabels]{enumitem}
\usepackage{graphicx}
\usepackage{dcolumn}
\usepackage{bm}
\usepackage[hidelinks]{hyperref}
\usepackage{tabularx}
\usepackage{mathtools}

\begin{document}

\preprint{APS/123-QED}

\title{Adiabatic evolution due to the conservative scalar self-force during orbital resonances}

\author{Zachary Nasipak}
\affiliation{NASA Goddard Space Flight Center, 8800 Greenbelt Road, Greenbelt, Maryland, 20771, USA}
\email{zachary.nasipak@nasa.gov}

\date{\today}

\begin{abstract}
We calculate the scalar self-force experienced by a scalar point-charge orbiting a Kerr black hole along $r\theta$-resonant geodesics. We use the self-force to calculate the averaged rate of change of the charge's orbital energy $\langle\dot{E}\rangle$, angular momentum $\langle\dot{L}_z\rangle$, and Carter constant $\langle\dot{Q}\rangle$, which together capture the leading-order adiabatic, secular evolution of the point-charge. Away from resonances, only the dissipative (time antisymmetric) components of the self-force contribute to $\langle\dot{E}\rangle$, $\langle\dot{L}_z\rangle$, and $\langle\dot{Q}\rangle$. We demonstrate, using a new numerical code, that during $r\theta$ resonances conservative (time symmetric) scalar perturbations also contribute to $\langle\dot{Q}\rangle$ and, thus, help drive the adiabatic evolution of the orbit. Furthermore, we observe that the relative impact of these conservative contributions to $\langle\dot{Q}\rangle$ is particularly strong for eccentric 2:3 resonances. These results provide the first conclusive numerical evidence that conservative scalar perturbations of Kerr spacetime are nonintegrable during $r\theta$ resonances. 
\end{abstract}

\maketitle


\section{Introduction}
\label{sec:intro}

Future space-based gravitational wave observatories, such as the Laser Interferometer Space Antenna (LISA) \cite{AmarETC13, ESA12, NASA11}, will extend gravitational wave science into the low-frequency, milli-Hertz (mHz) regime. Sensitivity to a new frequency band will facilitate the observation of new astrophysical sources, including \textit{extreme-mass-ratio inspirals} (EMRIs) \cite{AmarETC07}, binaries in which a stellar-mass compact object (mass $\mu \sim 10M_\odot$) gradually inspirals into a massive black hole (mass $M \sim 10^6 M_\odot$). EMRIs are characterized by their small mass-ratios $\epsilon = \mu/M \sim 10^{-7} - 10^{-4}$ and the multiperiodic structure of their long inspirals. A typical EMRI possesses three orbital frequencies\footnote{If we neglect the spin of the smaller compact object.} \cite{Schm02}---the azimuthal frequency $\Omega_\phi$ of the small body's revolution about the massive black hole, the radial frequency $\Omega_r$ of the small body's libration between pericenter and apocenter, and the polar frequency $\Omega_\theta$ of the small body's nutating orbital plane. The orbital frequencies slowly evolve as the small body completes $\gtrsim \epsilon^{-1}$ orbital cycles before merger. The orbital evolution is imprinted in the gravitational waves radiated by the binary, leading to signals filled with rich harmonic structure that persist for months to years in the mHz band. Consequently, the cumulative signal-to-noise ratios of these sources will range from tens to hundreds, providing unprecedented tests of general relativity and high-precision measurements of EMRI masses and spins \cite{BerrETC19, BakeETC19}.

An interesting feature of EMRIs is that, due to their evolving tri-periodic motion, many EMRIs will experience special orbital configurations known as \textit{orbital resonances}. Resonances occur when at least two frequencies of motion form a rational low-integer ratio (e.g., $\Omega_r/\Omega_\theta = 1/2$). For EMRIs, \textit{transient orbital $r\theta$ resonances}---resonances that form between $\Omega_r$ and $\Omega_\theta$---are particularly important.\footnote{Other orbital resonances are either astrophysically improbable, such as sustained $r\theta$ resonances \cite{Vand14b}, or they are expected to have a weak, immeasurable effect on EMRI gravitational wave signals, such as $r\phi$ and $\theta\phi$ resonances \cite{Hira11, Vand14a}. EMRIs can also experience other resonances, such as tidal resonances \cite{BongYangHugh19, BronCardStei22, GuptETC21, GuptETC22}, which we do not consider in this work.} They persist for a resonant timescale $T_\mathrm{res} \sim M \epsilon^{-1/2}$, and, depending on the orbital phase at which the EMRI enters the resonance, they can enhance or diminish the binary's gravitational wave emission \cite{FlanHind12, BerrETC16, FlanHughRuan14}. This alters the inspiral by speeding up or slowing down the system's adiabatic loss of orbital energy and angular momentum (and Carter constant), which leaves a measurable impact on the gravitational waveform \cite{SperGair21}. Almost all EMRIs are expected to encounter at least one $r\theta$ resonance as they emit observable mHz gravitational waves \cite{RuanHugh14}. Therefore, failing to accurately model these resonances can hamper the detection and characterization of EMRIs by future space-based gravitational wave detectors \cite{BerrETC16, BabaETC17, SperGair21}. In this work, we model the impact of different perturbative effects on the leading-order evolution of a binary as it passes through different $r\theta$-resonances.

\subsection{Modeling EMRIs via the self-force}

EMRIs are naturally modeled within the framework of perturbation theory and the self-force. In the self-force approach, the small body is treated as a perturbing particle orbiting in the stationary background spacetime associated with the massive black hole \cite{Mino03, PoisPounVega11}. As the small body orbits in this background, it interacts with its own perturbations, resulting in a \textit{gravitational self-force} (GSF) that provides $O(\epsilon)$ corrections to the motion. The conservative (time symmetric) perturbations induce a GSF that is responsible for nonsecular changes in the orbit, while the dissipative (time antisymmetric) perturbations lead to a GSF that drives the adiabatic inspiral of the small body and can be connected to the orbit-averaged gravitational wave fluxes out at infinity and down the massive black hole horizon. 

The cumulative impact of the self-force on EMRI dynamics can be further understood via a two-timescale analysis \cite{HindFlan08}. At leading adiabatic order, only the averaged first-order dissipative self-force contributes to the evolution of the orbital (and gravitational wave) phase, which accumulate like $\sim \epsilon^{-1}$. This defines the inspiral timescale $T_\mathrm{insp} \sim M\epsilon^{-1}$. At subleading post-1 adiabatic order, the oscillatory pieces of the first-order dissipative self-force and conservative self-force induce small shifts in the orbital parameters, which vary on the orbital timescale $T_\mathrm{orb} \sim M$. Additionally, the accumulation of the averaged second-order self-force contributes at this order. Altogether, the post-1 adiabatic effects produce $O(1)$ corrections to the cumulative phase. Therefore, to generate EMRI waveforms that meet LISA's subradian phase accuracy requirement, one must calculate the averaged dissipative components of the GSF to a precision $\gtrsim \epsilon^{-1}$, since their errors can grow secularly over the inspiral, while the oscillatory pieces of the GSF will have their errors suppressed by $O(\epsilon)$ relative to the leading adiabatic order and thus only need to be computed to a few digits of precision.


However, this picture is complicated for systems that pass through at least one orbital $r\theta$ resonance. Due to the presence of a new timescale $T_\mathrm{res}$, resonances produce a post-1/2 adiabatic correction that impacts the phase at $O(\epsilon^{-1/2})$. In the following section we review the source of this half-order correction and how it impacts the accuracy requirements for self-force calculations. 


\subsection{Action-angle variables and EMRI resonances}

To better understand the impact of these transient resonances, we can describe EMRI dynamics in terms of action-angle variables, as proposed in \cite{HindFlan08}. In this multiscale action-angle framework, the fast orbital evolution of the system (over timescales $T_\mathrm{orb}$) is captured by the angle variables $\vec{w} = (w_r, w_\theta, w_\phi)$, whose time-derivatives give us the frequencies of motion (at leading-order). The actions $\vec{J}$ then describe the slow-time evolution of the system over $T_\mathrm{insp}$. Orbital quantities, such as the orbital energy $E$, orbital angular momentum $L_z$, and the Carter constant $Q$ (defined in Sec.~\ref{sec:geo}), provide one suitable set of actions, leading to the equations of motion,
\begin{align} \label{eqn:eomE}
    \dot{w}_\mathcal{C} &= \Omega_\mathcal{C}(J) + O(\epsilon),
    &
    \dot{\mathcal{J}} &= \epsilon\mathcal{F}^\mathcal{J}(J, w) + O(\epsilon^2),
\end{align}
where $\dot{x} = dx/dt$; $\vec{J} = (E, L_z, Q)$; $\mathcal{J}$ is used to represent $E$, $L_z$, or $Q$; and $\mathcal{C}$ is used to represent $r$, $\theta$, or $\phi$.\footnote{In other words we use $\mathcal{J}$ to denote a function or coefficient that is related to $E$ \textit{or} $L_z$ \textit{or} $Q$, while $\vec{J}$ is used for quantities that depend on $E$ \textit{and} $L_z$ \textit{and} $Q$.} The $\mathcal{F}^\mathcal{J}$ are forcing functions that are constructed from the self-force and, at leading order, can be expressed as double Fourier expansions in the angle variables,
\begin{align}
    \mathcal{F}^\mathcal{J}(J,w) &\approx \sum_{k=-\infty}^\infty \sum_{n=-\infty}^\infty f^\mathcal{J}_{kn}(J) e^{-i(kw_\theta + nw_r)},
\end{align}
where the Fourier coefficients $f^\mathcal{J}_{kn}$ only depend on $\vec{J}$, and the forcing functions do not explicitly depend on $w_\phi$ due to the rotational Killing symmetry of Kerr spacetime. We can further separate our forcing functions into terms that only depend on the dissipative self-force, and those that only depend on the conservative, i.e., $\mathcal{F}^\mathcal{J} = \mathcal{F}_\mathrm{diss}^\mathcal{J} + \mathcal{F}_\mathrm{cons}^\mathcal{J}$. Based on the symmetries of the dissipative and conservative pieces, the dissipative contributions reduce to discrete cosine series,
\begin{align}
    \mathcal{F}_\mathrm{diss}^\mathcal{J}(J,w) &\approx f^\mathcal{J}_{00}(J) + 2\sum_{n=1}^\infty \mathrm{Re}[f^\mathcal{J}_{0n}(J)] \cos(nw_r) 
    \\ \notag
    & \quad\;
    + 2\sum_{k=1}^\infty \sum_{n\neq 0} \mathrm{Re}[f^\mathcal{J}_{kn}(J)] \cos(kw_\theta + nw_r)
    \\ \notag
    & \qquad\qquad\qquad\;\;
    + 2\sum_{k=1}^\infty \mathrm{Re}[f^\mathcal{J}_{k0}(J)] \cos(kw_\theta) ,
\end{align}
while the conservative contributions reduce to discrete sine series,
\begin{align}
    \mathcal{F}_\mathrm{cons}^\mathcal{J}(J,w) &\approx -2\sum_{n=1}^\infty \mathrm{Im}[f^\mathcal{J}_{0n}(J)] \sin(nw_r) 
    \\ \notag
    & \quad\;
    - 2\sum_{k=1}^\infty \sum_{n\neq 0} \mathrm{Im}[f^\mathcal{J}_{kn}(J)] \sin(kw_\theta + nw_r)
    \\ \notag
    & \qquad\qquad\qquad\;\;
    - 2\sum_{k=1}^\infty \mathrm{Im}[f^\mathcal{J}_{k0}(J)] \sin(kw_\theta).
\end{align}

Integrating \eqref{eqn:eomE} over a period $T_\mathrm{insp}$, the oscillatory pieces of the forcing functions will rapidly vary and destructively interfere with one another. Thus the leading-order adiabatic evolution is given by the averaged, nonoscillatory piece of each forcing function. Away from resonances, the only nonoscillatory terms are the zero modes from the dissipative contributions,
\begin{align} \label{eqn:avgNonRes}
    \langle \dot{\mathcal{J}} \rangle \approx \epsilon \langle\mathcal{F}^\mathcal{J}\rangle &\approx \epsilon\langle\mathcal{F}_\mathrm{diss}^\mathcal{J}\rangle \approx \epsilon f^\mathcal{J}_{00}(J).
\end{align}
However, as the system approaches a resonance, nonzero $(k,n)$-modes become stationary and contribute to the secular evolution of the system. This can be seen by expanding the angle variables around the exact moment of resonance $t_\mathrm{res}$, when $\Omega_r(t_\mathrm{res})/\Omega_\theta(t_\mathrm{res}) = \Omega_r^\mathrm{res}/\Omega_\theta^\mathrm{res} = \beta_r/\beta_\theta$
with $\beta_r,\beta_\theta \in \mathbb{Z}$. Introducing the condensed notation $x_{kn} = kx_\theta + nx_r$ for any arbitrary parameter $x$, the phase takes the expanded form
\begin{align} \label{eqn:angleExpansion}
    w_{kn}(t) &= w_{kn}^\mathrm{res} + \Omega_{kn}^\mathrm{res}\Delta t + \frac{1}{2}\dot{\Omega}_{kn}^\mathrm{res}\Delta t ^2 +  O(\Delta t ^3),
\end{align}
where $\Delta t = t - t_\mathrm{res}$ and $w_{kn}^\mathrm{res}=w_{kn}(t_\mathrm{res})$. Whenever 
\begin{equation} \label{eqn:resCondition}
    n\beta_r + k\beta_\theta = 0,
\end{equation}
$\Omega_{kn}^\mathrm{res} = 0$, and the phase becomes stationary (i.e., $w_{kn}(t) \approx w_{kn}^\mathrm{res}$) as long as $\frac{1}{2}\dot{\Omega}_{kn}^\mathrm{res}(t-t_\mathrm{res})^2 \lesssim 1$. Because $\dot{\Omega}_{kn}^\mathrm{res} \sim \epsilon$, this criteria is met when $|t-t_\mathrm{res}| \lesssim M\epsilon^{-1/2} \sim T_\mathrm{res}$. This defines the resonant timescale. Therefore any modes that satisfy the resonant condition of \eqref{eqn:resCondition} will contribute to the secular evolution of the system for a period $T_\mathrm{res}$,
\begin{align} \label{eqn:avgRes}
    \langle \dot{\mathcal{J}} \rangle \approx \epsilon \langle\mathcal{F}^{\mathcal{J}}\rangle &\approx \sum_{k\beta_\theta + n\beta_r = 0} \epsilon f^\mathcal{J}_{kn}(J) e^{-iw_{kn}(t_\mathrm{res})}.
\end{align}

Equation \eqref{eqn:avgRes} differs from \eqref{eqn:avgNonRes} in two key ways. First, the averages depend on the initial angles or phases at resonance $w_{kn}^\mathrm{res}$. Second, the averages include potential contributions from both dissipative and conservative perturbations. This poses a new problem for numerical models. Errors in the initial values of $w_{kn}$ or the values of $\mathrm{Im}[f^\mathcal{J}_{kn}]$ are usually suppressed away from resonances, but during a resonance they can accumulate and become magnified by $O(\epsilon^{-1/2})$. The exact magnification of the errors will depend on the relative strength of the Fourier coefficients $f^\mathcal{J}_{kn}$ to the dominant mode $f^\mathcal{J}_{00}$. Numerically evaluating these coefficients can, therefore, provide insight into how these errors grow during a resonance.

Flux-inspired mode-sum expressions for $\langle \dot{E}\rangle$, $\langle \dot{L}_z\rangle$, and $\langle \dot{Q}\rangle$ provide an efficient method for determining the dissipative contributions $\mathrm{Re}[f^\mathcal{J}_{kn}]$ using the asymptotic behavior of the perturbing gravitational field \cite{FlanHughRuan14}, but these results do not easily extend to conservative perturbations. While these conservative perturbations could still vanish at leading order during resonances \cite{FlanHind12}, i.e., $\langle\mathcal{F}_\mathrm{cons}^{\mathcal{J}}\rangle \approx 0$, previous authors have hypothesized that they will contribute to the average rate of change of the Carter constant, $\langle \dot{Q} \rangle$ \cite{IsoyETC13, FujiETC17, IsoyETC19}, breaking the integrability of the Hamiltonian conservative system at resonance \cite{BlanFlan22}. There have been attempts to extend asymptotic mode-sum methods to include conservative contributions \cite{IsoyETC13}, but the regularization of these conservative perturbations (see Sec.~\ref{sec:ssf}) complicates these procedures. Instead we can calculate $f^\mathcal{J}_{kn}$, and thereby the conservative and dissipative contributions to $\langle \dot{\mathcal{J}} \rangle$, if we know the values of the GSF along a resonance. However, while GSF calculations have been performed along generic orbits in Kerr spacetime, to date these GSF calculations have not been extended to $r\theta$ resonances \cite{Vand18}.

\subsection{Paper overview}

In this paper we make use of a scalar toy model to evaluate how, during $r\theta$ resonances, the conservative contributions to the self-force impact the adiabatic evolution of EMRIs. Using this model, we also study how numerical errors can propagate during these resonances. For our scalar model we consider a scalar point charge $q$ on a $r\theta$-resonant geodesic in Kerr spacetime. The charge experiences a geodesic \textit{scalar self-force} (SSF), which is akin to the GSF. Neglecting the GSF and treating $q^2/(\mu M)$ as a small parameter, the scalar self-forced equations of motion take the same perturbative form as \eqref{eqn:eomE}, but with $\epsilon$ replaced by $q/M$. Much like the GSF, the SSF ``pushes" the particle away from a geodesic as it sources the gradual evolution of the orbital parameters $E$, $L_z$, and $Q$. This work is an extension of previous results published by the author \cite{NasiEvan21}. In \cite{NasiEvan21}, we presented the first numerical results of the SSF along $r\theta$-resonant geodesics. However, the systematic uncertainty in our SSF results was too large to make definitive statements about the impact of the conservative SSF on the secular evolution of our two-body system. In this work, we have built a new \texttt{C++} code that is orders of magnitude faster than our previous \texttt{Mathematica} code and incorporates new algorithms that have greatly reduced our numerical and systematic errors, allowing us to accurately calculate $\langle \dot{E}\rangle$, $\langle \dot{L}_z\rangle$, and $\langle \dot{Q}\rangle$ via \eqref{eqn:eomE} and \eqref{eqn:avgRes}.

The paper is organized as follows. To set notation, in Sec.~\ref{sec:background} we review Kerr geodesics, the SSF, and the average rate of change of the orbital constants, $\langle\dot{E}\rangle$, $\langle\dot{L}_z\rangle$, and $\langle\dot{Q}\rangle$. For brevity, we will collectively refer to these average rates of change simply as adiabatic or secular averages, since they grow secularly with time. Furthermore, we will use $\langle\dot{\mathcal{J}}\rangle$ to represent all three secular averages. In Sec.~\ref{sec:comp}, we describe the methods employed by our new \texttt{C++} code for computing the SSF along $r\theta$ resonances and then calculating the resulting averaged secular evolution of the orbital quantities, $\langle\dot{\mathcal{J}}\rangle$. We highlight two new techniques used by our new code: (1) a Teukolsky solver that combines confluent Heun expansions, series of hypergeometric functions, and numerical integrators to solve the Teukolsky equation; and (2) a new application of mode-sum regularization that directly regularizes the Fourier harmonics that describe $\langle\dot{\mathcal{J}}\rangle$. In Sec.~\ref{sec:fluxes}, we demonstrate, with new numerical results, that the conservative perturbations do contribute to the secular evolution of $Q$, breaking the integrability of the conservative scalar system at resonance \cite{BlanFlan22}. We conclude with a discussion of these results in Sec.~\ref{sec:conclusion}. For this paper we use the metric signature $(-+++)$, the sign conventions, where applicable, of \cite{MisnThorWhee73}, and units such that $c=G=1$.

\section{Background}
\label{sec:background}

We employ the same resonant scalar model and SSF formalism used in \cite{NasiEvan21}. The zeroth-order background motion is given by a geodesic $x_p^\mu$ in Kerr spacetime $g_{\mu\nu}$ (Sec.~\ref{sec:geo}). The motion of the scalar charge $q$ sources a scalar field $\Phi^\mathrm{ret}$ (Sec.~\ref{sec:field}), and the charge interacts with a regular component of this field $\Phi^\mathrm{R}$, thus experiencing a SSF $F_\alpha$ (Sec.~\ref{sec:ssf}). The SSF then drives the secular evolution of the energy, angular momentum, and Carter constant that parametrize the charge's motion, $\langle\dot{E}\rangle$, $\langle\dot{L}_z\rangle$, and $\langle\dot{Q}\rangle$, respectively (Sec.~\ref{sec:ELQdot}). To establish notation, in the following section we provide a brief overview of how we construct these quantities.

\subsection{Bound geodesics in Kerr spacetime}
\label{sec:geo}

We consider a point particle with mass $\mu$ on a bound geodesic in a Kerr background $g_{\mu\nu}$. In Boyer-Lindquist coordinates $(t, r,\theta,\phi)$ the Kerr line element reads
\begin{align}
    ds^2 &= -\left(1-\frac{2Mr}{\Sigma}\right) dt^2 + \frac{\Sigma}{\Delta} dr^2
    \\ \notag
    & \qquad
    + \Sigma d\theta^2  - \frac{4Mar\sin^2\theta}{\Sigma} dtd\phi
    \\ \notag
    & \qquad \qquad
    + \frac{\sin^2\theta}{\Sigma}\left((r^2+a^2)^2-a^2\Delta\sin^2\theta\right) d\phi^2,
\end{align}
where $M$ is the Kerr mass parameter, $a$ the Kerr spin parameter, $\Sigma = r^2 + a^2\cos^2\theta$, and $\Delta = r^2 - 2Mr + a^2$. The worldline and four-velocity of the particle are denoted by $x_p^\mu = (t_p, r_p, \theta_p, \phi_p)$ and $u^\mu = dx_p^\mu/d\tau$, respectively, where $\tau$ is the particle's proper time. 

To solve for $x_p^\mu$ we leverage the three Killing symmetries of Kerr spacetime: the time Killing vector $\xi^\mu_{(t)}=g^\mu_{\phantom{\mu}t}$, the azimuthal Killing vector $\xi^\mu_{(\phi)}=g^\mu_{\phantom{\mu}\phi}$, and the Killing tensor $K^{\mu\nu}$ \cite{WalkPenr70}. (See Eq.~(C23) in \cite{NasiEvan21} for an explicit definition of $K^{\mu\nu}$.) Projecting these Killing symmetries onto the four-velocity of the particle provides us with four constants of motion: the particle mass $\mu$, the orbital energy ${E} = - \xi^\mu_{(t)} u_\mu = -u_t$, the $z$-component of the orbital angular momentum ${L}_z = \xi^\mu_{(\phi)} u_\mu = u_\phi$, and the (scaled) Carter constant ${Q} = K^{\mu\nu}u_\mu u_\nu - ({L}_z - a {E})^2 = {K}  - ({L}_z - a {E})^2 $ \cite{Cart68}.

With these conserved quantities, we obtain four first-order ordinary differential equations (ODEs) for $x_p^\mu$, which decouple when parametrized in terms of the Mino(-Carter) time parameter $\lambda$ \cite{Mino03, Cart68},
\begin{align} \label{eqn:geoT}
    \frac{dt_p}{d\lambda} &= V_{tr}(r_p) + V_{t\theta}(\theta_p),\\ \label{eqn:geoR}
    \frac{dr_p}{d\lambda} &= \pm \sqrt{V_r(r_p)},\\ \label{eqn:geoTheta}
    \frac{d\theta_p}{d\lambda} &= \pm \sqrt{V_\theta(\theta_p)},\\ \label{eqn:geoPhi}
    \frac{d\phi_p}{d\lambda} &= V_{\phi r}(r_p) + V_{\phi \theta}(\theta_p).
\end{align}
where $d\lambda = \Sigma^{-1} d\tau$, and the potential functions are given by
\begin{align} 
    V_{r}(r) &= \mathrlap{ P^2(r) - \Delta\left(r^2 + {K} \right),} 
    \\ \notag
    V_{\theta}(\theta) &= \mathrlap{{Q} - {L}_z^2 \cot^2\theta - a^2 (1 -{E}^2)\cos^2\theta,}
    \\\notag
    V_{tr}(r) &= \frac{r^2+a^2}{\Delta}P(r), 
    &
    V_{t\theta}(\theta) &= a{L}_z - a^2 {E} \sin^2\theta, 
    \\ \notag
    V_{\phi r}(r) &= \frac{a}{\Delta}P(r), 
    &
    V_{\phi \theta}(\theta) &= {L}_z \csc^2\theta - a {E},
\end{align}
with $P(r) = (r^2+a^2){E} - a {L}_z$.

To choose values of $({E}, {L}_z, {Q})$ that correspond to bound geodesics, we introduce the orbital parameters
\begin{gather}
    p = \frac{2 r_\mathrm{min} r_\mathrm{max}}{M(r_\mathrm{min} + r_\mathrm{max})}, \qquad e = \frac{r_\mathrm{max} - r_\mathrm{min} }{r_\mathrm{min} + r_\mathrm{max}}, 
    \\
    x = \cos\left( \frac{\pi}{2} - \theta_\mathrm{min}\right),
\end{gather}
where $p$ is the semilatus rectum, $e$ is the orbital eccentricity, and $x$ is (the projection of) the orbital inclination. The minimum radius $r_\mathrm{min}$, maximum radius $r_\mathrm{max}$, and minimum polar angle $\theta_\mathrm{min} = \pi - \theta_\mathrm{max}$ represent the turning points of the geodesic. We first choose values of $(p, e, x)$, then obtain the corresponding values of $({E}, {L}_z, {Q})$ using the methods of \cite{Schm02}. Once the constants of motion have been determined, we solve \eqref{eqn:geoT}-\eqref{eqn:geoPhi} using spectral integration methods \cite{HoppETC15, NasiOsbuEvan19}, which provide exponentially convergent numerical approximations of the geodesic solutions.

The resulting bound solutions can be separated into terms that are periodic and terms that grow secularly. The radial and polar motion are completely periodic, with Mino time periods,
\begin{align}
    \Lambda_r &= 2 \int_{r_\mathrm{min}}^{r_\mathrm{max}} \frac{dr}{\sqrt{V_r(r)}},
    &
    \Lambda_\theta &= 2 \int_{\theta_\mathrm{min}}^{\theta_\mathrm{max}} \frac{d\theta}{\sqrt{V_\theta(\theta)}},
\end{align}
and corresponding frequencies,
\begin{align}
    \Upsilon_r &= \frac{2\pi}{\Lambda_r} ,
    &
    \Upsilon_\theta &= \frac{2\pi}{\Lambda_\theta}. 
\end{align}
Therefore, we represent the radial and polar motion by
\begin{align} \label{eqn:radialSol}
    r_p(\lambda) &= \Delta r^{(r)}(\Upsilon_r\lambda) = \Delta r^{(r)}(\Upsilon_r\lambda + 2\pi),
    \\ \label{eqn:polarSol}   
    \theta_p(\lambda) &= \Delta \theta^{(\theta)}(\Upsilon_\theta\lambda) = \Delta \theta^{(\theta)}(\Upsilon_\theta\lambda + 2\pi). 
\end{align}
On the other hand, coordinate time $t_p$ and the azimuthal angle of the particle $\phi_p$ not only oscillate with respect to the radial and polar frequencies, but they also accumulate with rates
\begin{align}
    \Upsilon_t &= \Upsilon^{(r)}_t + \Upsilon^{(\theta)}_t,
    &
    \Upsilon_\phi &= \Upsilon^{(r)}_\phi + \Upsilon^{(\theta)}_\phi,
\end{align}
where
\begin{align}
    \Upsilon^{(r)}_t &= \frac{1}{\Lambda_r} \int_0^{\Lambda_r} V_{tr}(r_p) d\lambda,
    &
    \Upsilon^{(\theta)}_t &= \frac{1}{\Lambda_\theta} \int_0^{\Lambda_\theta} V_{t\theta}(\theta_p) d\lambda, 
    \\
    \Upsilon^{(r)}_\phi &= \frac{1}{\Lambda_r} \int_0^{\Lambda_r} V_{\phi r}(r_p) d\lambda,
    &
    \Upsilon^{(\theta)}_\phi &= \frac{1}{\Lambda_\theta} \int_0^{\Lambda_\theta} V_{\phi\theta}(\theta_p) d\lambda.
\end{align}
We represent their evolution as
\begin{align} \label{eqn:tSol}
    t_p(\lambda) &= \Upsilon_t \lambda + \Delta t^{(r)}(\Upsilon_r\lambda) + \Delta t^{(\theta)}(\Upsilon_\theta\lambda),
    \\ \label{eqn:phiSol}
    \phi_p(\lambda) &= \Upsilon_\phi \lambda + \Delta \phi^{(r)}(\Upsilon_r\lambda) + \Delta \phi^{(\theta)}(\Upsilon_\theta\lambda),
\end{align}
where $\Delta t^{(r)}$, $\Delta r^{(r)}$, and $\Delta \phi^{(r)}$ all share the same periodic structure, and likewise for $\Delta t^{(\theta)}$, $\Delta \theta^{(\theta)}$, and $\Delta \phi^{(\theta)}$.

The exact form of the oscillatory terms in \eqref{eqn:radialSol}, \eqref{eqn:polarSol}, \eqref{eqn:tSol}, and \eqref{eqn:phiSol} depends on the initial conditions of the bound geodesic. However, it is possible to relate geodesics that share the same orbital constants, but different initial conditions. To do so we first define a fiducial geodesic $\hat{x}^\mu(\lambda)$ with initial conditions (at $\lambda = 0)$
\begin{gather} \notag
    \hat{x}^\mu_p(0) = (\hat{t}_p(0),\hat{r}_p(0),\hat{\theta}_p(0),\hat{\phi}_p(0))=(0, r_\mathrm{min}, \theta_\mathrm{min}, 0)
    \\ \label{eqn:fiducial}
    \hat{u}^r(0) = \hat{u}^\theta(0) = 0.
\end{gather}
Then any geodesic with arbitrary initial conditions
\begin{gather} \notag
    {x}^\mu_p(0) =(t_p(0),r_p(0),\theta_p(0),{\phi}_p(0))=(t_0, r_0, \theta_0, \phi_0)
    \\
    {u}^r(0) = u^r_0,\qquad {u}^\theta(0) = {u}^\theta_0.
\end{gather}
can be expressed in terms of the fiducial solutions,
\begin{align}\label{eqn:geoTSol}
    t_p(\lambda) &= \Upsilon_t \lambda + \Delta \hat{t}^{(r)}(q_r + q_{r0}) - \Delta\hat{t}^{(r)}(q_{r0}) 
    \\ \notag
    & \qquad + \Delta \hat{t}^{(\theta)}(q_\theta + q_{\theta 0}) - \Delta \hat{t}^{(\theta)}(q_{\theta 0}) + t_0,
    \\
    r_p(\lambda) &= \Delta \hat{r}^{(r)}(q_r + q_{r0}),
    \\
    \theta_p(\lambda) &= \Delta \hat{\theta}^{(\theta)}(q_\theta + q_{\theta 0}),
    \\ \label{eqn:geoPhiSol}
    \phi_p(\lambda) &= \Upsilon_\phi \lambda + \Delta \hat{\phi}^{(r)}(q_r + q_{r0}) - \Delta \hat{\phi}^{(r)}(q_{r0})
    \\ \notag
    & \qquad + \Delta \hat{\phi}^{(\theta)}(q_\theta + q_{\theta 0}) - \Delta \hat{\phi}^{(\theta)}(q_{\theta 0}) + \phi_0,
\end{align}
where we have defined the phase variables $q_r = \Upsilon_r \lambda$ and $q_\theta = \Upsilon_\theta \lambda$.\footnote{The phase variables $q_r$ and $q_\theta$ are equivalent to the geodesic limit ($\epsilon\rightarrow 0$) of the angle variables $w_r$ and $w_\theta$ introduced in Sec.~\ref{sec:intro} if we had parametrized our equations in terms of $\lambda$ instead of $t$.}
The initial phases
$q_{r0}$ and $q_{\theta 0}$ are set by the initial conditions
$(r_0, u^r_0)$ and $(\theta_0, u^\theta_0)$, respectively, and all hatted quantities are constructed assuming the fiducial conditions of \eqref{eqn:fiducial}.

Together the Mino time frequencies also define the fundamental frequencies of the bound geodesic with respect to coordinate time,
\begin{align}
    \Omega_r &= \frac{\Upsilon_r}{\Upsilon_t}
    &
    \Omega_\theta &= \frac{\Upsilon_\theta}{\Upsilon_t}
    &
    \Omega_\phi &= \frac{\Upsilon_\phi}{\Upsilon_t}.
\end{align}
When at least two of these frequencies are commensurate, the geodesic is said to be resonant. In this work, we are interested in $r\theta$ resonances, where $\Omega_r/\Omega_\theta = \beta_r/\beta_\theta$ with $\beta_r,\beta_\theta \in \mathbb{Z}$. For $r\theta$ resonances, we define the resonant Mino frequency, resonant Mino period, and resonant fundamental frequency
\begin{equation} \label{eqn:UpsilonRes}
    \Upsilon = \frac{\Upsilon_r}{\beta_r} = \frac{\Upsilon_\theta}{\beta_\theta}, \qquad \Lambda = \frac{2\pi}{\Upsilon}, \qquad \Omega = \frac{\Upsilon}{\Upsilon_t},
\end{equation}
respectively.

An important feature of nonresonant geodesics is that they are \textit{ergodic} or space-filling. Given an infinite amount of Mino time, a nonresonant orbit will uniformly sample every point on the two-torus spanned by the phase variables $q_r$ and $q_\theta$. In practice, this means that any nonresonant geodesic is directly related to a fiducial geodesic up to some trivial translation in time and azimuth. To change the initial conditions, we can simply pick new values of $t_0$ and $\phi_0$. Due to the Killing symmetries of Kerr, time and azimuthal translations do not impact quantities such as the SSF or the secular rate of change of the orbital quantities, e.g., $\langle \dot{E} \rangle$. Therefore, we can choose fiducial initial conditions when we model nonresonant geodesics without loss of generality.

In contrast, resonant geodesics are not ergodic but follow restricted tracks through coordinate space (and on the $q_r$-$q_\theta$ two-torus). Furthermore, different choices of initial conditions $q_{r0}$ and $q_{\theta 0}$ can also lead to unique trajectories (e.g., Fig.~2 in \cite{NasiEvan21}). Unlike nonresonant geodesics, $r\theta$ resonances will only map to fiducial geodesics if they simultaneously pass through $r_\mathrm{min}$ and $\theta_\mathrm{min}$. This only occurs if the condition
\begin{align} \label{eqn:fiducialResCondition}
    \frac{q_{\theta 0}}{\beta_\theta} - \frac{q_{r 0}}{\beta_r} = 2\pi N
\end{align}
is satisfied for some integer $N$. Generally \eqref{eqn:fiducialResCondition} is not met. Nevertheless, two $r\theta$ resonances can be mapped onto one another provided they share the same initial resonant phase (modulo $2\pi$)
\begin{align} \label{eqn:q0}
    q_0 = \frac{q_{\theta 0}}{\beta_\theta} - \frac{q_{r 0}}{\beta_r},
\end{align}
and the same orbital parameters (e.g., $E$, $L_z$, and $Q$). Therefore, when constructing the SSF or $\langle \dot{Q} \rangle$ (see Secs.~\ref{sec:ssf} and \ref{sec:ELQdot}), we must take into account their dependence on the initial phases. In practice, due to the interdependence of $q_{r0}$ and $q_{\theta 0}$ in \eqref{eqn:q0}, we set $q_{r0} = 0$ and parametrize our initial conditions in terms of $q_{\theta 0}$, without loss of generality, when dealing with $r\theta$ resonances.

\subsection{Scalar perturbations of Kerr spacetime}
\label{sec:field}

We now consider that our point-mass $\mu$, orbiting on a geodesic $x_p^\mu$, also possesses a scalar charge $q$. The charge sources a scalar field $\Phi$ (per unit charge $q$) that satisfies the Klein-Gordon equation
\begin{align} \label{eqn:KG}
    q g^{\mu\nu}\nabla_\mu \nabla_\nu \Phi(x) = -4\pi\rho(x),
\end{align}
where $\rho$ is the scalar charge density,
\begin{align} \label{eqn:chargeDensity}
    \rho(x) = q\frac{\delta(r-r_p)\delta(\cos\theta-\cos\theta_p)\delta(\phi-\phi_p)}{V_{tr}(r)+V_{t\theta}(\theta)}.
\end{align}

Transforming to the frequency domain, we can solve \eqref{eqn:KG} via separation of variables \cite{Cart68, BrilETC72, Teuk73}. We then obtain the physical retarded solution $\Phi^\mathrm{ret}$ by imposing causal boundary conditions at the black hole horizon $r_+ = M + \sqrt{M^2 - a^2}$ and infinity. When $\Phi^\mathrm{ret}$ is reconstructed back into the time-domain, the only frequencies that contribute are those that correspond to the discrete frequency spectrum of the charge's bound motion,
\begin{equation}
    \omega_{mkn} = m\Omega_\phi + k\Omega_\theta + n \Omega_r.
\end{equation}
As a result, the scalar field can be expressed as a discrete mode-sum,
\begin{align} \label{eqn:phiRet}
    \Phi^\mathrm{ret}(x) = \sum_{jmkn} R^\mathrm{ret}_{jmkn}(r) S_{jmkn}(\theta) e^{im\phi} e^{-i\omega_{mkn}t},
\end{align}
where we have introduced the compact notation,
\begin{equation}
    \sum_{jmkn} = \sum_{j = 0}^\infty \sum_{m=-j}^j \sum_{k=-\infty}^\infty \sum_{n=-\infty}^\infty.
\end{equation}

The polar dependence in \eqref{eqn:phiRet} is captured by the scalar spheroidal harmonics $S_{jmkn}(\theta)$ \cite{PresTeuk73, BreuRyanWall77}, which satisfy
\begin{align} \label{eqn:swshODE}
    &\Bigg[\frac{d^2}{d^2\theta} - \cot\theta \frac{d}{d\theta} - a^2\omega_{mkn}^2\sin^2\theta
    \\ \notag
    & \qquad + 2ma\omega_{mkn} - \frac{m^2}{\sin^2\theta} + \lambda_{jmkn}\Bigg]S_{jmkn}(\theta) = 0,
\end{align}
where $\lambda_{jmkn}$ is the spheroidal eigenvalue. In the limit $a\omega_{mkn}\rightarrow 0$, $\lambda_{jmkn} \rightarrow l(l+1)$ and \eqref{eqn:swshODE} reduces to the spherical harmonic equation. Therefore, the $S_{jmkn}$ can be expressed as rapidly convergent sums of spherical harmonics,
\begin{align} \label{eqn:swsh}
    S_{jmkn}(\theta)e^{im\phi} = \sum_{l = 0}^\infty b^{l}_{jmkn} Y_{lm}(\theta,\phi),
\end{align}
where the coupling coefficients $b^{l}_{jmkn}$ satisfy a three-term recursion relation \cite{Hugh00, WarbBara10}.

The radial dependence in \eqref{eqn:phiRet} is captured by the Teukolsky solutions $R^\mathrm{ret}_{jmkn}(r)$, which satisfy the inhomogeneous spin-0 radial Teukolsky equation
\begin{equation} \label{eqn:teukODE}
    \left[\frac{d^2}{dr^2} + G^\mathrm{T}(r) \frac{d}{dr} +  U^\mathrm{T}_{jmkn}(r) \right]R_{jmkn}(r)= Z_{jmkn}(r),
\end{equation}
with causal boundary conditions \cite{Teuk73}, where
\begin{align}
    G^\mathrm{T}(r) &= \frac{2(r-M)}{\Delta}, 
    \\
    U^\mathrm{T}_{jmkn}(r) &= \frac{[(r^2+a^2)\omega_{mkn}-ma]^2}{\Delta^2}-\frac{\lambda_{jmkn}}{\Delta},
\end{align}
and $Z_{jmkn}$ is the radial component of the mode-decomposed source, $\rho$, given in \eqref{eqn:chargeDensity}. The causal boundary conditions are captured by the unit-normalized ingoing and upgoing homogeneous solutions, which have the respective asymptotic behaviors,
\begin{align}
    R^-_{jmkn}(r\rightarrow r_+) &\sim e^{-ip_{mkn} r_*},   
    \\
    R^+_{jmkn}(r \rightarrow \infty) &\sim \frac{e^{i\omega_{mkn} r_*}}{r},
\end{align}
where $p_{mkn} = \omega_{mkn} - ma/(2Mr_+)$ is the horizon-shifted frequency, and $r_*$ is the tortoise coordinate defined by the differential relation $dr_*/dr = (r^2+a^2)/\Delta$.

While we could reconstruct $\Phi^\mathrm{ret}$ using \eqref{eqn:phiRet}, the resulting field exhibits Gibbs ringing in the source region $r_\mathrm{min}\leq r \leq r_\mathrm{max}$ due to the pointlike distributional source in \eqref{eqn:chargeDensity}. Therefore, we circumvent the Gibbs phenomenon by applying the method of extended homogeneous solutions \cite{BaraOriSago08, HoppEvan10, Warb15}.
Outside the source region ($r < r_\mathrm{min}$ and $r > r_\mathrm{max}$) $R^\mathrm{ret}_{jmkn}$ simply reduces to the homogeneous solutions $R^\pm_{jmkn}$ multiplied by complex normalization coefficients $C^\pm_{jmkn}$,
\begin{align}
    R^\mathrm{ret}_{jmkn}(r < r_\mathrm{min}) &= C^-_{jmkn} R^-_{jmkn}(r),
    \\
    R^\mathrm{ret}_{jmkn}(r > r_\mathrm{max}) &= C^+_{jmkn} R^+_{jmkn}(r).
\end{align}
The $C^\pm_{jmkn}$, also known as Teukolsky amplitudes, are calculated using the standard method of variation of parameters, as outlined in \cite{NasiOsbuEvan19,NasiEvan21}. Importantly, varying the initial phases of the scalar charge's orbit will change $C^\pm_{jmkn}$ by an overall phase factor \cite{DrasFlanHugh05, FlanHughRuan14}
\begin{align} \label{eqn:fiducalCjmkn}
    C^\pm_{jmkn} = e^{i\xi_{mkn}(t_0, q_{r0},q_{\theta 0}, \phi_0)} \hat{C}^\pm_{jmkn},
\end{align}
where the fiducial coefficients $\hat{C}^\pm_{jmkn}$ are calculated assuming fiducial initial conditions \eqref{eqn:fiducial}, and the phase factor $\xi_{mkn}$ is given by
\begin{align} \notag
    \xi_{mkn}&(t_0, q_{r0},q_{\theta 0}, \phi_0) = 
    \\ 
    &  m(\Delta\hat{\phi}^{(r)}(q_{r0}) + \Delta\hat{\phi}^{(\theta)}(q_{\theta 0}) - \phi_0)
    - kq_{\theta 0}
    \\ \notag
    & \qquad  - \omega_{mkn}(\Delta\hat{t}^{(r)}(q_{r0}) + \Delta\hat{t}^{(\theta)}(q_{\theta 0}) - t_0) - n q_{r0}.
\end{align}
Consequently, in the vacuum regions of spacetime $\Phi_\mathrm{ret}$ can be reconstructed from an exponentially convergent sum over the homogeneous radial solutions $R^\pm_{jmkn}$,
\begin{align} \label{eqn:PhiMinus}
    \Phi^\mathrm{ret}(r < r_\mathrm{min}) &= \sum_{l=0}^\infty \sum_{m=-l}^l \psi^-_{lm}(t,r) Y_{lm}(\theta,\phi),
    \\
    &= \Phi^-(x),
    \\ \label{eqn:PhiPlus}
    \Phi^\mathrm{ret}(r > r_\mathrm{max}) &= \sum_{l=0}^\infty \sum_{m=-l}^l \psi^+_{lm}(t,r) Y_{lm}(\theta,\phi),
    \\
    &= \Phi^+(x),
\end{align}
where we have introduced the extended homogeneous mode functions,
\begin{align} \label{eqn:psilm}
    \psi^\pm_{lm}(t,r) &= \sum_{k=-\infty}^\infty \sum_{n=-\infty}^\infty \sum_{j = |m|}^\infty  \psi_{ljmkn}^\pm (r) e^{-i\omega_{mkn}t},
    \\ \label{eqn:psiljmkn}
    \psi_{ljmkn}^\pm (r) &= b^l_{jmkn} C^\pm_{jmkn} R^\pm_{jmkn}(r).
\end{align}
To obtain $\Phi^\mathrm{ret}$ over the entire radial domain, we simply extend our homogeneous solutions into the source region, e.g.,
\begin{align} \label{eqn:PhiEHS}
    \Phi^\mathrm{ret}(t,r,\theta,\phi) &= \Phi^-(t,r,\theta,\phi)\Theta(r_p(t)-r) 
    \\\notag
    &\qquad +  \Phi^+(t,r,\theta,\phi)\Theta(r-r_p(t)). 
\end{align}
While $\psi^\pm_{lm}(t,r)$ and $\psi_{ljmkn}^\pm (r)$ are not formal solutions of our inhomogeneous Teukolsky wave equations, summing over all of the extended harmonic modes results in a convergent field solution that is free of Gibbs ringing and accurately represents $\Phi^\mathrm{ret}$ over the entire spacetime domain up to the charge's worldline $x^\mu_p$.

\subsection{Scalar self-force}
\label{sec:ssf}

Now we take into account the backreaction of the scalar field on the charge $q$. This produces a SSF $F_\alpha$ (per unit charge squared) that drives the inspiral of the particle \cite{Quin00},
\begin{equation} \label{eqn:ssfEOM}
    u^\mu \nabla_\mu (\mu u^\alpha) = q^2 F^\alpha.
\end{equation}
While the field $\Phi^\mathrm{ret}$ formally diverges along the charge's worldline (where the SSF is evaluated), only a regular component of the field contributes to the SSF. This regular contribution is completely captured by the Detweiler-Whiting regular field $\Phi^\mathrm{R}$ \cite{DetwWhit03}, 
\begin{align}
    F^\alpha = \lim_{x\rightarrow x_p} g^{\alpha\beta} \nabla_\beta \Phi^\mathrm{R},
\end{align}
though it is often convenient to further decompose $\Phi^\mathrm{R}$ into its conservative and dissipative pieces, $\Phi^\mathrm{R}= \Phi^\mathrm{cons} + \Phi^\mathrm{diss}$, resulting in conservative and dissipative contributions to the self-force,
\begin{align}
    F^\mathrm{cons}_\alpha &= \nabla_\alpha \Phi^\mathrm{cons},
    &
    F^\mathrm{diss}_\alpha &= \nabla_\alpha \Phi^\mathrm{diss},
\end{align}
respectively. The dissipative scalar perturbations are associated with the radiative field $\Phi^\mathrm{diss} = \frac{1}{2}(\Phi^\mathrm{ret} - \Phi^\mathrm{adv})$ that drives the inspiral of the scalar charge, where $\Phi^\mathrm{adv}$ is the advanced field solution of \eqref{eqn:KG}. While $\Phi^\mathrm{ret}$ and $\Phi^\mathrm{adv}$ diverge along the worldline, their singular behaviors perfectly cancel when constructing $\Phi^\mathrm{diss}$, resulting in a smooth, well-defined field along the worldline. The remaining conservative scalar perturbations $\Phi^\mathrm{cons} = \Phi^\mathrm{R} - \Phi^\mathrm{diss}$ only source nonsecular changes in the motion when the system is not in resonance. In this work, we investigate whether this behavior also extends to resonances.

To calculate these conservative perturbations, we first construct the regular field from the difference $\Phi^\mathrm{R} = \Phi^\mathrm{ret} - \Phi^\mathrm{S}$, where $\Phi^\mathrm{S}$ is the Detweiler-Whiting singular field. Like $\Phi^\mathrm{ret}$, $\Phi^\mathrm{S}$ satisfies \eqref{eqn:KG}, but with nonradiative boundary conditions that result in a solution that captures the local, singular behavior of the field. Therefore, $\Phi^\mathrm{S}$ also diverges at the location of the charge, requiring a regularization procedure that delicately handles the subtraction of $\Phi^\mathrm{ret}$ and $\Phi^\mathrm{S}$. In this work we employ mode-sum regularization \cite{BaraOri00, BaraOri03a}, in which the divergent quantities $F^{\mathrm{ret}}_{\alpha}= \nabla_\alpha \Phi^\mathrm{ret}$ and $F^{\mathrm{S}}_{\alpha} = \nabla_\alpha \Phi^\mathrm{S}$ are decomposed onto a spherical harmonic basis, resulting in finite multipole moments (spherical harmonic $l$-modes) that can be subtracted mode by mode,
\begin{equation} \label{eqn:modeSumReg}
    F_\alpha = \lim_{x\rightarrow \pm x_p} \sum_{l=0}^\infty \left(F^{\mathrm{ret},l}_{\alpha\pm} -
    F^{\mathrm{S},l}_{\alpha\pm} \right).
\end{equation}
The $\pm$ notation takes into account that $F^{\mathrm{ret},l}_{\alpha}$ and $F^{\mathrm{S},l}_{\alpha}$ are discontinuous at $x_p^\mu$; therefore, their values depend on the radial direction from which we approach the worldline. While we only need to regularize $F_\alpha^\mathrm{cons}$, in practice we first compute $F_\alpha$ via \eqref{eqn:modeSumReg} and then use it to construct $F_\alpha^\mathrm{cons}$ and $F_\alpha^\mathrm{diss}$.

An advantage of this mode-sum approach is that $F^{\mathrm{S},l}_{\alpha\pm}$ can be expressed as a series of $l$-independent \textit{regularization parameters}, which are constructed by expanding $\Phi^\mathrm{S}$ in the local neighborhood of $x^\mu_p$ \cite{BaraOri03a, DetwWhit03, HeffOtteWard14},
\begin{align} \label{eqn:regExpansion}
    F^{\mathrm{S},l}_{\alpha\pm} &= \pm \frac{1}{2}A_\alpha (2l + 1) + B_\alpha 
    \\ \notag
    & \qquad + \sum_{n = 1}^{\infty}
    \prod_{k=1}^n\frac{D^{(2n)}_\alpha}{(2l+2k+1)(2l-2k+1)}.
\end{align}
The higher-order parameters, $D_\alpha^{(2n)}$, are not strictly needed for convergence. They vanish when summed over all $l$-modes,
\begin{equation}
    \sum_{l = 0}^{\infty}
    \prod_{k=1}^n\frac{D^{(2n)}_\alpha}{(2l+2k+1)(2l-2k+1)} = 0,
\end{equation}
but each additional $D_\alpha^{(2n)}$ that we incorporate in our mode-sum regularization increases the rate at which the sum in \eqref{eqn:modeSumReg} converges. Including parameters up to $n = n_\mathrm{max}$ results in \eqref{eqn:modeSumReg} converging like $l^{-2(n_\mathrm{max}+1)}$. This is particularly useful in our numerical calculations, where we are forced to truncate the sum at some finite $l=l_\mathrm{max}$. In this work, we make use of $A_\alpha$, $B_\alpha$, and $D^{(2)}_\alpha$, which were analytically derived by Heffernan \cite{Heff21} and are available on Zenodo \cite{HeffZeno22}.

Using our results from Sec.~\ref{sec:field}, it is straightforward to construct the time, radial, and azimuthal components of $F^{\mathrm{ret},l}_{\alpha\pm}$,
\begin{align} \label{eqn:FtRet}
    F^{\mathrm{ret},l}_{t\pm} &= \sum_{m=-l}^l F^{\mathrm{ret},lm}_{t\pm} =\sum_{m=-l}^l \partial_t \psi_{lm}^\pm(t,r) Y_{lm}(\theta,\phi), \\
    \label{eqn:FrRet}
    F^{\mathrm{ret},l}_{r\pm} &= \sum_{m=-l}^l F^{\mathrm{ret},lm}_{r\pm} =\sum_{m=-l}^l \partial_r \psi_{lm}^\pm(t,r) Y_{lm}(\theta,\phi), \\
    \label{eqn:FphiRet}
    F^{\mathrm{ret},l}_{\phi\pm} &= \sum_{m=-l}^l F^{\mathrm{ret},lm}_{\phi\pm} =\sum_{m=-l}^l i m \psi_{lm}^\pm(t,r) Y_{lm}(\theta,\phi).
\end{align}
Assembling the polar component is more complicated. Taking the polar derivative of \eqref{eqn:PhiMinus}-\eqref{eqn:PhiEHS} results in a series of spherical harmonic derivatives,
\begin{align} \label{eqn:FthetaRetNaive}
    F^{\mathrm{ret},l}_{\theta\pm} &= \sum_{m=-l}^l \psi_{lm}^\pm(t,r) \partial_\theta Y_{lm}(\theta,\phi),
\end{align}
instead of the spherical harmonic basis needed for our regularization scheme. Unfortunately, simply reexpanding \eqref{eqn:FthetaRetNaive} onto a basis of spherical harmonics is computationally impractical due to the strong coupling between $\partial_\theta Y_{lm}$ and $Y_{lm}$ \cite{Warb15}. 

To circumvent this issue, we follow the methods of \cite{Warb15} and multiply $\Phi^\mathrm{ret}$ by a suitable window function $f(x)$ so that the combination $f \partial_\theta Y_{lm}$ can be reexpressed as a finite series of spherical harmonics. If the windowed, unregularized self-force field,
\begin{align} \notag
    \tilde{F}^\mathrm{ret}_\theta(x) &= \nabla_\theta\left[f(x)\Phi^\mathrm{ret}(x)\right] 
    \\
    &= f(x)\partial_\theta\Phi^\mathrm{ret}(x) + \Phi^\mathrm{ret}(x)\partial_\theta f(x),
\end{align}
reduces to $F^\mathrm{ret}_\theta(x)$ as we approach the worldline (i.e., $\tilde{F}^\mathrm{ret}_\theta \rightarrow {F}^\mathrm{ret}_\theta$ as $x^\mu \rightarrow x^\mu_p$), then we can calculate $F_\theta$ via the mode-sum regularization of $\tilde{F}^\mathrm{ret}_\theta(x)$,\footnote{Alternatively, we can think of this window function method as a process for choosing a different extension of the self-force operator away from the particle's worldline, i.e., $\nabla_\alpha \rightarrow f(x) \nabla_\alpha$, as done in \cite{Vand18}.}
\begin{equation}
    F_\theta = \lim_{x\rightarrow \pm x_p}\sum_{l=0}^\infty \left( \tilde{F}^{\mathrm{ret},l}_{\theta\pm} - \tilde{B}_\theta - \frac{\tilde{D}_\theta^{(2)}}{(2l-1)(2l+3)} \right).
\end{equation}
where we have made use of the fact that $\tilde{A}_\theta = \tilde{A}_\phi = 0$ along all geodesics in Kerr spacetime. For a general choice of $f(x)$, the new regularization parameters $\tilde{B}_\theta$ and $\tilde{D}_\theta^{(2)}$ will differ from the original analytically known parameters ${B}_\theta$ and ${D}_\theta^{(2)}$, because our window function will also change the structure of the singular field. To ensure that $\tilde{B}_\theta = B_\theta$ and $\tilde{D}_\theta^{(2)} = D^{(2)}_\theta$, we must choose $f(x)$ so that the singular structure of $\Phi^\mathrm{S}$ is preserved at its first few leading orders. (See Appendix \ref{app:reg} for more details.) This is achieved if $f(\cos\theta) = 1 + O(\cos\theta - \cos\theta_p)^4$.

Therefore, we introduce the window function
\begin{align}
	f(\theta, \theta_p) &= \sum_{j=0}^3 \alpha_j (\theta_p) \cos^j\theta \sin\theta,
\end{align}
where
\begin{align}
	\alpha_0(\theta_p) &= \frac{2-7\cos^2\theta_p+8\cos^4\theta_p(1-\cos^2\theta_p)}{2\sin^7\theta_p},
	\\
	\alpha_1(\theta_p) &= \frac{3\cos^3\theta_p(1+4\cos^2\theta_p)}{2\sin^7\theta_p},
	\\
	\alpha_2(\theta_p) &= \frac{1-8\cos^2\theta_p(1+\cos^2\theta_p)}{2\sin^7\theta_p},
	\\
	\alpha_3(\theta_p) &= \frac{\cos\theta_p(3+2\cos^2\theta_p)}{2\sin^7\theta_p}.
\end{align}
With this window function, we get a finite coupling between the spherical harmonics and their derivatives,
\begin{equation}
	f(\theta, \theta_p) \partial_\theta Y_{lm}(\theta, \phi) = \sum_{n=-4}^4 \beta^{(n)}_{lm}(\theta_p) Y_{l+n,m}(\theta,\phi),
\end{equation}
where $\beta^{(n)}_{lm}$ is defined in Appendix \ref{app:reg}. As a result,
\begin{align} \label{eqn:FthetaRet}
	\tilde{F}^{\mathrm{ret},l}_{\theta\pm} &= \sum_{m=-l}^l  {F}^{\mathrm{ret},lm}_{\theta\pm}
	\\ \notag
	&= \sum_{m=-l}^l \sum_{n=-4}^4 \beta^{(-n)}_{l+n,m}(\theta_p) \psi^\pm_{l+n,m}(t,r)
	Y_{lm}(\theta,\phi),
\end{align}
which is amenable to mode-sum regularization.

The final step before mode-sum regularizing the SSF is to construct the retarded self-force multipole contributions along the worldline of our point charge. To do so, we follow the methods outlined in \cite{NasiEvan21}. First, we assume that our point charge is following a fiducial resonant geodesic $\hat{x}^\mu_p$ given by \eqref{eqn:geoTSol}-\eqref{eqn:geoPhiSol} with $t_0=q_{r0}=q_{\theta0}=\phi_0=0$. We then construct $F^{\mathrm{ret},l}_{\alpha\pm}$ via \eqref{eqn:FtRet}-\eqref{eqn:FphiRet} and \eqref{eqn:FthetaRet} and evaluate both ${F}^{\mathrm{ret},l}_{\alpha\pm}$ and ${F}^{\mathrm{S},l}_{\alpha\pm}$ along $\hat{x}^\mu_p$, e.g.,
\begin{align*}
    F^{\mathrm{ret},l}_{\alpha\pm}(t,r,\theta,\phi) \rightarrow F^{\mathrm{ret},l}_{\alpha\pm}(\hat{t}_p(\lambda),\hat{r}_p(\lambda),\hat{\theta}_p(\lambda),\hat{\phi}_p(\lambda)).
\end{align*}
When evaluating the mode functions along the worldline, the terms that accumulate linearly with $\lambda$ cancel, allowing us to instead parametrize $F^{\mathrm{ret},l}_{\alpha\pm}$, $F^{\mathrm{S},l}_{\alpha\pm}$, and $F_\alpha$ in terms of the phase variables $q_r$ and $q_\theta$ \cite{Vand18, NasiEvan21}. Therefore, in a slight abuse of notation, we use $\hat{F}^{\mathrm{ret},l}_{\alpha\pm}(q_r,q_\theta)$, $\hat{F}^{\mathrm{S},l}_{\alpha\pm}(q_r,q_\theta)$, and $\hat{F}_\alpha(q_r,q_\theta)$ to denote the retarded self-force multipole contributions, the singular self-force multipole contributions, and the regularized SSF evaluated along a fiducial geodesic, respectively.

To assemble the dissipative and conservative contributions to the self-force, we leverage the symmetries of Kerr spacetime to relate $\hat{F}^\mathrm{diss}_\alpha$ and $\hat{F}^\mathrm{cons}_\alpha$ to $\hat{F}_\alpha$ \cite{HindFlan08, NasiEvan21},
\begin{align} \notag
    \hat{F}^\mathrm{diss}_\alpha(q_r, q_\theta) &= \frac{1}{2}\left[\hat{F}_\alpha(q_r, q_\theta) - {\epsilon}_{(\alpha)} \hat{F}_\alpha(2\pi - q_r, 2\pi - q_\theta) \right],
    \\ \label{eqn:ssfCons}
    \hat{F}^\mathrm{cons}_\alpha(q_r, q_\theta) &= \frac{1}{2}\left[\hat{F}_\alpha(q_r, q_\theta) + {\epsilon}_{(\alpha)} \hat{F}_\alpha(2\pi - q_r, 2\pi - q_\theta) \right],
\end{align}
where ${\epsilon}_{(\alpha)} \doteq (-1, 1, 1, -1)$.

Using these fiducial self-force quantities, we can also evaluate the SSF for a resonant geodesic with nonfiducial initial conditions $q_{r0}\neq0$ and $q_{\theta0}\neq0$ via the shifting relation \cite{NasiEvan21},
\begin{align}
    F_\alpha(q_r,q_\theta; q_{r0}, q_{\theta0}) = \hat{F}_\alpha(q_r + q_{r0}, q_\theta + q_{\theta 0}).
\end{align}
The shifting relation holds for ${F}^{\mathrm{ret},l}_{\alpha\pm}(q_r,q_\theta; q_{r0}, q_{\theta0})$ and ${F}^{\mathrm{S},l}_{\alpha\pm}(q_r,q_\theta; q_{r0}, q_{\theta0})$, as well. Therefore, by computing the fiducial SSF quantities along a two-dimensional grid spanned by $q_r$ and $q_\theta$, we efficiently capture the evolution of the SSF along a whole family of resonant geodesics that share the same orbital constants and frequencies, but differ in their initial conditions.

\subsection{Secular evolution of $E$, $L_z$, and $Q$}
\label{sec:ELQdot}

Once we determine the SSF, we examine its impact on the evolution of the charge. First, we reexpress the equations of motion \eqref{eqn:ssfEOM} in terms of $\mu$, $E$, $L_z$, and $Q$,
\begin{gather} \label{eqn:dmudtau}
    \frac{d\mu}{d\tau} = - q^2 u^\alpha F_\alpha,
    \\
    \frac{dE}{d\tau} = -q^2 a_t,
    \qquad \qquad
    \frac{dL_z}{d\tau} = q^2 a_\phi,
    \\
    \frac{dQ}{d\tau} = 2q^2 K^{\mu\nu} u_\mu a_\nu  - 2q^2(L_z-aE)\left(a_\phi + a a_t \right),
\end{gather}
where $a^\mu$ is the self-acceleration of the scalar charge,
\begin{align} \label{eqn:4acceleration}
    \mu a^\mu = (g^{\mu\nu} + u^\mu u^\nu)F_\nu = F^\mu - \frac{u^\mu}{q^2} \frac{d\mu}{d\tau}.
\end{align}

The leading-order, adiabatic evolution of our system is captured by the secular rate of change of these orbital quantities, $\langle\dot{\mu}\rangle$, $\langle \dot{E} \rangle$, $\langle \dot{L}_z \rangle$, and $\langle \dot{Q} \rangle$. The brackets denote an orbit-average with respect to coordinate time,
\begin{align}
    \langle X \rangle = \frac{1}{T} \int_0^T X(t) dt,
\end{align}
where $T$ is the orbital period. We can integrate \eqref{eqn:dmudtau} analytically,\footnote{The evolution of the mass is given by $\mu(t) =\mu_0 - q \Phi^\mathrm{R}[x^\mu_p(t)]$, where $\mu_0$ is an integration constant commonly referred to as the charge's bare mass.} leading to the trivial result $\langle \dot{\mu} \rangle = 0$.  Following \cite{DrasHugh06}, we reexpress the remaining orbit-averages as integrals over Mino time. Therefore, for a scalar charge $q$ on a $r\theta$-resonant geodesic with initial orbital phase $q_{\theta 0}$,\footnote{Recall from Sec.~\ref{sec:geo} that, due to the coupling between the radial and polar motion, we can set $q_{r0}=0$ without loss of generality, provided we allow $q_{\theta 0}$ to be nonzero.} the secular evolution due to the SSF is given by
\begin{align} \label{eqn:Jdot}
    \langle \dot{\mathcal{J}} \rangle &= \frac{q^2}{\mu\Lambda \Upsilon_t} \int_0^\Lambda \hat{I}_\mathcal{J}(\Upsilon_r\lambda, \Upsilon_\theta\lambda+q_{\theta 0}) d\lambda,
\end{align}
where ${\mathcal{J}}$ is used to represent $E$, $L_z$, or $Q$, and the integrands are given by
\begin{gather} \label{eqn:IE}
    \hat{I}_{E} = -\hat{\Sigma}_p \hat{a}_t,
    \qquad \qquad 
    \hat{I}_{L_z} = \hat{\Sigma}_p \hat{a}_\phi,
    \\ \label{eqn:IQ}
    \hat{I}_{Q} = 2\hat{\Sigma}_p \Big[\hat{K}^{\mu\nu} \hat{u}_\mu \hat{a}_\nu - (L_z-aE)\left(\hat{a}_\phi + a \hat{a}_t \right)\Big].
\end{gather}
Note that all hatted quantities are evaluated along a fiducial geodesic that shares the same frequencies as the scalar charge's orbit. 

As discussed in Sec.~\ref{sec:intro}, the averages in \eqref{eqn:Jdot} will vary with the choice of initial phase $q_{\theta 0}$. This is in contrast to nonresonant orbits, which have no phase dependence. Therefore, we define the double average
\begin{equation} \label{eqn:doubleAvg}
    \langle \langle \dot{\mathcal{J}} \rangle \rangle_0 = \frac{1}{2\pi\Lambda} \int_0^{2\pi} \int_0^{\Lambda} \dot{\mathcal{J}}(\lambda; q_{\theta 0}) d\lambda\, dq_{\theta 0}
\end{equation}
to capture the piece of the secular evolution that is independent of $q_{\theta 0}$. Computing this double average is equivalent to computing $\langle\dot{\mathcal{J}}\rangle$ by naively assuming that our resonant geodesics are nonresonant. The phase dependence of our resonant averages is then captured by the \textit{residual averages},
\begin{align} \label{eqn:resAvg}
   \langle \delta \dot{\mathcal{J}} \rangle = \langle \dot{\mathcal{J}} \rangle - \langle \langle \dot{\mathcal{J}} \rangle \rangle_0.
\end{align}

To compute these averages, we first take into account that our orbit averages in \eqref{eqn:Jdot} will depend on both $F^\mathrm{diss}_\alpha$ and $F^\mathrm{cons}_\alpha$. Therefore, any quantity that depends on the self-force can be decomposed into a conservative contribution and a dissipative contribution, e.g.,
\begin{align}
    \langle \delta \dot{Q} \rangle = \langle \delta \dot{Q} \rangle^\mathrm{diss} + \langle \delta \dot{Q} \rangle^\mathrm{cons}.
\end{align}
Henceforth, quantities with a ``cons'' label are calculated using only $F^\mathrm{cons}_\alpha$, while any quantity with a ``diss'' label is calculated using only $F^\mathrm{diss}_\alpha$. Furthermore, we can decompose quantities into their contributions from both $F^{\mathrm{ret},l}_{\alpha\pm}$ and $F^{\mathrm{S},l}_{\alpha\pm}$, e.g.,
\begin{align}
    \langle \delta \dot{Q} \rangle^\mathrm{cons} =\left[\sum_{l=0}^\infty \left( \langle \delta \dot{Q} \rangle^{\mathrm{ret},l}_{\pm} - \langle \delta \dot{Q} \rangle^{\mathrm{S},l}_\pm \right)\right] - \langle \delta \dot{Q} \rangle^{\mathrm{diss}}.
\end{align}
These decompositions allow us to effectively mode-sum regularize the secular quantities themselves, rather than regularizing the self-force and then computing the averaged rates of change. For example, we find that regularizing $\langle \delta \dot{Q} \rangle^\mathrm{cons}$ significantly reduces systematic uncertainties in our final results, as we will discuss in Sec.~\ref{sec:comp}.

Based on the symmetries of the conservative perturbations [see \eqref{eqn:ssfCons}] and global flux-balance arguments \cite{Galt82, Mino03, QuinWald99}, $\langle \dot{E} \rangle^\mathrm{cons} = \langle \dot{L}_z\rangle^\mathrm{cons} = \langle \langle \dot{Q} \rangle \rangle_0^\mathrm{cons} = 0$. This is in agreement with nonresonant orbits, where conservative perturbations have no impact on the leading-order secular evolution. However, as discussed in \cite{IsoyETC13,FujiETC17,IsoyETC19}, during $r\theta$ resonances these same symmetry and flux-balance arguments no longer guarantee that $\langle \delta \dot{Q} \rangle^\mathrm{cons}$ will vanish. Therefore, we compute $\langle \delta \dot{Q} \rangle^\mathrm{cons}$ for a scalar charge on several different $r\theta$-resonant orbits.

\section{Numerical methods and implementation}
\label{sec:comp}

To carry out our calculations we developed a new numerical code in \texttt{C++}, which we will refer to as CPP. CPP consists of a \textit{driver program} that calculates, for a scalar charge following a resonant geodesic, the average rate of change of its orbital energy, angular momentum, and Carter constant---$\langle \dot{E} \rangle$, $\langle \dot{L}_z \rangle$, and $\langle \dot{Q} \rangle$---due to the SSF. The {driver program} calls on eight separate modules, each of which implements a different piece of the self-force problem outlined in Sec.~\ref{sec:background}:
\begin{enumerate}
    \item a \textit{geodesic module} that determines the background motion of a perturbing particle,
    
    \item a \textit{harmonic module} that evaluates the frequency-domain harmonics of the self-force experienced by a perturbing particle,
    
    \item a \textit{spheroidal harmonic module} that solves for the spheroidal harmonics,
    
    \item a \textit{radial Teukolsky module} that solves for the homogeneous radial Teukolsky solutions,
    
    \item a \textit{source integration module} that calculates the normalization constants (Teukolsky amplitudes) due to the presence of a perturbing point-source, 
    
    \item a \textit{self-force module} that sums over the harmonics to construct unregularized modes of the self-force,
    
    \item a \textit{secular evolution module} that determines the averaged rate of change of the background orbital quantities due to a perturbing force, and
    
    \item \label{item:last} a \textit{regularization module} that regularizes divergent self-force data.
\end{enumerate}

One advantage of writing this new code is that CPP, which only relies on floating-point arithmetic at double precision, is much faster and less memory-intensive than our old arbitrary-precision \texttt{Mathematica} code, which we used in previous investigations of the SSF \cite{NasiOsbuEvan19, NasiEvan21}. From here on, we refer to the \texttt{Mathematica} code as MMA. With CPP, a typical SSF calculation (for a single resonance with all initial conditions taken into account) completes in $\sim 300$ CPU hours on a laptop with a 2.4 GHz 8-core Intel Core i9 processor. As a result, SSF calculations complete in less than a day when mode computations are distributed across the laptop's cores with MPI \cite{MPI}. For comparison, comparable SSF calculations took over 15,000 CPU hours with MMA, and had to be performed in parallel on a cluster for several weeks. Another benefit of CPP is that is relies on freely available, open-access software---such as GSL \cite{GSL,GSL09} and Boost \cite{BOOST}---making it far more accessible than MMA, which relies on proprietary software.

Another advantage is that, due to its modular design, CPP can easily be generalized to the problem of calculating the GSF in radiation gauges. In fact, our spheroidal harmonic, radial Teukolsky, and source integration modules have already been generalized to handle gravitational perturbations. However, for the purposes of this work, we will only focus on the numerical implementation of the SSF problem. In Sec.~\ref{sec:driver} we describe the CPP driver program and how it generates the results presented in Sec.~\ref{sec:fluxes}, while detailed descriptions of the CPP modules are found in Appendix \ref{app:modules}. 

CPP implements many of the same numerical methods used by MMA and outlined in \cite{NasiOsbuEvan19, NasiEvan21}, but we highlight two key differences between these two codes. The first is that in MMA we determine the radial Teukolsky solutions using semianalytic series of hypergeometric functions derived by Mano, Suzuki, and Takasugi (MST) \cite{ManoSuzuTaka96a, ManoSuzuTaka96b} that are evaluated at high levels of precision. In contrast, CPP computes the radial Teukolsky solutions using a combination of routines that are better adapted to the limits of double precision, including adaptive step-size numerical integration, asymptotic expansions, and the MST series. (See \ref{app:teukModule}.) This required the development of a special functions \texttt{C++} library that computes a number of functions in the complex domain, particularly hypergeometric functions.

The second key difference is that in MMA we first regularize the SSF, and then we use the regularized data $F_\alpha$ to compute $\langle \dot{E} \rangle$, $\langle \dot{L}_z \rangle$, $\langle \dot{Q} \rangle$ as functions of the initial resonant phase $q_{\theta 0}$. Therefore, uncertainties and errors introduced to $F_\alpha$ by our MMA regularization procedure are propagated to $\langle \dot{E} \rangle$, $\langle \dot{L}_z \rangle$, $\langle \dot{Q} \rangle$. In CPP we first calculate the mode contributions to the secular rates of change using the unregularized SSF modes $\hat{F}^{\mathrm{ret},l}_{\alpha\pm}$, and then we regularize the coefficients of the Fourier series that represent our orbit-averaged quantities. (See \ref{app:secularEvolutionModule} and \ref{app:regModule}.) Directly regularizing the Fourier coefficients of $\langle \dot{E} \rangle$, $\langle \dot{L}_z \rangle$, $\langle \dot{Q} \rangle$ significantly reduces estimated uncertainties due to mode-sum regularization. We expand on this new regularization scheme in Sec.~\ref{sec:driver}.

\subsection{Driver program}
\label{sec:driver}

\subsubsection{Input}

The driver program requires the following initial inputs:
\begin{enumerate}
    \item The orbital parameters $(a, e, x)$ that describe the background resonant geodesic of our scalar charge,
    \item the resonant ratio $\beta_r/\beta_\theta$ of the radial and polar frequencies;
    \item the geodesic sampling number $N_\mathrm{geo} = 2^{n_\mathrm{geo}}$, where $n_\mathrm{geo}\in \mathbb{Z}^+$; 
    \item the SSF sampling number $N_\mathrm{SSF} = 2^{n_\mathrm{SSF}}$, where $n_\mathrm{SSF}\in \mathbb{Z}^+ < n_\mathrm{geo}$; and
    \item the maximum SSF $l$-mode $l_\mathrm{max}$.
\end{enumerate}
In this work, we typically set $n_\mathrm{geo} = 12$, $n_\mathrm{SSF} = 8$, and $l_\mathrm{max}=25$.

\subsubsection{Computing the background geodesic}

Given this input, the program first determines the value of $p$ that produces the resonance $\Upsilon_r/\Upsilon_\theta=\beta_r/\beta_\theta$. After completing the orbital inputs $(a,p,e,x)$, the program constructs the fiducial geodesic functions described in Sec.~\ref{sec:geo} using the geodesic module outlined in Appendix \ref{app:geoModule}. The functions are evaluated at $N_\mathrm{geo}$ evenly-spaced points in $q_r$ and $q_\theta$, resulting in the discretely sampled values $\Delta\hat{x}_i=\{\Delta \hat{t}^{(r)}_{i_r}, \Delta \hat{t}^{(\theta)}_{i_\theta}, \Delta \hat{r}^{(r)}_{i_r},\Delta \hat{\theta}^{(\theta)}_{i_\theta},\Delta \hat{\phi}^{(r)}_{i_r}, \Delta \hat{\phi}^{(\theta)}_{i_\theta}\}$, where $i_r,i_\theta = 0, 1, \dots, N_\mathrm{geo}-1$.

\subsubsection{Constructing SSF data}

After the program constructs the geodesic data, it calculates the unregularized SSF $l$-modes $\hat{F}^{\mathrm{ret},l}_{\alpha\pm}$ [see \eqref{eqn:modeSumReg}] using the SSF module described in Appendix \ref{app:sumModule}. The program samples these modes on an $N_\mathrm{SSF} \times N_\mathrm{SSF}$ grid in $q_r$ and $q_\theta$, leading to the discrete mode values $\hat{F}^{\mathrm{ret},l}_{\alpha\pm,j_rj_\theta}$, where the grid points are indexed by the integers $j_r,j_\theta = 0, 1, \dots, N_\mathrm{SSF}-1$.

After computing $\hat{F}^{\mathrm{ret},l}_{\alpha\pm,j_rj_\theta}$ for $0\leq l \leq l_\mathrm{max}$, the program constructs a number of quantities that are related to the SSF. First it evaluates the regularization parameters on the same $N_\mathrm{SSF} \times N_\mathrm{SSF}$ grid as the SSF using the regularization module in Appendix \ref{app:regModule}. Like $\hat{F}^{\mathrm{ret},l}_{\alpha\pm,j_rj_\theta}$, the discretely sampled parameters are referred to as $\hat{A}_{\alpha,j_rj_\theta}$, $\hat{B}_{\alpha,j_rj_\theta}$, and $\hat{D}^{(2)}_{\alpha,j_rj_\theta}$. Next, it constructs the $l$-mode contributions to the dissipative SSF using \eqref{eqn:ssfCons},
\begin{align} \notag
    \hat{F}^{\mathrm{diss},l}_{\alpha\pm,j_rj_\theta}
    &= \frac{1}{2}\Big(\hat{F}^{\mathrm{ret},l}_{\alpha\pm,j_rj_\theta} 
    -\epsilon_{(\alpha)} \hat{F}^{\mathrm{ret},l}_{\alpha\pm;N_\mathrm{SSF}-j_r-1,N_\mathrm{SSF}-1-j_\theta} \Big),
\end{align} 
which do not require any regularization.
After that, the program constructs the partially regularized self-force quantities,
\begin{align} \label{eqn:Fret-A}
    \hat{F}^{\mathrm{ret}-A,l}_{\alpha\pm,j_rj_\theta} &= \hat{F}^{\mathrm{ret},l}_{\alpha\pm,j_rj_\theta} \mp \frac{1}{2}\hat{A}_{\alpha,j_rj_\theta}(2l+1).
\end{align}
By removing the leading-order singular behavior in \eqref{eqn:Fret-A}, $\hat{F}^{\mathrm{ret}-A,l}_{\alpha+,j_rj_\theta}$ and $\hat{F}^{\mathrm{ret}-A,l}_{\alpha-,j_rj_\theta}$ should agree to machine precision in the absence of additional numerical error. Therefore, taking their difference gives us an estimate of the overall numerical error in $\hat{F}^{\mathrm{ret},l}_{\alpha\pm,j_rj_\theta}$. Furthermore, using \eqref{eqn:Fret-A}, the program constructs a third self-force quantity,
\begin{align} \label{eqn:FalphaLG}
    \hat{F}^{\mathrm{ret}-A,l}_{\alpha\lessgtr,j_rj_\theta} &= 
    \begin{cases}
        \hat{F}^{\mathrm{ret}-A,l}_{\alpha+,j_rj_\theta} & j_r > N_\mathrm{split},
        \\
        \hat{F}^{\mathrm{ret}-A,l}_{\alpha-,j_rj_\theta} & j_r < N_\mathrm{split},
        \\
        \frac{1}{2}\left(\hat{F}^{\mathrm{ret}-A,l}_{\alpha-,j_rj_\theta}+\hat{F}^{ \mathrm{ret}-A,l}_{\alpha+,j_rj_\theta} \right) & j_r = N_\mathrm{split},
    \end{cases}
\end{align}
where $N_\mathrm{split}$ is the value of $j_r$ that minimizes the difference $|\Delta \hat{r}^{(r)}_{j_r} - \sqrt{r_\mathrm{min} r_\mathrm{max}}|$.\footnote{This condition was found through numerical experimentation.} For eccentric orbits and high frequencies, we encounter large cancellations between the extended homogeneous functions when constructing the self-force. As observed by previous authors (e.g., \cite{VandShah15}), these cancellations become more severe the farther one extends the homogeneous functions into the radial libration region of the perturbing point source. Consequently, $\hat{F}^{\mathrm{ret}-A,l}_{\alpha+,j_rj_\theta}$ tends to accumulate larger numerical errors when the scalar charge is closer to $r_\mathrm{min}$, while $\hat{F}^{\mathrm{ret}-A,l}_{\alpha-,j_rj_\theta}$ suffers from larger numerical errors when the charge is closer to $r_\mathrm{max}$. Therefore, $\hat{F}^{\mathrm{ret}-A,l}_{\alpha\lessgtr,j_rj_\theta}$ makes use of each solution in the region where its numerical error is best mitigated. Similarly, the program computes the new quantity $\hat{F}^{\mathrm{diss},l}_{\alpha\lessgtr,j_rj_\theta}$ by replacing $\hat{F}^{\mathrm{ret}-A,l}_{\alpha\pm,j_rj_\theta}$ with $\hat{F}^{\mathrm{diss},l}_{\alpha\pm,j_rj_\theta}$ in \eqref{eqn:FalphaLG}. Finally, the driver program constructs the following contributions to the conservative SSF,
\begin{align}
    \hat{F}^{\mathrm{cons}+\mathrm{S}-A,l}_{\alpha\pm,j_rj_\theta}&=\hat{F}^{\mathrm{ret}-A,l}_{\alpha\pm,j_rj_\theta}-\hat{F}^{\mathrm{diss},l}_{\alpha\pm,j_rj_\theta},
    \\
    \hat{F}^{\mathrm{cons}+\mathrm{S}-A,l}_{\alpha\lessgtr,j_rj_\theta}&=\hat{F}^{\mathrm{ret}-A,l}_{\alpha\lessgtr,j_rj_\theta}-\hat{F}^{\mathrm{diss},l}_{\alpha\lessgtr,j_rj_\theta}.      
\end{align}

\subsubsection{Evaluating the dissipative averages}

Next, the program uses the secular evolution module in Appendix \ref{app:secularEvolutionModule} to construct the orbit averages $\langle \dot{\mathcal{J}} \rangle^\mathrm{diss} = \langle \langle \dot{\mathcal{J}}\rangle\rangle_0^\mathrm{diss}+\langle \delta \dot{\mathcal{J}}\rangle^\mathrm{diss}$ defined in \eqref{eqn:doubleAvg} and \eqref{eqn:resAvg}. Rather than directly evaluating $\langle \dot{\mathcal{J}} \rangle^\mathrm{diss}$ for each value of $q_{\theta 0}$, the secular evolution module represents each average as a Fourier series with respect to $q_{\theta 0}$ and, given some self-force data, directly computes the Fourier coefficients of each series.

For the dissipative averages, the Fourier coefficients are purely real and the Fourier sums reduce to discrete cosine series,
\begin{align} \label{eqn:XDiss0}
    \langle \langle \dot{\mathcal{J}} \rangle \rangle^\mathrm{diss}_0 &= \frac{q^2}{\mu} \tilde{\mathcal{F}}_{0}^{\mathcal{J},\mathrm{diss}},
    \\
    \label{eqn:deltaXDiss}
    \langle \delta \dot{\mathcal{J}} \rangle^\mathrm{diss} &= \frac{2q^2}{\mu}\sum_{k = 1}^{K_\mathcal{J}} \tilde{\mathcal{F}}_{k}^{\mathcal{J},\mathrm{diss}} \cos{kq_{\theta 0}},
    \\ \label{eqn:FourierDissSum}
    \tilde{\mathcal{F}}_{k}^{\mathcal{J},\mathrm{diss}} &= \sum_{l=0}^{L_\mathcal{J}} \tilde{\mathcal{F}}_{k,l}^{\mathcal{J},\mathrm{diss}}.
\end{align}
The secular evolution module constructs three estimates for the Fourier coefficients $\tilde{\mathcal{F}}_{k,l}^{\mathcal{J},\mathrm{diss}}$ using the self-force $l$-modes $\hat{F}^{\mathrm{diss},l}_{\alpha\pm}$ and $ \hat{F}^{\mathrm{diss},l}_{\alpha\lessgtr}$. The driver program then uses the values of $L_\mathcal{J}$ and $K_\mathcal{J}$ and the set of coefficients $\tilde{\mathcal{F}}_{k,l}^{\mathcal{J},\mathrm{diss}}$ that minimize the total estimated uncertainty in $\langle \langle \dot{\mathcal{J}} \rangle \rangle^\mathrm{diss}_0$ and $\langle \delta \dot{\mathcal{J}} \rangle^\mathrm{diss}$. 

This uncertainty arises due to several sources of numerical error: (1) $\tilde{\sigma}^{\mathcal{J},\mathrm{SSF}}_{k,l\pm}$, the estimated error in $\tilde{\mathcal{F}}_{k,l}^{\mathcal{J},\mathrm{diss}}$ due to the numerical error in $\hat{F}^{\mathrm{diss},l}_{\alpha}$; (2) $\tilde{\sigma}^{\mathcal{J},\mathrm{diss}}_{k,l\pm}$, the estimated error in $\tilde{\mathcal{F}}_{k,l}^{\mathcal{J},\mathrm{diss}}$ introduced by the numerical methods in the secular evolution module; (3) $\tilde{\sigma}^{\mathcal{J},\mathrm{diss}}_{k,\mathrm{trunc}}$, the estimated error from truncating the sum over $l$ in \eqref{eqn:FourierDissSum}; and (4) ${\sigma}^{\mathcal{J},\mathrm{diss}}_{\mathrm{trunc}}$, the estimated error from truncating the sum over $k$ in \eqref{eqn:deltaXDiss}. Based on the convergence criteria established in our SSF module, we expect our dissipative SSF data to be accurate to a precision of $\sim \epsilon_\mathrm{tol} = 10^{-10}$. Assuming that this produces a similar error of error in the Fourier coefficients leads to the conservative estimate
\begin{align}  \label{eqn:sigmaKdissSSF}
    \tilde{\sigma}^{\mathcal{J},\mathrm{SSF}}_{k,l\pm} \approx (5\times \epsilon_\mathrm{tol}) \times |\tilde{\mathcal{F}}_{0,l\pm}^{\mathcal{J},\mathrm{diss}}|.
\end{align}
The uncertainty $\tilde{\sigma}^{\mathcal{J},\mathrm{diss}}_{k,l\pm}$, on the other hand, is computed by the secular evolution module as described in Appendix \ref{app:secularEvolutionModule}. Meanwhile, the program estimates $\tilde{\sigma}^{\mathcal{J},\mathrm{diss}}_{k,\mathrm{trunc}}$ to be the maximum of $\tilde{\sigma}^{\mathcal{J},\mathrm{diss}}_{k,L_\mathcal{J}}$, $\tilde{\sigma}^{\mathcal{J},\mathrm{SSF}}_{k,L_\mathcal{J}}$, and $|\tilde{\mathcal{F}}_{k,L_\mathcal{J}}^{\mathcal{J},\mathrm{diss}} + \tilde{\mathcal{F}}_{k,L_\mathcal{J}+1}^{\mathcal{J},\mathrm{diss}}|$. To get an intermediate estimate for the uncertainty in $\tilde{\mathcal{F}}_{k}^{\mathcal{J},\mathrm{diss}}$, denoted by $\tilde{\sigma}^{\mathcal{J},\mathrm{diss}}_{k}$, the driver program use the standard method of propagation of error and adds the contributing errors in quadrature,
\begin{align} \label{eqn:sigmaKdiss}
    \left(\tilde{\sigma}^{\mathcal{J},\mathrm{diss}}_{k}\right)^2 &= \left(\tilde{\sigma}^{\mathcal{J},\mathrm{diss}}_{k,\mathrm{trunc}}\right)^2 
    \\
    \notag
    & \qquad \qquad+ \sum_{l=0}^{L_\mathcal{J}}\left[ \left(\tilde{\sigma}^{\mathcal{J},\mathrm{diss}}_{k,l}\right)^2 + \left(\tilde{\sigma}^{\mathcal{J},\mathrm{SSF}}_{k,l}\right)^2\right].
\end{align}
Then the final truncation error ${\sigma}^{\mathcal{J},\mathrm{diss}}_{\mathrm{trunc}}$ is taken to be the maximum value of $\tilde{\sigma}^{\mathcal{J},\mathrm{diss}}_{K_\mathcal{J}}$ and $|\tilde{\mathcal{F}}^{\mathcal{J},\mathrm{diss}}_{K_\mathcal{J}}|$. Consequently, the program approximates the uncertainty in $\langle \delta \dot{\mathcal{J}} \rangle^\mathrm{diss}$ to be
\begin{align} \label{eqn:sigmaDeltaXDiss}
    \left(\sigma^{\mathrm{diss}}_{\delta \mathcal{J}}\right)^2 &= \frac{4q^4}{\mu^2}\left[ \left(\tilde{\sigma}^{\mathcal{J},\mathrm{diss}}_{\mathrm{trunc}}\right)^2
    + \sum_{k = 1}^{K_\mathcal{J}} \left\vert\tilde{\sigma}^{\mathcal{J},\mathrm{diss}}_{k}\cos kq_{\theta0}\right\vert^2 \right],
\end{align}
while the uncertainty in $\langle \langle \dot{\mathcal{J}} \rangle \rangle^\mathrm{diss}_0$ is given by $\sigma^{\mathrm{diss}}_{\mathcal{J}_0} = q^2 \tilde{\sigma}^{\mathcal{J},\mathrm{diss}}_{0}/\mu$. We find that the self-force data $ \hat{F}^{\mathrm{diss},l}_{\alpha\lessgtr}$ tends to best minimize the uncertainties $\sigma^{\mathrm{diss}}_{\mathcal{J}_0}$ and $\sigma^{\mathrm{diss}}_{\delta \mathcal{J}}$.

\subsubsection{Evaluating the conservative averages}

The conservative averages are efficiently described by discrete sine series,
\begin{align} \label{eqn:deltaXCons}
    \langle \delta \dot{\mathcal{J}} \rangle^\mathrm{cons} &= \frac{2q^2}{\mu}\mathrm{Im}\left[\sum_{k = 1}^{K_\mathcal{J}} \tilde{\mathcal{F}}^{\mathcal{J},\mathrm{cons}}_k \sin{kq_{\theta 0}}\right],
\end{align}
where the Fourier coefficients are purely imaginary and $\langle \langle \dot{\mathcal{J}}\rangle\rangle_0^\mathrm{cons} = 0$. Once again, the driver program first 
uses the secular evolution module to calculate the $l$-mode contributions to $\tilde{\mathcal{F}}^{\mathcal{J},\mathrm{cons}}_k$ 
using the SSF data $\hat{F}^{\mathrm{cons}+\mathrm{S}-A,l}_{\alpha\pm}$ and $\hat{F}^{\mathrm{cons}+\mathrm{S}-A,l}_{\alpha\lessgtr}$. Like the dissipative case, this leads to three estimates for the $l$-dependent Fourier coefficients $\tilde{\mathcal{F}}_{k,l}^{\mathcal{J},\mathrm{cons}+\mathrm{S}-A}$. However, unlike the dissipative case, naively summing over $\tilde{\mathcal{F}}_{k,l}^{\mathcal{J},\mathrm{cons}+\mathrm{S}-A}$ leads to a divergent result and, thus, these coefficients need to be regularized. (See Figs.~\ref{fig:consConvergence} and \ref{fig:consConvergence2}.) Much like the SSF itself, these Fourier coefficients are amenable to mode-sum regularization. Therefore, the program uses the secular evolution module to generate Fourier coefficient regularization parameters, $\tilde{\mathcal{F}}_{k}^{\mathcal{J},B}$ and $\tilde{\mathcal{F}}_{k}^{\mathcal{J},D}$, by replacing SSF data with the regularization parameters $\hat{B}_{\alpha}$ and $\hat{D}^{(2)}_{\alpha}$ as input.

\begin{figure*}[th!]
    \centering
    \includegraphics[width=0.995\linewidth]{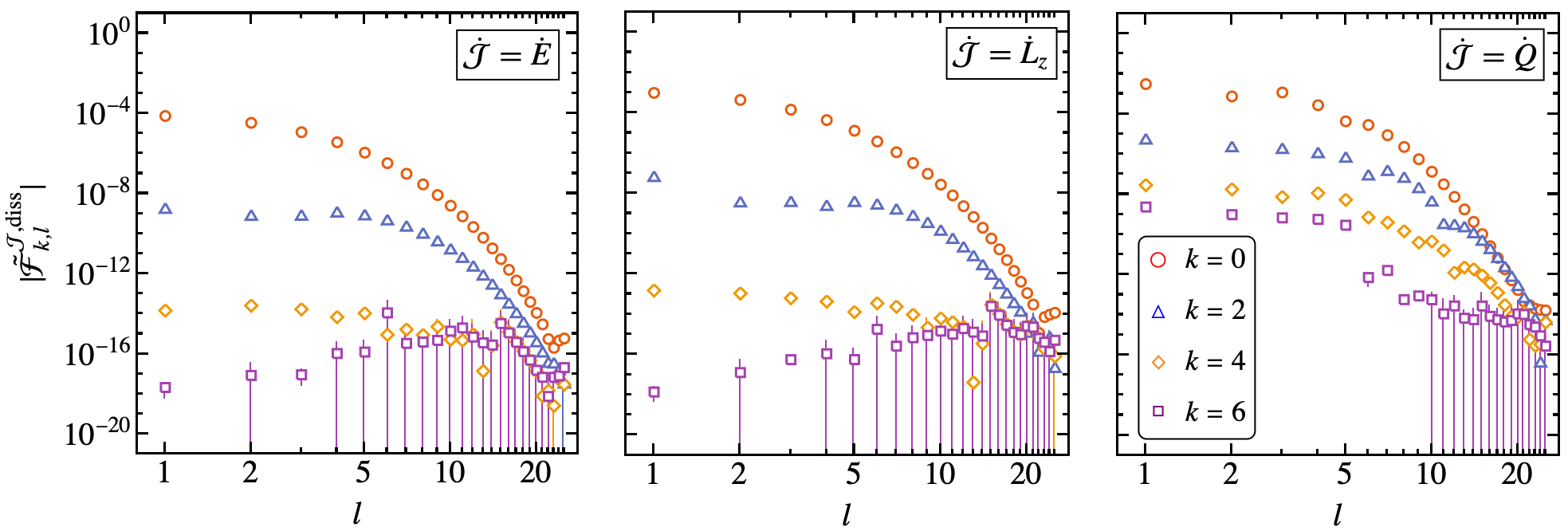}
    \caption{The $l$-mode convergence of the Fourier coefficients $\tilde{\mathcal{F}}_{k,l}^{\mathcal{J},\mathrm{diss}}$ that describe the dissipative averages $\langle \dot{\mathcal{J}}\rangle^\mathrm{diss}$---where $\mathcal{J}$ represents $E$, $L_z$, or $Q$---for a 2:3 resonant orbit defined by the parameters $(a/M,p,e,x)=(0.9,6.643,0.2,\cos\pi/4)$. Because the odd $k$-modes vanish, we only plot the even modes $k=(0,2,4,6)$. The magnitude of the coefficients rapidly decay with both $l$ and $k$. Therefore, we see that modes with $k>2$ tend to be dominated by numerical error, especially at higher values of $l$.}
    \label{fig:dissConvergence}
\end{figure*}

The program then submits $\tilde{\mathcal{F}}_{k,l\pm}^{\mathcal{J},\mathrm{cons}+\mathrm{S}-A}$, $\tilde{\mathcal{F}}_{k,l\lessgtr}^{\mathcal{J},\mathrm{cons}+\mathrm{S}-A}$, $\tilde{\mathcal{F}}_{k}^{\mathcal{J},B}$, and $\tilde{\mathcal{F}}_{k}^{\mathcal{J},D}$ to the regularization module described in Appendix \ref{app:regModule}. This module produces three estimates for the regularized Fourier coefficients $\tilde{\mathcal{F}}^{\mathcal{J},\mathrm{cons}}_k$. The regularization module also outputs estimated uncertainties for these regularized coefficients, which we refer to as $\tilde{\sigma}^{\mathcal{J},\mathrm{cons}}_{k}$. While the conservative averages are subject to all of the same errors as the dissipative averages, the uncertainty introduced by regularization, $\tilde{\sigma}^{\mathcal{J},\mathrm{cons}}_{k}$, dominates other sources of error, such as the error introduced by the secular evolution module, $\tilde{\sigma}^{\mathcal{J},\mathrm{cons}}_{k,l}$, and the numerical error introduced by the SSF data, $\tilde{\sigma}^{\mathcal{J},\mathrm{SSF}}_{k,l}$. Therefore, the program simply approximates the uncertainty in $\langle \delta \dot{\mathcal{J}} \rangle^\mathrm{cons}$ as
\begin{align} \label{eqn:sigmaXCons}
    \left(\sigma^{\mathrm{cons}}_{\delta \mathcal{J}}\right)^2 &= \frac{4q^4}{\mu^2}\Bigg[ \left(\tilde{\sigma}^{\mathcal{J},\mathrm{cons}}_{\mathrm{trunc}}\right)^2
    + \sum_{k = 1}^{K_\mathcal{J}}\left\vert \tilde{\sigma}^{\mathcal{J},\mathrm{cons}}_{k}\sin{kq_{\theta 0}}\right\vert^2 \Bigg],
\end{align}
where, as before, the uncertainty due to truncating our Fourier series, $\tilde{\sigma}^{\mathcal{J},\mathrm{cons}}_{\mathrm{trunc}}$, is taken to be the maximum value of $\tilde{\sigma}^{\mathcal{J},\mathrm{cons}}_{K_\mathcal{J}}$ and $|\tilde{\mathcal{F}}^{\mathcal{J},\mathrm{cons}}_{K_\mathcal{J}}|$. Likewise, the program uses the value of $K_\mathcal{J}$ and the SSF data that minimize $\sigma^{\mathrm{cons}}_{\delta \mathcal{J}}$. We find that this is usually best satisfied by $\hat{F}^{\mathrm{cons}+\mathrm{S}-A,l}_{\alpha\lessgtr}$ and $K_\mathcal{J} = 2$.

\section{Secular growth of the orbital constants during resonances}
\label{sec:fluxes}

We use the CPP code outlined in Sec.~\ref{sec:comp} to calculate the average rate of change of $E$, $L_z$, and $Q$ due to the SSF for the resonant orbits listed in Table \ref{tab:sources}. From here on we use $\mathcal{J}$ to denote $E$, $L_z$, or $Q$. In Sec.~\ref{sec:val1} we verify the accuracy of our dissipative averages, $\langle \dot{\mathcal{J}}\rangle^\mathrm{diss}$, by analyzing their convergence and comparing them to the asymptotic behavior of the scalar field. Then in Sec.~\ref{sec:val2} we validate the conservative averages, $\langle\delta \dot{\mathcal{J}}\rangle^\mathrm{cons}$, by analyzing their convergence from mode-sum regularization and verifying that $\langle\delta \dot{E}\rangle^\mathrm{cons}$ and $\langle \delta \dot{L}_z\rangle^\mathrm{cons}$ vanish as we expect from flux-balance arguments. After validating our new methods and code, we demonstrate in Sec.~\ref{sec:QdotCons} that $\langle \delta \dot{Q}\rangle^\mathrm{cons}$ does not vanish for the resonances considered in this work. Furthermore we estimate the error that is introduced to the leading-order evolution if we neglect these conservative contributions. To simplify notation, we set $M=q=1$ for the remainder of this section.

\renewcommand{\arraystretch}{1.5}
\begin{table}[h!]
    \centering
    \caption{The orbital parameters of the resonant geodesics studied in Sec.~\ref{sec:fluxes}. The values of $p$ are truncated at four significant digits for brevity. The integers $\beta_r$ and $\beta_\theta$ are defined by the relation $\beta_r/\beta_\theta = \Upsilon_r/\Upsilon_\theta$. }
    \label{tab:sources}
    \begin{tabular*}{\linewidth}{c @{\extracolsep{\fill}} ccccc}
        \hline
        \hline
        label & $a/M$ & $p$ & $e$ & $x$ & $\beta_r/\beta_\theta$ \\
        \hline
        $e02.12$ & 0.9 & 4.508 & 0.2 & $\cos\pi/4$ & $1/2$ \\
        $e02.23$ & 0.9 & 6.643 & 0.2 & $\cos\pi/4$ & $2/3$ \\
        $e05.12$ & 0.9 & 4.607 & 0.5 & $\cos\pi/4$ & $1/2$ \\
        $e05.23$ & 0.9 & 6.707 & 0.5 & $\cos\pi/4$ & $2/3$ \\
        \hline
        \hline
    \end{tabular*}
\end{table}

\subsection{Validation tests: Dissipative perturbations}
\label{sec:val1}

First we verify that the coefficients $\tilde{\mathcal{F}}_{k,l}^{\mathcal{J},\mathrm{diss}}$ converge exponentially with $l$. We find that the $k=0$ and $k=2$ modes consistently exhibit exponential convergence for each orbit in Table \ref{tab:sources}, while higher-order $k$-modes are dominated by numerical noise at larger values of $l$. To illustrate this behavior, in Fig.~\ref{fig:dissConvergence} we plot the magnitude of $\tilde{\mathcal{F}}_{k,l}^{\mathcal{J},\mathrm{diss}}$ for the $e02.23$ orbit in Table \ref{tab:sources}. The error bars represent the estimated uncertainty in each coefficient. Note that we do not include any odd $k$-modes, because they vanish due to the symmetries of Kerr spacetime. We see that the higher modes possess large uncertainties and do not exponentially decay but instead vary around $\sim 10^{-14}$. Ultimately, this behavior is consistent with the numerical accuracy of our SSF results, which we expect to be accurate down to $\sim 10^{-13}$ based on our code's mode-sum convergence criteria outlined in Appendix \ref{app:sumModule}. Furthermore, we see that the mode content of the dissipative averages is largely captured by the $k=0$ and $k=2$ Fourier modes, while the higher modes are suppressed by several orders of magnitude. Therefore, our numerical errors in the higher modes do not have a significant impact on our calculations of $\langle \dot{\mathcal{J}} \rangle^\mathrm{diss}$.

Next, we validate our calculations of $\langle \dot{\mathcal{J}} \rangle^\mathrm{diss}$ using a standard flux-balance comparison \cite{Galt82, Mino03, QuinWald99}. Due to global conservation laws, the average rate of change of $E$ and $L_z$---which we can interpret as the local work and torque performed on our charge by the SSF---must be balanced by the flux of energy and angular momentum radiated by the scalar field at infinity, $\langle \dot{E} \rangle^\infty_\mathrm{rad}$ and $\langle \dot{L}_z \rangle^\infty_\mathrm{rad}$, and the horizon, $\langle \dot{E} \rangle^\mathcal{H}_\mathrm{rad}$ and $\langle \dot{L}_z \rangle^\mathcal{H}_\mathrm{rad}$, 
\begin{gather} \label{eqn:EdotBalance}
    \langle \dot{E} \rangle = - \left(\langle \dot{E} \rangle^\mathcal{H}_\mathrm{rad} + \langle \dot{E} \rangle^\infty_\mathrm{rad}\right) = - \langle \dot{E} \rangle_\mathrm{rad},
    \\
    \langle \dot{L}_z \rangle = -\left(\langle \dot{L}_z \rangle^\mathcal{H}_\mathrm{rad} + \langle \dot{L}_z \rangle^\infty_\mathrm{rad}\right) = -\langle \dot{L}_z \rangle_\mathrm{rad}.
\end{gather}
While a flux-balance law does not exist for the Carter constant, we can still relate $\langle \dot{Q} \rangle^\mathrm{diss}$ to the asymptotic behavior of the scalar field at the horizon and infinity \cite{DrasFlanHugh05,SagoETC06,FlanHughRuan14,IsoyETC13}, which we refer to as $\langle \dot{Q} \rangle^\mathcal{H}_\mathrm{rad}$ and $\langle \dot{Q} \rangle^\infty_\mathrm{rad}$,\footnote{Despite the suggestive naming, we emphasize that $\langle \dot{Q} \rangle^\mathcal{H}_\mathrm{rad}$ and $\langle \dot{Q} \rangle^\infty_\mathrm{rad}$ are not fluxes.} leading to a similar condition
\begin{gather} \label{eqn:QdotBalance}
    \langle \dot{Q} \rangle = - \left(\langle \dot{Q} \rangle^\mathcal{H}_\mathrm{rad} + \langle \dot{Q} \rangle^\infty_\mathrm{rad}\right) = - \langle \dot{Q} \rangle_\mathrm{rad},
\end{gather}

\renewcommand{\arraystretch}{1.5}
\begin{table*}[h!t]
    \centering
    \caption{The third and fifth columns list values for the Fourier coefficients $\tilde{\mathcal{F}}_{k}^{E,\mathrm{diss}}$ and $\tilde{\mathcal{F}}_{k}^{L_z,\mathrm{diss}}$---which describe the dissipative averages $\langle \dot{E} \rangle^\mathrm{diss}$ and $\langle \dot{L}_z \rangle^\mathrm{diss}$, respectively---for the resonances listed in Table \ref{tab:sources}. In parentheses we report the last significant digit of each coefficient and its estimated error. For example, $-2.184(3\pm3)\times 10^{-9}$ is equivalent to $(-2.1843\times 10^{-9}) \pm (3\times 10^{-13})$, while $(0.5\pm 2.8) \times 10^{-13}$ is equivalent to $(5 \times 10^{-14}) \pm (2.8 \times 10^{-13})$. We include additional significant figures in our error estimates when they are larger than the magnitude of our coefficient values. The fourth and sixth columns then report the absolute difference between $\tilde{\mathcal{F}}_{k}^{\mathcal{J},\mathrm{diss}}$ and $\tilde{\mathcal{F}}_{k}^{\mathcal{J},\mathrm{rad}}$, the Fourier coefficients of the asymptotic averages given in \eqref{eqn:XdotRad}. Note that to properly compare these coefficients we add them together due to the minus sign in \eqref{eqn:EdotBalance}.}
    \label{tab:fluxBalance}
    \begin{tabular*}{\linewidth}{c @{\extracolsep{\fill}} ccccc}
        \hline
        \hline
         orbit & $k$ & $\tilde{\mathcal{F}}_{k}^{E,\mathrm{diss}}$ & $|\tilde{\mathcal{F}}_{k}^{E,\mathrm{diss}}+\tilde{\mathcal{F}}_{k}^{E,\mathrm{rad}}|$ & $\tilde{\mathcal{F}}_{k}^{L_z,\mathrm{diss}}$ & $|\tilde{\mathcal{F}}_{k}^{L_z,\mathrm{diss}}+\tilde{\mathcal{F}}_{k}^{L_z,\mathrm{rad}}|$\\
         \hline
         $e02.12$ & 0 & $-5.53491867(1\pm9)\phantom{.0}\times10^{-4\phantom{0}}$ & $2.1\times 10^{-13}$ & $-4.14427631(8\pm5)\phantom{.0}\times10^{-3\phantom{0}}$ & $2.3\times10^{-12}$ \\
         & 2 & $-2.184(3\pm3)\phantom{000.000}\times 10^{-9\phantom{0}}$ & $5.2\times10^{-14}$ & $-3.028(1\pm2)\phantom{00.0000}\times10^{-8\phantom{0}}$ & $2.1\times 10^{-13}$  \\
         & 4 & $-(1\pm4)\phantom{.00000.00000}\times 10^{-13}$  & $1.3 \times 10^{-13}$ & $\phantom{+}(0.08\pm2.07)\phantom{000000}\times10^{-13}$ & $2.7\times10^{-15}$ \\
         \hline
         $e02.23$ & 0 & $-1.317244272(8\pm7)\phantom{.}\times10^{-4\phantom{0}}$ & $1.2\times 10^{-14}$ & $-1.674686519(5\pm8)\phantom{.}\times10^{-3\phantom{0}}$ & $4.4\times10^{-13}$ \\
         & 2 & $\phantom{+}4.2406(7\pm7)\phantom{.00000}\times 10^{-9\phantom{0}}$ & $1.4\times10^{-14}$ & $\phantom{+}6.6172(1\pm8)\phantom{.00000}\times10^{-8\phantom{0}}$ & $6.1\times 10^{-14}$  \\
         & 4 & $-(3\pm7)\phantom{.00.40674272}\times 10^{-14}$  & $2.3 \times 10^{-15}$ & $\phantom{+}(2\pm8)\phantom{.00000.00000}\times10^{-13}$ & $4.1\times10^{-15}$ \\
         \hline
         $e05.12$ & 0 & $-5.86478(5\pm5)\phantom{.0000}\times10^{-4\phantom{0}}$ & $3.3\times 10^{-10}$ & $-3.56366(6\pm2)\phantom{.0000}\times10^{-3\phantom{0}}$ & $1.4\times10^{-9\phantom{0}}$ \\
         & 2 & $-5.(2\pm1)\phantom{.000000000}\times 10^{-8\phantom{0}}$ & $3.1\times10^{-10}$ & $-7.5(8\pm3)\phantom{.00000000}\times10^{-7\phantom{0}}$ & $1.2\times 10^{-9\phantom{0}}$  \\
         & 4 & $-(0.01\pm3.57)\phantom{000000}\times 10^{-9\phantom{0}}$  & $4.8 \times 10^{-12}$ & $\phantom{+}(0.02\pm8.57)\phantom{000000}\times10^{-9\phantom{0}}$ & $9.9\times10^{-12}$ \\
         \hline
         $e05.23$ & 0 & $-1.293567(9\pm2)\phantom{.000}\times10^{-4\phantom{0}}$ & $1.6\times 10^{-13}$ & $-1.3291823(9\pm2)\phantom{.00}\times10^{-3\phantom{0}}$ & $1.3\times10^{-11}$ \\
         & 2 & $\phantom{+}5.09(2\pm3)\phantom{.0000000}\times 10^{-8\phantom{0}}$ & $6.3\times10^{-12}$ & $\phantom{+}7.516(8\pm6)\phantom{.000000}\times10^{-7\phantom{0}}$ & $4.4\times 10^{-12}$  \\
         & 4 & $-(0.3\pm2.8)\phantom{00000000}\times 10^{-12}$  & $4.6 \times 10^{-14}$ & $\phantom{+}(3\pm6)\phantom{.00.00000000}\times10^{-11}$ & $4.6\times10^{-14}$ \\
         & 6 & $-(0.01\pm5.49)\phantom{000000}\times 10^{-12}$  & $1.4 \times 10^{-14}$ & $\phantom{+}(0.005\pm4.971)\phantom{0000}\times10^{-12}$ & $9.1\times10^{-16}$ \\
         \hline
         \hline
    \end{tabular*}
\end{table*}

\renewcommand{\arraystretch}{1.5}
\begin{table*}
    \centering
    \caption{The third and fifth columns list values for the Fourier coefficients $\tilde{\mathcal{F}}_{k}^{Q,\mathrm{diss}}$ and $\mathrm{Im}[\tilde{\mathcal{F}}_{k}^{Q,\mathrm{cons}}]$---which describe the averages $\langle \dot{Q} \rangle^\mathrm{diss}$ and $\langle \dot{Q} \rangle^\mathrm{cons}$, respectively---for the resonances listed in Table \ref{tab:sources}. We report the estimated uncertainty in each coefficient using the same conventions described in Table \ref{tab:fluxBalance}. The fourth column reports the absolute difference between $\tilde{\mathcal{F}}_{k}^{Q,\mathrm{diss}}$ and $\tilde{\mathcal{F}}_{k}^{Q,\mathrm{rad}}$.}
    \label{tab:Qbalance}
    \begin{tabular*}{\linewidth}{c @{\extracolsep{\fill}} cccc}
        \hline
        \hline
         orbit & $k$ & $\tilde{\mathcal{F}}_{k}^{Q,\mathrm{diss}}$ & $|\tilde{\mathcal{F}}_{k}^{Q,\mathrm{diss}}+\tilde{\mathcal{F}}_{k}^{Q,\mathrm{rad}}|$ & $\mathrm{Im}[\tilde{\mathcal{F}}_{k}^{Q,\mathrm{cons}}]$ \\
         \hline
         $e02.12$ & 0 & $-1.32198762(0\pm2)\phantom{.}\times10^{-2\phantom{0}}$ & $8.7\times10^{-12}$ & 0 \\
         & 2 & $\phantom{+}6.345(0\pm7)\phantom{.00000}\times10^{-8\phantom{0}}$ & $1.9\times10^{-12}$ & $\phantom{+}2.5(8\pm1)\phantom{..0}\times 10^{-8\phantom{0}}$ \\
         & 4 & $-(0.8\pm6.6)\phantom{0000000}\times10^{-12}$ & $1.8\times10^{-13}$ & $\phantom{+}(2\pm7)\phantom{..00.0} \times 10^{-11}$ \\
         \hline
         $e02.23$ & 0 & $-6.33151483(7\pm3)\phantom{.}\times10^{-3\phantom{0}}$ & $1.1\times10^{-13}$ & 0 \\
         & 2 & $-2.0993(2\pm3)\phantom{.0000}\times10^{-7\phantom{0}}$ & $1.5\times10^{-13}$ & $-7.3(1\pm2)\phantom{..}\times 10^{-8\phantom{0}}$ \\
         & 4 & $-(3\pm3)\phantom{.00000.0000}\times10^{-12}$ & $8.9\times10^{-14}$ & $-(0.1\pm1.2)\phantom{.} \times 10^{-10}$ \\
         \hline
         $e05.12$ & 0 & $-1.177592(6\pm9)\phantom{.00}\times10^{-2\phantom{0}}$ & $1.1\times10^{-9\phantom{0}}$ & 0 \\
         & 2 & $\phantom{+}2.0(8\pm1)\phantom{.0000000}\times10^{-6\phantom{0}}$ & $1.6\times10^{-10}$ & $\phantom{+}(1\pm3)\phantom{..0.0}\times 10^{-6\phantom{0}}$ \\
         & 4 & $-(0.1\pm3.0)\phantom{0000000}\times10^{-8\phantom{0}}$ & $7.8\times10^{-10}$ & 
         $-$ \\
         \hline
         $e05.23$ & 0 & $-5.059826(2\pm3)\phantom{.00}\times10^{-3\phantom{0}}$ & $1.0\times10^{-10}$ & 0 \\
         & 2 & $-2.470(7\pm2)\phantom{.00000}\times10^{-6\phantom{0}}$ & $1.6\times10^{-10}$ & $-1.(0\pm2)\phantom{..0}\times 10^{-6\phantom{0}}$ \\
         & 4 & $-(3\pm2)\phantom{.0000.00000}\times10^{-10}$ & $2.9\times10^{-11}$ & $-(0.4\pm2.4)\phantom{.} \times 10^{-7\phantom{0}}$ \\
         \hline
         \hline
    \end{tabular*}
\end{table*}

For a scalar charge on a resonant orbit, these radiative averages reduce to
\begin{align} \label{eqn:XdotRad}
    \langle \dot{\mathcal{J}} \rangle^{\mathcal{H}/\infty}_\mathrm{rad} &= \frac{1}{4\pi} \sum_{j=0}^\infty \sum_{m=-j}^j \sum_{N=-\infty}^\infty \langle \dot{\mathcal{J}} \rangle^{\mathcal{H}/\infty}_{jmN},
    \\ \notag
    \langle \dot{\mathcal{J}} \rangle^\mathcal{H}_{jmN} &= \sum_{(k,n)_N} \sum_{(k',n')_N} \mathcal{A}^{\mathcal{J}}_{mkn}
    p_{mk'n'} \varpi_+^2 C^-_{jmkn} \bar{C}^-_{jmk'n'},
    \\ \notag
    \langle \dot{\mathcal{J}} \rangle^\infty_{jmN} &= \sum_{(k,n)_N} \sum_{(k',n')_N} \mathcal{A}^{\mathcal{J}}_{mkn}
    \omega_{mk'n'} C^+_{jmkn} \bar{C}^+_{jmk'n'},
\end{align}
where $\varpi^2_+ = {r_+^2+a^2}$, an overbar denotes complex conjugation, and $\sum_{(k,n)_N}$ refers to a sum over all integer $k$ and $n$ values that satisfy $k\beta_\theta + n\beta_r = N$. The coefficients are related to the frequencies and orbital constants via $\mathcal{A}^{E}_{mkn} = \omega_{mkn}$, $\mathcal{A}^{L_z}_{mkn} = m$, and
\begin{align}
    \frac{1}{2}\mathcal{A}^{Q}_{mkn} &= k\Upsilon_\theta + \omega_{mkn} \left(a L_z - a^2 E - \Upsilon^{(\theta)}_t\right) 
    \\ \notag
    & \qquad \qquad \qquad \qquad - m \left(L_z - aE - \Upsilon^{(\theta)}_\phi \right).
\end{align}
Just like $\langle \dot{E} \rangle$, $\langle \dot{L}_z \rangle$, and $\langle \dot{Q} \rangle$, these averages depend on $q_{\theta 0}$, and therefore are efficiently described by Fourier series. The calculation of their Fourier coefficients, which we denote as $\tilde{\mathcal{F}}_{k}^{\mathcal{J},\mathrm{rad}}$, is described in Appendix \ref{app:fourierFluxes}.

Thus, we calculate $\langle \dot{\mathcal{J}} \rangle_\mathrm{rad}$ and the corresponding Fourier coefficients $\tilde{\mathcal{F}}_{k}^{\mathcal{J},\mathrm{rad}}$ for the resonances listed in Table \ref{tab:sources}. We then compare these coefficients to $\tilde{\mathcal{F}}_{k}^{\mathcal{J},\mathrm{diss}}$ via \eqref{eqn:EdotBalance}-\eqref{eqn:QdotBalance}. In Tables \ref{tab:fluxBalance} and \ref{tab:Qbalance}, we report the values of $\tilde{\mathcal{F}}_{k}^{\mathcal{J},\mathrm{diss}}$ and the absolute error between $\tilde{\mathcal{F}}_{k}^{\mathcal{J},\mathrm{diss}}$ and $\tilde{\mathcal{F}}_{k}^{\mathcal{J},\mathrm{rad}}$ for each resonance. We report all $k$-modes that our code uses when evaluating $\langle \dot{\mathcal{J}} \rangle^\mathrm{diss}$. Interestingly, there are instances where our code includes modes in which the magnitude of $\tilde{\mathcal{F}}_{k}^{\mathcal{J},\mathrm{diss}}$ is less than its uncertainty. While the values of these individual modes are highly uncertain, our code still incorporates these modes because they improve our overall uncertainty estimate for $\langle \dot{\mathcal{J}} \rangle^\mathrm{diss}$. From Table \ref{tab:fluxBalance}, we see that the absolute errors always fall under the estimated uncertainty of our results, demonstrating that our coefficients are not only accurate, but that our estimated uncertainties account for the dominant sources of numerical error in our dissipative data.


\subsection{Validation tests: Conservative perturbations}
\label{sec:val2}

\begin{figure*}[p]
    \centering
    \includegraphics[width=0.95\linewidth]{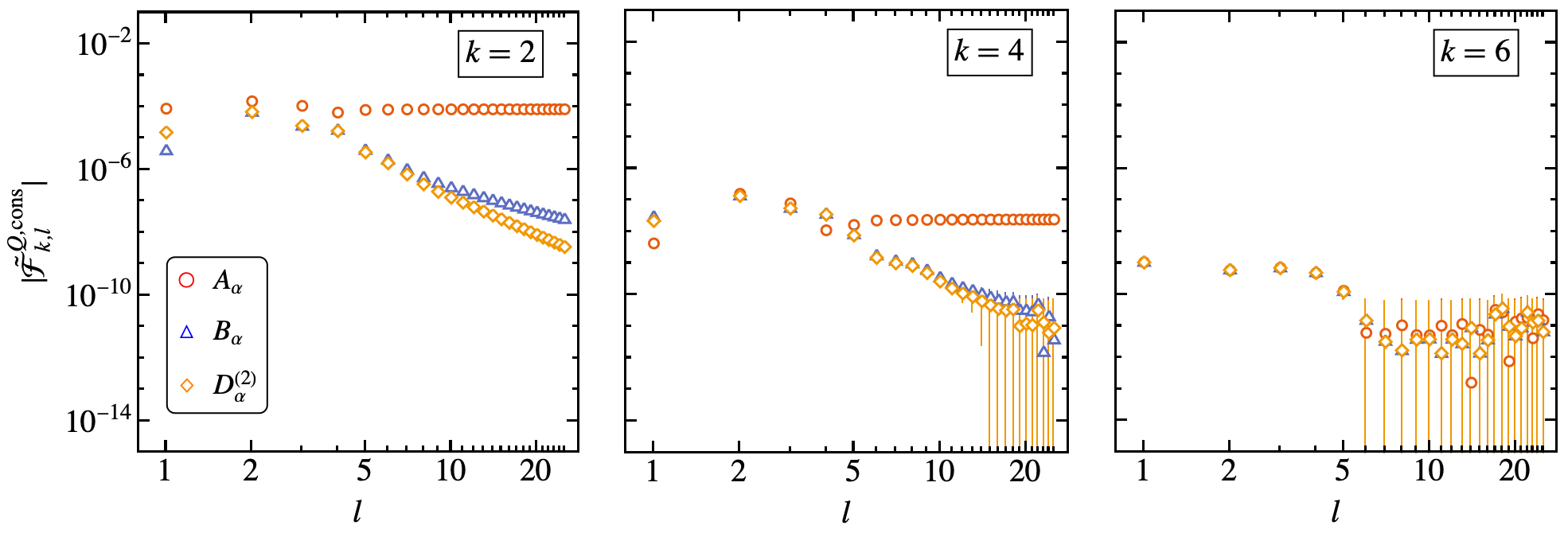}
    \caption{The $l$-mode convergence of the Fourier coefficients that describe $\langle \dot{Q}\rangle^\mathrm{cons}$ for the $e02.23$ orbit in Table \ref{tab:sources}. The red circles represent $|\tilde{\mathcal{F}}_{k,l}^{Q,\mathrm{cons+S}-A}|$, which decays like $l^0$; the blue triangles represent $|\tilde{\mathcal{F}}_{k,l}^{Q,\mathrm{cons+S}-A}-\tilde{\mathcal{F}}_{k,l}^{Q,B}|$, which decays like $l^{-2}$; and the orange diamonds represent $|\tilde{\mathcal{F}}_{k,l}^{Q,\mathrm{cons+S}-A}-\tilde{\mathcal{F}}_{k,l}^{Q,B}-\tilde{\mathcal{F}}_{k,l}^{Q,D}/(2l-1)/(2l+3)|$, which decays like $l^{-4}$. Because the odd $k$-modes vanish, we only plot the even modes $k=(2,4,6)$. The error bars represent the estimated uncertainty in each coefficient.}
    \label{fig:consConvergence}
\end{figure*}

\begin{figure*}[p]
    \centering
    \includegraphics[width=0.95\linewidth]{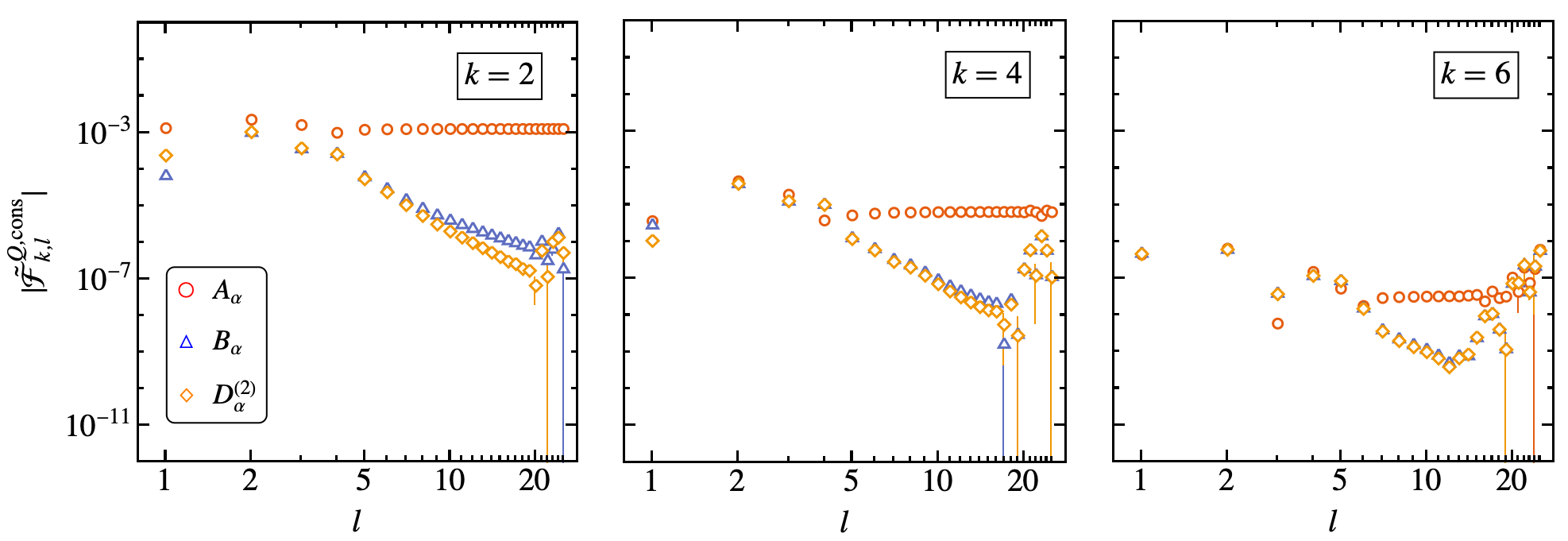}
    \caption{Similar to Fig.~\ref{fig:consConvergence}, the $l$-mode convergence of the Fourier coefficients that describe $\langle \dot{Q}\rangle^\mathrm{cons}$ as one includes more regularization parameters, but this time for the $e05.23$ orbit in Table \ref{tab:sources}.}
    \label{fig:consConvergence2}
\end{figure*}

To validate the conservative averages calculated by our new code, we test the $l$-mode convergence of the Fourier coefficients $\tilde{\mathcal{F}}_{k}^{\mathcal{J},\mathrm{cons}}$ that describe $\langle \dot{\mathcal{J}} \rangle^\mathrm{cons}$. Mode-sum regularization of the conservative perturbations leads to the algebraic convergence of $\tilde{\mathcal{F}}_{k,l}^{\mathcal{J},\mathrm{cons}}$ with $l$. As we include the $A_\alpha$, $B_\alpha$, and $D^{(2)}_\alpha$ parameters in our mode-sum regularization, we expect the regularized values of $\tilde{\mathcal{F}}_{k,l}^{\mathcal{J},\mathrm{cons}}$ to fall-off like $l^0$, $l^{-2}$ and $l^{-4}$, respectively. The rates of convergence are demonstrated in Figs.~\ref{fig:consConvergence} and \ref{fig:consConvergence2}, where we plot $|\tilde{\mathcal{F}}_{k,l}^{Q,\mathrm{cons+S}-A}|$ (red circles), $|\tilde{\mathcal{F}}_{k,l}^{Q,\mathrm{cons+S}-A}-\tilde{\mathcal{F}}_{k,l}^{Q,B}|$ (blue triangles), and $|\tilde{\mathcal{F}}_{k,l}^{Q,\mathrm{cons+S}-A}-\tilde{\mathcal{F}}_{k,l}^{Q,B}-\tilde{\mathcal{F}}_{k,l}^{Q,D}/(2l-1)/(2l+3)|$ (orange diamonds). Figure \ref{fig:consConvergence} plots the coefficients for the $e02.23$ resonance, while Fig.~\ref{fig:consConvergence2} plots the coefficients for $e05.23$. The error bars display the estimated uncertainty in each coefficient. Note that this estimated uncertainty only takes into account errors due to truncating our (formally infinite) self-force mode-sums, not the errors due to catastrophic cancellations when evaluating the sums. We find that, much like the dissipative coefficients, the $k=2$ and $k=4$ conservative coefficients largely exhibit the expected decay rates in $l$, though the $k=4$ modes become contaminated by numerical error at higher values of $l$. The higher $k$-modes are almost entirely dominated by numerical error but are also orders of magnitude smaller than the $k=2$ mode. Thus, they can be neglected without introducing significant error to $\langle \dot{\mathcal{J}} \rangle^\mathrm{cons}$. For low eccentricities, the numerical error is well captured by our uncertainty estimates and therefore can largely be attributed to the truncation of the SSF mode-sum. On the other hand, for higher eccentricities, our uncertainty estimates do not capture the poor convergence at higher $l$-modes and thus are most likely due to the large cancellations in the extended homogeneous solutions, as discussed in Sec.~\ref{sec:driver}. Nonetheless, in the absence of these errors, the conservative coefficients converge as expected.

\begin{figure*}[p]
    \centering
    \includegraphics[width=0.45\linewidth]{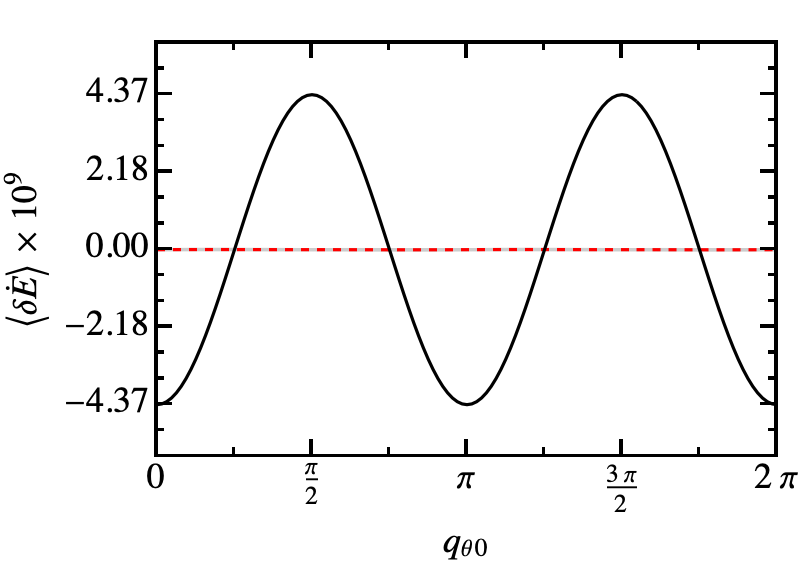}
    \includegraphics[width=0.45\linewidth]{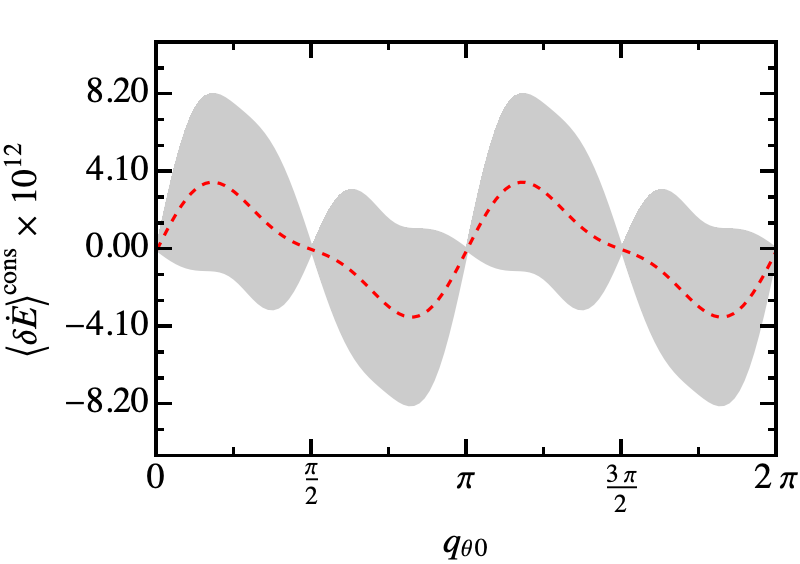}
    \caption{Residual variations $\langle \delta\dot{E}\rangle^\mathrm{diss}$ (solid black line) and $\langle \delta\dot{E}\rangle^\mathrm{cons}$ (red dashed line) as functions of the initial resonant phase $q_{\theta 0}$ for the $e02.12$ resonance in Table \ref{tab:sources}. The shaded (gray) region estimates the uncertainty in $\langle \delta\dot{E}\rangle^\mathrm{cons}$. The plot on the left shows both $\langle \delta\dot{E}\rangle^\mathrm{diss}$ and $\langle \delta\dot{E}\rangle^\mathrm{cons}$, while the plot on the right only depicts $\langle \delta\dot{E}\rangle^\mathrm{cons}$ and its region of uncertainty.}
    \label{fig:eVar}
\end{figure*}

Finally, we produce one last validation test for our conservative data using the conservative averages $\langle \delta\dot{E} \rangle^\mathrm{cons}$ and $\langle \delta\dot{L}_z \rangle^\mathrm{cons}$. Based on the flux-balance laws in \eqref{eqn:EdotBalance}, the secular evolution of the energy and angular momentum should be driven by purely dissipative perturbations. Therefore, $\langle \delta\dot{E} \rangle^\mathrm{cons}$ and $\langle \delta\dot{L}_z \rangle^\mathrm{cons}$ must vanish for all values of $q_{\theta 0}$. We find that $\langle \dot{E} \rangle^\mathrm{cons} = \langle \dot{L}_z \rangle^\mathrm{cons} = 0$ within the estimated uncertainties of our calculations for all of the resonances listed in Table \ref{tab:sources}. As an example, in Fig.~\ref{fig:eVar} we plot $\langle \delta\dot{E} \rangle^\mathrm{diss}$ and $\langle \delta\dot{E} \rangle^\mathrm{cons}$, along with their estimated uncertainties, for the $e02.12$ resonance. While our code produces nonzero values for $\langle\delta \dot{E} \rangle^\mathrm{cons}$, these values are always orders of magnitude smaller than the dissipative average $\langle\delta \dot{E} \rangle^\mathrm{diss}$. More importantly, they are also smaller than the estimated uncertainty in $\langle \delta\dot{E} \rangle^\mathrm{cons}$ and, consequently, consistent with zero. We observe similar behavior in $\langle \delta\dot{L}_z \rangle^\mathrm{cons}$ and across all of the orbits in Table \ref{tab:sources}.

\begin{figure*}[ht!]
    \centering
    \includegraphics[width=0.45\linewidth]{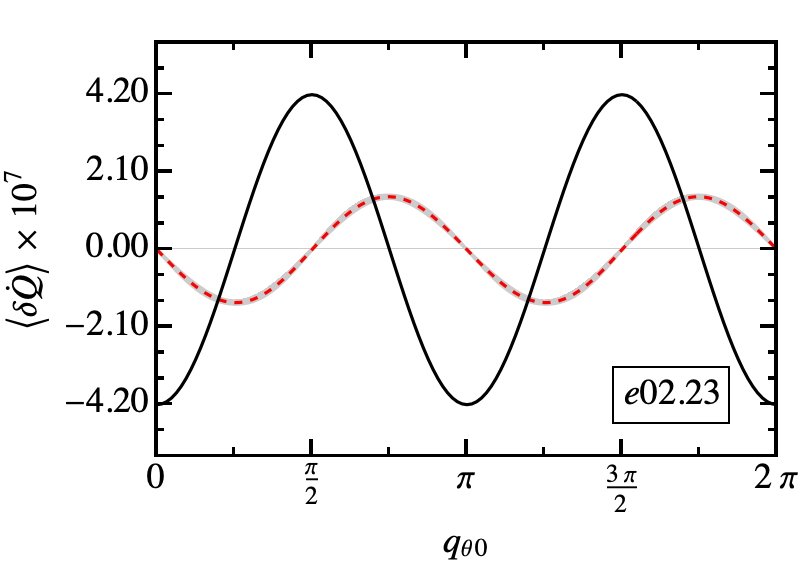}
    \includegraphics[width=0.45\linewidth]{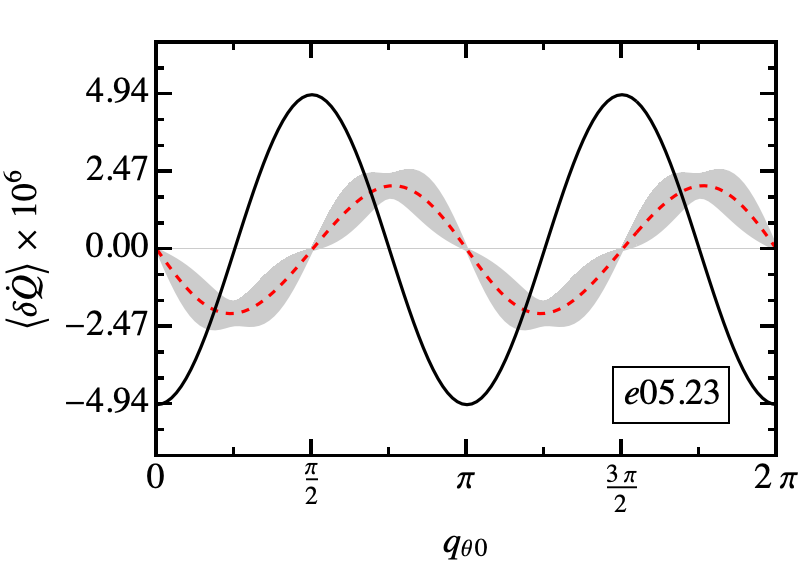}
    \includegraphics[width=0.45\linewidth]{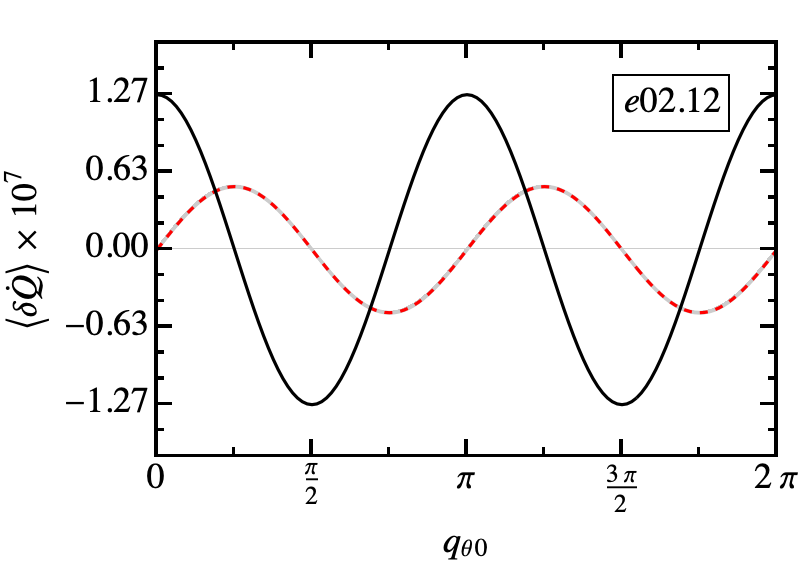}
    \includegraphics[width=0.45\linewidth]{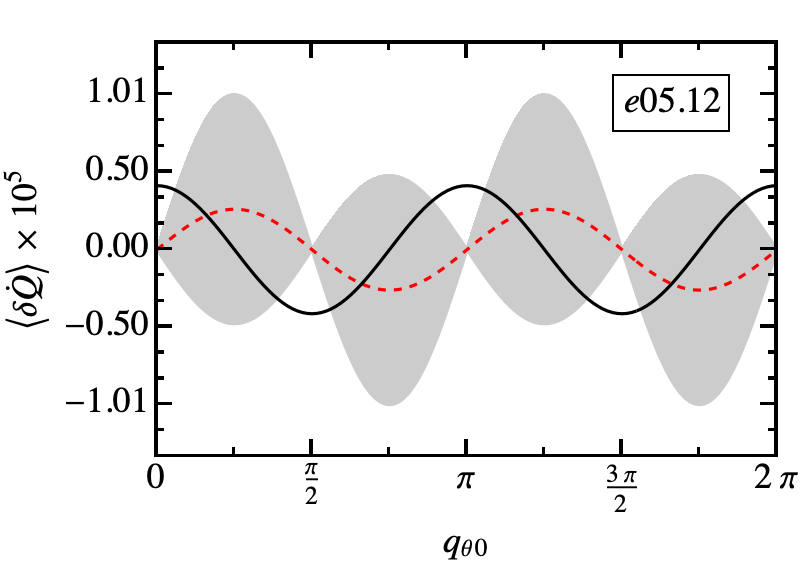}
    \caption{Residual variations $\langle \delta\dot{Q}\rangle^\mathrm{diss}$ (solid black line) and $\langle \delta\dot{Q}\rangle^\mathrm{cons}$ (red dashed line) as functions of the initial resonant phase $q_{\theta 0}$ for the resonances in Table \ref{tab:sources}. The shaded (gray) region estimates the uncertainty in $\langle \delta\dot{Q}\rangle^\mathrm{cons}$.}
    \label{fig:QdotVar}
\end{figure*}

\begin{figure*}[ht!]
    \centering
    \includegraphics[width=0.45\linewidth]{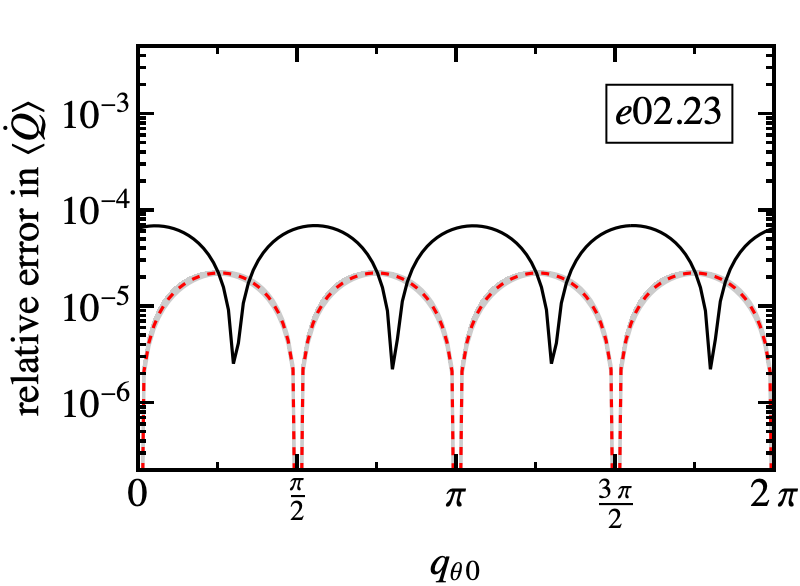}
    \includegraphics[width=0.45\linewidth]{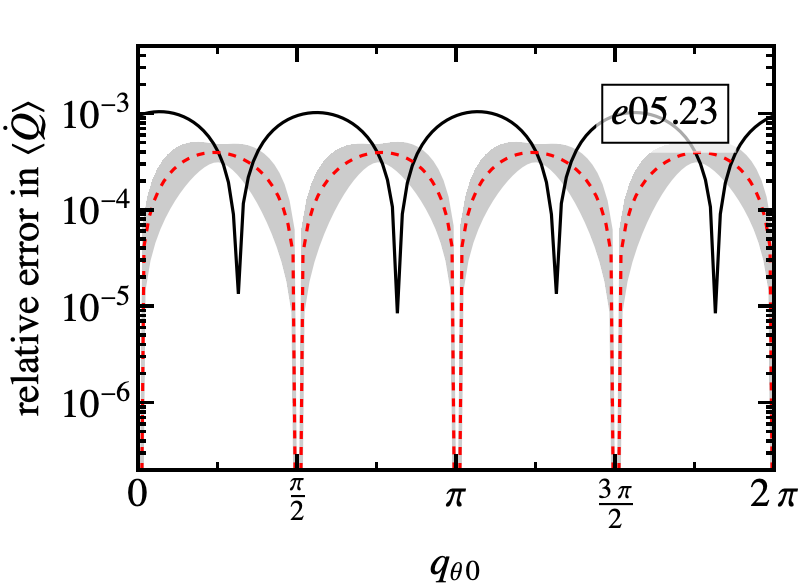}
    \includegraphics[width=0.45\linewidth]{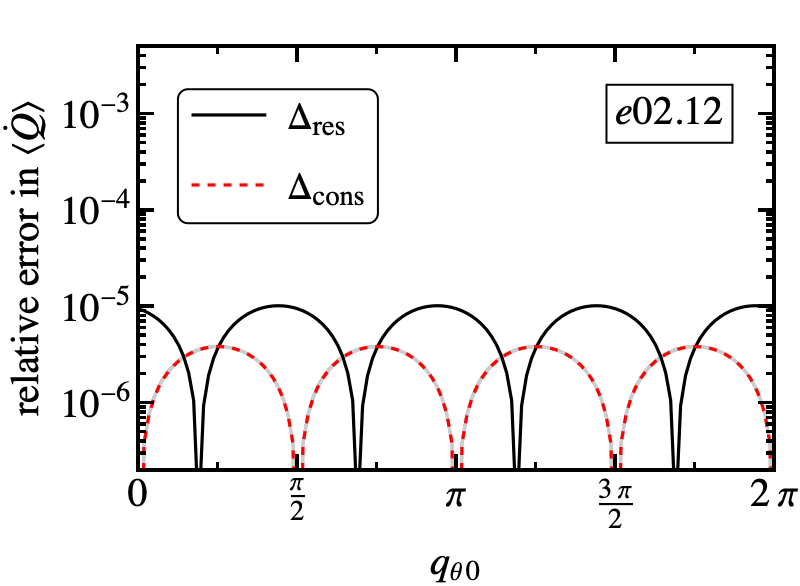}
    \includegraphics[width=0.45\linewidth]{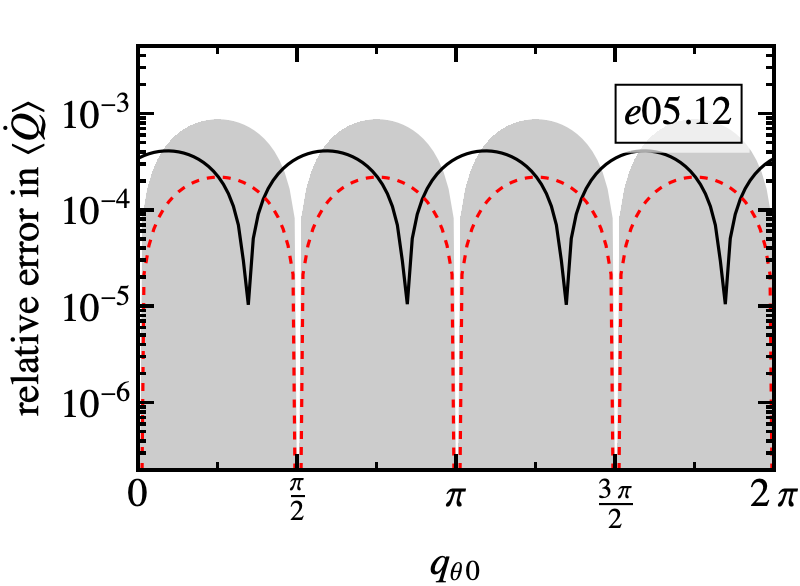}
    \caption{Relative errors in $\langle \dot{Q}\rangle$ if we neglect $\langle \delta\dot{Q}\rangle$ (solid black lines) or if we neglect $\langle \delta\dot{Q}\rangle^\mathrm{cons}$. These relative errors, denoted by $\Delta_\mathrm{res}$ and $\Delta_\mathrm{cons}$, are defined in \eqref{eqn:DeltaRes} and \eqref{eqn:DeltaCons}, respectively. The shaded (gray) region gives the range of errors that are consistent with $\Delta_\mathrm{cons}$ due to the estimated uncertainty in $\langle \delta\dot{Q}\rangle^\mathrm{cons}$.}
    \label{fig:relErrorQ}
\end{figure*}

\subsection{Conservative contributions to $\langle\dot{Q}\rangle$}
\label{sec:QdotCons}

In Table \ref{tab:Qbalance} we report the values of $\mathrm{Im}[\tilde{\mathcal{F}}_{k}^{Q,\mathrm{cons}}]$ for the resonances in Table \ref{tab:sources}. (Recall that $\tilde{\mathcal{F}}_{k}^{Q,\mathrm{cons}}$ is purely imaginary.) We find nonzero values of $\mathrm{Im}[\tilde{\mathcal{F}}_{k}^{Q,\mathrm{cons}}]$ for the $e02.12$, $e02.23$, $e05.23$ resonances. On the other hand, all of the conservative coefficients of the $e05.12$ resonance are consistent with zero due to the large numerical errors in our calculation of the SSF at higher eccentricities. We observe that $|\tilde{\mathcal{F}}_{2}^{Q,\mathrm{cons}}|$ is always comparable in magnitude to $|\tilde{\mathcal{F}}_{2}^{Q,\mathrm{diss}}|$, with the coefficients differing roughly by a factor of two across all of the resonances. Furthermore, $\mathrm{Im}[\tilde{\mathcal{F}}_{k}^{Q,\mathrm{cons}}]$ and $\tilde{\mathcal{F}}_{2}^{Q,\mathrm{diss}}$ differ in sign for the 1:2 resonances, but share the same sign for the 2:3 resonances. For both $\langle \delta \dot{Q} \rangle^\mathrm{diss}$ and $\langle \delta \dot{Q} \rangle^\mathrm{cons}$, increasing the eccentricity increases the magnitude of the coefficients, a behavior that mirrors the gravitational case \cite{FlanHughRuan14}. Decreasing the semilatus rectum $p$ of the resonances also produces larger coefficients, but the $k=2$ mode actually makes a larger relative contribution to $\langle \dot{Q} \rangle$ in the 2:3 resonances when compared to the 1:2 resonances. In other words, while $|\tilde{\mathcal{F}}_{2}^{Q,\mathrm{cons}}|+|\tilde{\mathcal{F}}_{2}^{Q,\mathrm{diss}}|$ is greater for the 1:2 resonances, the ratio $(|\tilde{\mathcal{F}}_{2}^{Q,\mathrm{cons}}|+|\tilde{\mathcal{F}}_{2}^{Q,\mathrm{diss}}|)/|\tilde{\mathcal{F}}_{0}^{Q,\mathrm{diss}}|$ is greater for the 2:3 resonances.

In Fig.~\ref{fig:QdotVar} we use our coefficients to evaluate and plot $\langle \delta \dot{Q} \rangle^\mathrm{diss}$ (solid black lines) and $\langle \delta \dot{Q} \rangle^\mathrm{cons}$ (dashed red lines) as functions of $q_{\theta 0}$, along with the estimated uncertainty in $\langle \delta \dot{Q} \rangle^\mathrm{cons}$ (gray shaded region). Again, this demonstrates that $\langle \delta \dot{Q} \rangle^\mathrm{cons}$ is generally nonvanishing for all but the $e05.12$ resonance. Furthermore, comparing these plots with those of previous investigations of the SSF during resonances \cite{NasiEvan21}, we see a drastic improvement in our uncertainty estimate.\footnote{The conservative variations reported in this paper are opposite in sign to those plotted in our previous paper \cite{NasiEvan21}. We believe this is due to a plotting mistake in our previous paper, where we mistakenly set $q_{\theta 0} \rightarrow - q_{\theta0}$.} In fact, the uncertainties in $\langle \delta \dot{Q} \rangle^\mathrm{cons}$ for the low eccentricity resonances are so small that the regions of uncertainty (the gray shaded regions) are smaller than the dashed red line plotting $\langle \delta \dot{Q} \rangle^\mathrm{cons}$. Even if our error estimates were off by an order of magnitude, we would still observe nonvanishing values of $\langle \delta \dot{Q} \rangle^\mathrm{cons}$, clearly demonstrating that so-called ``conservative" scalar perturbations do not necessarily conserve the integrability of our perturbed system, but contribute to the leading-order secular evolution of the scalar charge.

Finally, we consider the relative error that is introduced to $\langle \dot{Q} \rangle$ if we neglect the impact of $q_{\theta 0}$ on the secular evolution,
\begin{align} \label{eqn:DeltaRes}
    \Delta_\mathrm{res} = \left\vert 1- \frac{\langle\langle \dot{Q} \rangle\rangle_0}{\langle \dot{Q} \rangle}\right\vert = \left\vert \frac{\langle \delta \dot{Q} \rangle^\mathrm{diss} + \langle \delta \dot{Q} \rangle^\mathrm{cons}}{\langle \dot{Q} \rangle}\right\vert,
\end{align}
and the relative error due to neglecting potential contributions from conservative perturbations,
\begin{align} \label{eqn:DeltaCons}
    \Delta_\mathrm{cons} = \left\vert 1- \frac{\langle\langle \dot{Q} \rangle\rangle_0 + \langle \delta \dot{Q} \rangle^\mathrm{diss}}{\langle \dot{Q} \rangle}\right\vert = \left\vert \frac{\langle \delta \dot{Q} \rangle^\mathrm{cons}}{\langle \dot{Q} \rangle}\right\vert.
\end{align}
In Fig.~\ref{fig:relErrorQ} we plot $\Delta_\mathrm{res}$ (solid black lines) and $\Delta_\mathrm{cons}$ (red dashed lines) for all of the resonances listed in Table \ref{tab:sources}. The gray shaded regions reflect the range of potential errors one may have due to the estimated uncertainty in $\langle \dot{Q} \rangle^\mathrm{cons}$. The quantity $\Delta_\mathrm{res}$ effectively estimates the post-1/2 adiabatic error that the resonance introduces to the orbital phase of the inspiraling scalar charge. Therefore, we see that neglecting the conservative perturbations induces an error that is equal in magnitude to the standard post-1/2 adiabatic error.

\section{Conclusion}
\label{sec:conclusion}

In this work we calculated the averaged time rate of change of the Carter constant, $\langle \dot{Q} \rangle$, due to the SSF for a scalar charge following several different $r\theta$-resonant geodesics. To perform these calculations, we built a new code in \texttt{C++} that is several orders of magnitude faster than the old \texttt{Mathematica} code used in previous investigations of the SSF. This code makes use of several new algorithms, including a module that numerically integrates the Teukolsky equation using a combination of MST and confluent Heun expansions to populate boundary data, the creation of a window function that enables the application of new analytically known regularization parameters, and a new regularization scheme that uses mode-sum regularization to directly regularize the Fourier coefficients of our secular averages, $\langle \dot{E} \rangle$, $\langle \dot{L}_z \rangle$, and $\langle \dot{Q} \rangle$. We found that our results are consistent with previous investigations of the SSF and satisfy known conservation and balance laws, thus validating our new code. 

Using our results, we demonstrated that during these resonances the secular evolution of the Carter constant due to purely conservative perturbations, $\langle \dot{Q} \rangle^\mathrm{cons}$, does not vanish, but its value varies with the initial phase of the resonance, $q_{\theta 0}$, as predicted by \cite{IsoyETC13,IsoyETC19}. This is the first conclusive numerical evidence that during $r\theta$ resonances motion in Kerr spacetime is no longer integrable under conservative scalar perturbations \cite{BlanFlan22}. This is in contrast to nonresonant orbits, where $\langle \dot{Q} \rangle^\mathrm{cons}$ vanishes exactly, and the perturbed conservative system remains integrable. 

Additionally, we find that $\langle \dot{Q} \rangle^\mathrm{cons}$ is comparable in magnitude to $\langle \delta \dot{Q} \rangle^\mathrm{diss}$, the residual variation in the secular evolution of $Q$ under purely dissipative perturbations. Essentially, ignoring the conservative perturbations at leading order will introduce a post-1/2 adiabatic error to the orbital phase of the inspiraling scalar charge, and this error appears to be equal in magnitude to the error introduced by ignoring resonances altogether. Therefore, it is vital that these conservative contributions are computed accurately, since any errors introduced to $\langle \dot{Q} \rangle^\mathrm{cons}$ will accumulate over the resonance and the rest of the inspiral. This will be a challenge for highly eccentric orbits, which are susceptible to large numerical errors that arise from catastrophic cancellations in the self-force mode sum, just as we saw for the $e05.12$ resonance studied in this work. Frequency-domain effective source regularization or hyperboloidal compactification may be able to ameliorate these issues, but these remain topics for future work.

Furthermore, it remains to be seen if $\langle \dot{Q} \rangle^\mathrm{cons}$ will also vanish when driven by purely gravitational perturbations. Many of the methods and code that were designed for this work can be generalized to the gravitational case. Therefore, moving forward we will compute $\langle \dot{Q} \rangle$ due to the GSF for a point mass on a resonant geodesic. If, surprisingly, $\langle \dot{Q} \rangle^\mathrm{cons}$ does vanish in the gravitational case, this may hint at some additional unknown symmetries that are not captured by our scalar perturbation model. 

\begin{acknowledgements}
This research was supported by an appointment to the NASA Postdoctoral Program at the NASA Goddard Space Flight Center, administered by Oak Ridge Associated Universities under contract with NASA. The author also thanks Charles R.~Evans and Christopher Munna for useful discussions and their helpful feedback.
\end{acknowledgements}

\appendix

\section{Window function}
\label{app:reg}

We aim to find a suitable window function $f(\cos\theta)$ that allows us to easily recast $F^{\mathrm{ret},l}_{\theta\pm}$ in \eqref{eqn:FthetaRetNaive} onto a basis of spherical harmonics [i.e., \eqref{eqn:FthetaRet}] without changing the value of the self-force along the worldline and without altering our mode-sum regularization procedure.

First we examine how applying an arbitrary window function will alter the regularization parameters in \eqref{eqn:regExpansion}. These regularization parameters are derived from the multipoles of the singular SSF contribution, which are defined by
\begin{equation} \label{eqn:FSl}
	F_\alpha^{\mathrm{S},l}(t,r) = \sum_{m=-l}^l \int_0^{4\pi} d\Omega\, F_\alpha^\mathrm{S}(t,r,\theta,\phi) \bar{Y}_{lm}(\theta,\phi),
\end{equation}
where $d\Omega=d\cos\theta d\phi$ and an overbar denotes complex conjugation. As shown in \cite{Heff14,HeffETC17,Heff21}, $F_\alpha^\mathrm{S}$ can be rewritten as an expansion away from the worldline,
\begin{equation}
	F_\alpha^\mathrm{S}(t,r,\theta,\phi) = \sum_{n=1}^\infty
	{B_\alpha^{(3n-2)}}\rho^{-2n-1} \epsilon^{n-3},
\end{equation}
where $\rho^2 = (\hat{g}_{\alpha\beta}\hat{u}^\alpha \Delta x^\beta)^2 + \hat{g}_{\alpha\beta}
\Delta x^\alpha \Delta x^\beta$; a hat denotes a quantity that is evaluated along the worldline $x_p^\alpha$; $\Delta x^\alpha = x^\alpha - x_p^\alpha$; and the expansion terms have the form $B^{(k)}_\alpha = b_{a_1 a_2 \dots a_k} \Delta x^{a_1}\Delta x^{a_2} \dots \Delta x^{a_k}$, where $b_{a_1 a_2 \dots a_k}$ are coefficients that do not depend on $t$, $r$, $\theta$, or $\phi$. The variable $\epsilon \propto \Delta x$ is a bookkeeping variable for tracking the singular ``order'' of each term.

To construct the regularization parameters, one inserts this expansion into \eqref{eqn:FSl} and takes the limit as $x^\alpha \rightarrow x^\alpha_p$, leading to the regularization quantities,
\begin{align} \notag
	F_\alpha^{[-2]} &=  \epsilon^{-2} \lim_{\Delta r\rightarrow 0^\pm} \sum_{m=-l}^l \int d\Omega\, \bar{Y}_{lm} \frac{B_\alpha^{(1)}}{\rho^3},
	\\
	F_\alpha^{[-1]} &= \epsilon^{-1} \sum_{m=-l}^l \int d\Omega\, \bar{Y}_{lm}
	\left[\frac{B_\alpha^{(4)}}{\rho^5} \right]_{\Delta r= 0},
	\\ \notag
	F_\alpha^{[-2n+1]} &= \epsilon^{(2n-1)} \sum_{m=-l}^l \int d\Omega\, \bar{Y}_{lm}
	\left[\frac{B_\alpha^{(6n+4)}}{\rho^{4n+5}} \right]_{\Delta r= 0}.
\end{align}
After evaluating these integrals, the regularization parameters are then given by
\begin{gather*}
    F_\alpha^{[-2]} =\frac{1}{2} A^\pm_\alpha(2l+1), \qquad  F_\alpha^{[-1]} = B^\pm_\alpha,
    \\
    F_\alpha^{[-2n+1]} = D^{(2n)}_\alpha \left[\prod_{k=1}^n (2l-1-2k)(2l+1+2k)\right]^{-1}.
\end{gather*}

Now consider the effect of multiplying the SSF by some window function $f(\cos\theta)$. Then, to preserve our regularization scheme, our regularization parameters must be calculated by similarly weighting the singular SSF contribution $F_\alpha^\mathrm{S}(t,r,\theta,\phi)$ by this window function, leading to the new parameters
\begin{align} \notag
	\tilde{F}_\alpha^{[-2]} &=  \epsilon^{-2} \lim_{\Delta r\rightarrow 0^\pm}\sum_{m=-l}^l \int d\Omega\, \bar{Y}_{lm}\frac{f B_\alpha^{(1)}}{\rho^3},
	\\
	\tilde{F}_\alpha^{[-1]} &= \epsilon^{-1} \sum_{m=-l}^l \int d\Omega\, \bar{Y}_{lm}
	\left[\frac{f B_\alpha^{(4)}}{\rho^5} \right]_{\Delta r=0},
	\\ \notag
	\tilde{F}_\alpha^{[-2n+1]} &= \epsilon^{(2n-1)} \sum_{m=-l}^l \int d\Omega\, \bar{Y}_{lm}
	\left[\frac{fB_\alpha^{(6n+4)}}{\rho^{4n+5}} \right]_{\Delta r= 0},
\end{align}
If we expand our window function $f(\cos\theta)$ around the worldline, then we
have $f(z=\cos\theta) = \hat{f} + (\partial_z \hat{f}) (z-z_p) \epsilon + \frac{1}{2}
(\partial_z^2 \hat{f})(z-z_p)^2 \epsilon^2 + O(\epsilon^3)$. Inserting this expansion
into the integrals above, any terms $O(\epsilon)$ or greater will vanish as
we approach the worldline. Therefore, if $\hat{f} = 1$, $\partial_z \hat{f} = 0$, $\partial_z^2 \hat{f} = 0$, then we have that $\tilde{A}^\pm_\alpha = A^\pm_\alpha + O(\epsilon)$ and $\tilde{B}^\pm_\alpha = B^\pm_\alpha + O(\epsilon)$.

However, if we want to make use of the higher-order regularization parameter
$D^{(2)}_\theta$ (assuming it is known), then we must choose a window function that satisfies the four conditions $\hat{f}= 1 $ and $\partial_z \hat{f} = 0$,
$\partial_z^2 \hat{f} = 0$, \emph{and} $\partial_z^3 \hat{f} = 0$. One particular window function that satisfies these conditions is
\begin{align}
	f(\theta; \theta_p) &=
	\sum_{k=0}^3 \alpha_k(\theta_p) \cos^k\theta\sin\theta ,
\end{align}
where
\begin{align}
	\alpha_ 0(\theta_p) &= \frac{2-7\cos^2\theta_p+8\cos^4\theta_p(1-\cos^2\theta_p)}{2\sin^7\theta_p},
	\\
	\alpha_ 1(\theta_p) &= \frac{3\cos^3\theta_p(1+4\cos^2\theta_p)}{2\sin^7\theta_p},
	\\
	\alpha_ 2(\theta_p) &= \frac{1-8\cos^2\theta_p(1+\cos^2\theta_p)}{2\sin^7\theta_p},
	\\
	\alpha_ 3(\theta_p) &= \frac{\cos\theta_p(3+2\cos^2\theta_p)}{2\sin^7\theta_p}.
\end{align}
By applying this window function to $F^{\mathrm{ret},l}_{\theta\pm}$, we can re-expand the derivatives of spherical harmonics in terms of spherical harmonics using the following relations,
\begin{align} \notag
	\sin\theta \partial_\theta Y_{lm} &= -(l+1)C_{lm}
	Y_{l-1,m} + l C_{l+1,m} Y_{l+1,m},
	\\ 
	\cos\theta Y_{lm} &= C_{lm} Y_{l-1,m}
	+ C_{l+1,m} Y_{l+1,m},
\end{align}
where
\begin{equation}
    C_{lm} = \sqrt{\frac{l^2-m^2}{(2l+1)(2l-1)}}.
\end{equation}
Combining these relations, we find that
\begin{equation}
	\cos^k\theta \sin\theta \partial_\theta Y_{lm} = \sum_{n=-k-1}^{k+1} 
	\zeta^{(n,k)}_{lm} Y_{l+n,m}.
\end{equation}
where
\begin{equation}
	\tilde{\zeta}^{(n,k)}_{lm} = \tilde{\zeta}^{(n-1,k-1)}C_{l+n,m} + 
	\tilde{\zeta}^{(n+1,k-1)}C_{l+n+1,m},
\end{equation}
with (initial) conditions
\begin{gather*}
	\tilde{\zeta}^{(-1,0)}_{lm} = -(l+1)C_{lm}, 
	\qquad \qquad 
	\tilde{\zeta}^{(0,0)}_{lm} = 0, 
	\\
	\tilde{\zeta}^{(+1,0)}_{lm} = l C_{l+1,m},
\end{gather*}
and the requirement that $\tilde{\zeta}^{(n,k)}_{lm}$ vanishes if $|n|>k+1$.

With this window function, we get a finite coupling between the
derivative of the spherical harmonics and the spherical harmonics themselves
\begin{align}
	f(\theta;\theta_p) \partial_\theta Y_{lm}(\theta,\phi) = \sum_{n=-4}^{4} \beta^{(n)}_{lm}(\theta_p) Y_{l+n,m}(\theta,\phi) ,
\end{align}
where
\begin{align}
	\beta^{(n)}_{lm}(\theta_p) = \sum_{k=0}^3 \alpha_k(\theta_p)\tilde{\zeta}^{(n,k)}_{lm}.
\end{align}
Making use of this expansion, our windowed self-force quantity $\tilde{F}^{\mathrm{ret},l}_{\theta\pm} = f {F}^{\mathrm{ret},l}_{\theta\pm} $, when expressed on a basis of spherical harmonics, takes
the form
\begin{equation}
	\tilde{F}^{\mathrm{ret},l}_{\theta\pm} = \sum_{m=-l}^l \left( \sum_{n=-4}^4 \beta^{(-n)}_{l+n,m} \psi^\pm_{l+n,m} 
	\right)Y_{lm},
\end{equation}
which is amenable to mode-sum regularization.

\section{Decription of CPP modules}
\label{app:modules}

\subsection{Geodesic module}
\label{app:geoModule}

The geodesic module performs the following numerical routines:
\begin{enumerate}
    \item Given the orbital parameters $(a,p,e,x)$ and sample number $N_\mathrm{geo} = 2^{n_\mathrm{geo}}$, where $n_\mathrm{geo}\in \mathbb{Z}^+$, we use spectral integration methods \cite{HoppETC15,NasiOsbuEvan19} to calculate the orbital constants $(E, L_z, Q)$, the Mino time frequencies $\Upsilon_\alpha$, and the fiducial geodesic functions $\Delta\hat{x}=\{\Delta \hat{t}^{(r)}, \Delta \hat{t}^{(\theta)}, \Delta \hat{r}^{(r)},\Delta \hat{\theta}^{(\theta)},\Delta \hat{\phi}^{(r)}, \Delta \hat{\phi}^{(\theta)}\}$ defined in Sec.~\ref{sec:geo}.
    
    \item We then sample our functions at $N_\mathrm{geo}$ evenly spaced points on the intervals $q_r \in [0,2\pi)$ and $q_\theta \in [0,2\pi)$, e.g.,
    \begin{align} \label{eqn:deltaRSample}
        \Delta\hat{r}^{(r)}_{j_r} &= \Delta\hat{r}^{(r)}(q_{r, j_r}), & q_{r,j_r} = \frac{2\pi j_r}{N_\mathrm{geo}},
        \\ \label{eqn:deltaThetaSample}
        \Delta\hat{\theta}^{(\theta)}_{j_\theta} &= \Delta\hat{\theta}^{(\theta)}(q_{\theta, j_\theta}), & q_{\theta,j_\theta} = \frac{2\pi j_\theta}{N_\mathrm{geo}},
    \end{align}
    where $j_{r/\theta} = 0, 1, \dots, N_\mathrm{geo}-1$.
    
    \item Finally, we output the discretely sampled geodesic functions, which we denote as $\Delta\hat{x}_j=\{\Delta \hat{t}^{(r)}_{j_r}, \Delta \hat{t}^{(\theta)}_{j_\theta}, \Delta \hat{r}^{(r)}_{j_r},\Delta \hat{\theta}^{(\theta)}_{j_\theta},\Delta \hat{\phi}^{(r)}_{j_r}, \Delta \hat{\phi}^{(\theta)}_{j_\theta}\}$, along with $(E, L_z, Q)$ and $\Upsilon_\alpha$.
\end{enumerate}

\subsection{SSF module}
\label{app:sumModule}

Given output from the geodesic module, the SSF module constructs the unregularized SSF modes $\hat{F}^{\mathrm{ret},l}_{\alpha\pm}$:
\begin{enumerate}
    \item As initial input, the user must specify a maximum spherical multipole mode $l=l_\mathrm{max}$ and a sample number $N_\mathrm{SSF} = 2^{n_\mathrm{SSF}}$ where $n_\mathrm{SSF}\in \mathbb{Z}^+ \leq n_\mathrm{geo}$. The module then generates $\hat{F}^{\mathrm{ret},lm}_{\alpha\pm}$ for $-l_\mathrm{max} \leq m \leq l_\mathrm{max}$. The coupling between the spherical and spheroidal harmonics makes it much more efficient to pick a value of $m$ and calculate several $l$-modes at once, rather than to calculate modes on an individual $(l,m)$-basis. Typically we set $l_\mathrm{max}=25$ and $n_\mathrm{SSF}=8$. We also parallelize this part of the calculation, distributing the $m$-modes across separate cores.
    
    \item Given a specific value of $m$, the module generates the modes $\hat{F}^{\mathrm{ret},lm}_{\alpha\pm}$ for $|m| \leq l \leq l_\mathrm{max}$ using \eqref{eqn:FtRet}-\eqref{eqn:FphiRet} and \eqref{eqn:FthetaRet}. These modes are evaluated on a two-dimensional $N_\mathrm{SSF} \times N_\mathrm{SSF}$ grid spanned by
    $q_r \in [0,2\pi)$ and $q_\theta \in [0,2\pi)$. As a result, the code outputs discretely sampled modes
    \begin{align} \label{eqn:discreteSSF}
        \hat{F}^{\mathrm{ret},lm}_{\alpha\pm,j_rj_\theta} = \hat{F}^{\mathrm{ret},lm}_{\alpha\pm}(q_{r,j_r}, q_{\theta,j_\theta}),
    \end{align}
    where $q_{r,j_r}$ and $q_{\theta,j_\theta}$ are given by \eqref{eqn:deltaRSample} and \eqref{eqn:deltaThetaSample} but with $N_\mathrm{geo}$ replaced by $N_\mathrm{SSF}$.
    
    \item We then construct the modes $\hat{F}^{\mathrm{ret},lm}_{\alpha\pm}$ for $|m| \leq l \leq l_\mathrm{max}$ from the extended homogeneous function (and its derivatives) $\psi_{lm}^\pm$, $\partial_t\psi_{lm}^\pm$, $\partial_r\psi_{lm}^\pm$ via \eqref{eqn:psilm}. This requires a three-fold summation of the harmonic functions $\psi_{ljmkn}^\pm$ and $\partial_r\psi_{ljmkn}^\pm$ [see \eqref{eqn:psiljmkn}] over the mode numbers $j$, $k$, and $n$. We denote this summation by
    \begin{align} \label{eqn:FljmknSum}
        \hat{F}^{\mathrm{ret},lm}_{\alpha\pm,j_rj_\theta}
        &= \sum_{k=-\infty}^\infty \sum_{n=-\infty}^\infty \sum_{j=|m|}^\infty\hat{F}^{\mathrm{ret},ljmkn}_{\alpha\pm,j_rj_\theta},
    \end{align}
    where the harmonics $\hat{F}^{\mathrm{ret},ljmkn}_{\alpha\pm,j_rj_\theta}$ are computed via the harmonic module in Appendix \ref{app:harmonicModule}.
    
    \item \label{enum:kModeSum} Beginning with the outermost sum in \eqref{eqn:FljmknSum}, given the values $l_\mathrm{max}$ and $m$, we compute
    \begin{align}
        \hat{F}^{\mathrm{ret},lm}_{\alpha\pm,j_rj_\theta} &\approx \sum_{k=k_0}^{k_\mathrm{init}} \hat{F}^{\mathrm{ret},lmk}_{\alpha\pm,j_rj_\theta} 
        \\ \notag
        & \qquad + \sum_{k=k_\mathrm{init}+1}^{k_\mathrm{max}} \hat{F}^{\mathrm{ret},lmk}_{\alpha\pm,j_rj_\theta}
        \\ \notag
        & \qquad \qquad + \sum_{k=k_\mathrm{min}}^{k_0-1} \hat{F}^{\mathrm{ret},lmk}_{\alpha\pm,j_rj_\theta},
    \end{align}
    for all values $|m| \leq l \leq l_\mathrm{max}$.
    To execute this sum we first set $k_0 = -m - 4$ and $k_\mathrm{init}=-m+4$. Since the $(l,m,k)$-modes tend to peak near $k=-m$ \cite{FujiHikiTago09}, initially summing over this range tends to capture the most dominant modes before we start testing for convergence of the sum.
        
    We then continue the sum for $k > k_\mathrm{init}$ and $k < k_0$. The sums are truncated based on the two convergence criteria,
    \begin{enumerate}
        \item $\left\vert \frac{\hat{F}^{\mathrm{ret},lmk}_{\alpha\pm,j_rj_\theta}}{\hat{B}_{\alpha,j_rj_\theta}} \right\vert < \epsilon_\mathrm{tol}$ and $\left\vert {\hat{F}^{\mathrm{ret},lmk}_{\alpha\pm,j_rj_\theta}} \right\vert < \left\vert {\hat{F}^{\mathrm{ret},lm,k+\Delta k}_{\alpha\pm,j_rj_\theta}} \right\vert$,
        \label{enum:convergenceA}
        \item $\left\vert {\hat{F}^{\mathrm{ret},lmk}_{\alpha\pm,j_rj_\theta}} \right\vert \leq \epsilon_\mathrm{DBL}$, 
        \label{enum:convergenceB}
    \end{enumerate}
    where $\Delta k = 1$ when $k < k_0$ and $\Delta k = -1$ when $k > k_\mathrm{init}$, $\epsilon_\mathrm{DBL}$ refers to the precision to which doubles are represented by our compiler, and $\hat{B}_{\alpha,j_rj_\theta}$ denotes the regularization parameter $B_\alpha$ [see \eqref{eqn:regExpansion}] evaluated along the fiducial geodesic at the discrete points $(q_{r,j_r}, q_{\theta,j_\theta})$. Furthermore, in this work we set $\epsilon_\mathrm{tol} = 10^{-10}$. We truncate the sums when all of the modes $k \in [k_\mathrm{max}-5, k_\mathrm{max}]$ and $k \in [k_\mathrm{min}, k_\mathrm{min}+5]$ satisfy at least one of the convergence criteria set above for all $l$, $j_r$, and $j_\theta$. 
        
    We require six $k$-modes to satisfy the convergence criteria before truncating the sum. Note that $\hat{F}^{\mathrm{ret},ljmkn}_{\alpha\pm,j_rj_\theta}$ vanishes when $l+m+k = $ odd for most self-force components. In this case, only half of the modes are useful for testing convergence. Consequently, even though six modes must satisfy the convergence criteria, most of the time only three of those six modes are nonvanishing.
        
    Additionally, we normalize our results by the coefficient $\hat{B}_{\alpha,j_rj_\theta}$ when testing for convergence, because this is the last divergent piece of the singular self-force contribution that we must subtract from our harmonic modes.
        
    Furthermore, for larger eccentricity orbits, the mode sum is prone to catastrophic cancellation. When this occurs, precision loss can prevent the convergence criteria from being met for a sufficient number of neighboring $k$-modes. To prevent the sum from never truncating in this scenario, we set the hard limits $k_\mathrm{min} \geq k_\mathrm{0}-k_\mathrm{cutoff}$ and $k_\mathrm{max} \leq k_\mathrm{init}+k_\mathrm{cutoff}$ with $k_\mathrm{cutoff} = l_\mathrm{max}+50$.
        
    \item Given $l_\mathrm{max}$, $m$, and $k$, we compute
    \begin{align}
        \hat{F}^{\mathrm{ret},lmk}_{\alpha\pm,{j_rj_\theta}} &\approx \sum_{n=n_0}^{n_\mathrm{init}} \hat{F}^{\mathrm{ret},lmkn}_{\alpha\pm,{j_rj_\theta}} 
        \\ \notag
        & \qquad + \sum_{n=n_\mathrm{init}+1}^{n_\mathrm{max}} \hat{F}^{\mathrm{ret},lmkn}_{\alpha\pm,{j_rj_\theta}}
        \\ \notag
        & \qquad \qquad + \sum_{n=n_{\mathrm{min}[\omega]}}^{n_0-1} \hat{F}^{\mathrm{ret},lmkn}_{\alpha\pm,{j_rj_\theta}},
    \end{align}
    for all values $|m| \leq l \leq l_\mathrm{max}$. The sums are constructed so that we only sum over positive frequencies. Negative frequencies are determined from the symmetries of the mode functions and, therefore, are incorporated into the values of $\hat{F}^{\mathrm{ret},lmkn}_{\alpha\pm,{j_rj_\theta}}$ via the harmonic module (see Appendix \ref{app:harmonicModule}). Thus, $n_{\mathrm{min}[\omega]}$ is given by the requirement that $\omega_{mkn_{\mathrm{min}[\omega]}}\geq 0$ and $\omega_{mk,n_{\mathrm{min}[\omega]}-1}< 0$. The $(l,m,k,n)$-modes tend to peak in magnitude near lower values of $n$ and $\omega_{mkn}$. Therefore, if $n_{\mathrm{min}[\omega]} \geq -8$, then we set $n_0=n_{\mathrm{min}[\omega]}+1$, else $n_0 = -8$. We then set $n_\mathrm{init} = n_0 + 16$. Based on trial and error, we find that these choices of $n_0$ and $n_\mathrm{init}$ tend to capture the modes with the largest magnitudes in our initial sum. 
        
    We then continue the sum for $n > n_\mathrm{init}$ and $n < n_0$. The $n$-mode summation is truncated using the same convergence criteria as the $k$-mode summation [see Step (\ref{enum:kModeSum})], only we replace ${\hat{F}^{\mathrm{ret},lmk}_{\alpha\pm,j_rj_\theta}}$ with ${\hat{F}^{\mathrm{ret},lmkn}_{\alpha\pm,j_rj_\theta}}$ and $\Delta k$ with $\Delta n$. We truncate the sums when all of the modes $n \in [n_\mathrm{max}-4, n_\mathrm{max}]$ and $n \in [n_\mathrm{min}, n_\mathrm{min}+4]$ satisfy either (\ref{enum:convergenceA}) or (\ref{enum:convergenceB}) for all $l$, $j_r$, and $j_\theta$. Like the $k$-mode sum, we also set the hard limits $n_\mathrm{min} \geq n_{\mathrm{min}[\omega]}$ and $n_\mathrm{max} \leq n_\mathrm{init}+n_\mathrm{threshold}$ with $n_\mathrm{threshold} = 120(1-e^2)^{-3/2} + |k|$, where $e$ is the eccentricity of the orbit. The formula for $n_\mathrm{threshold}$ was determined through numerical experimentation.
        
    \item Given $l_\mathrm{max}$, $m$, $k$, and $n$, we compute
    \begin{align}
        \hat{F}^{\mathrm{ret},lmkn}_{\alpha\pm,{j_rj_\theta}} \approx \sum_{j=|m|}^{j_\mathrm{max}} \hat{F}^{\mathrm{ret},ljmkn}_{\alpha\pm,{j_rj_\theta}},
    \end{align}
    for all values $|m| \leq l \leq l_\mathrm{max}$, where $j_\mathrm{max}$ is set by the alternate convergence criteria
    \begin{equation*}
        \left\vert \frac{\hat{F}^{\mathrm{ret},lj_\mathrm{max}mkn}_{\alpha\pm,{j_rj_\theta}}}{\hat{F}^{\mathrm{ret},llmkn}_{\alpha\pm,{j_rj_\theta}}} \right\vert < \epsilon_\mathrm{coupling},
    \end{equation*}
        which must be satisfied for all $l$, $j_r$, and $j_\theta$. Additionally, we require $l + j_\mathrm{max} = \mathrm{even}$, since all $l + j_\mathrm{max} = \mathrm{odd}$ modes vanish. In this work we set $\epsilon_\mathrm{coupling} = 10^{-18}$. The convergence of these modes is fairly uniform, leading to the simplified convergence condition.
        
    \item Given $l_\mathrm{max}$, $j$, $m$, $k$, and $n$, we compute $\hat{F}^{\mathrm{ret},ljmkn}_{\alpha\pm,{j_rj_\theta}}$ for all values $|m| \leq l \leq l_\mathrm{max}$ using the harmonic module described in Appendix \ref{app:harmonicModule}.
    
    \item Once all $\hat{F}^{\mathrm{ret},lm}_{\alpha\pm}$ are calculated, we sum over the $m$-modes,
    \begin{align}
        \hat{F}^{\mathrm{ret},l}_{\alpha\pm,j_rj_\theta} = \sum_{m=-l}^l \hat{F}^{\mathrm{ret},lm}_{\alpha\pm,j_rj_\theta},
    \end{align}
    giving us discretely sampled unregularized SSF $l$-modes, $\hat{F}^{\mathrm{ret},l}_{\alpha\pm,j_rj_\theta}$, in the range $0 \leq l \leq l_\mathrm{max}$.
\end{enumerate}

\subsection{Harmonic module}
\label{app:harmonicModule}

The harmonic module produces the individual self-force harmonics
\begin{align} \label{eqn:harmonicT}
    \hat{F}^{\mathrm{ret},ljmkn}_{t\pm} &= -i\omega_{mkn}\psi_{ljmkn}^\pm(r) Y_{lm}(\theta,\phi)e^{-i\omega_{mkn}t},
    \\
    \hat{F}^{\mathrm{ret},ljmkn}_{r\pm} &= \partial_r\psi_{ljmkn}^\pm(r) Y_{lm}(\theta,\phi)e^{-i\omega_{mkn}t},
    \\
    \hat{F}^{\mathrm{ret},ljmkn}_{\theta\pm} &= \sum_{n=-4}^{4}\beta^{(-n)}_{l+n,m}(\hat{\theta}_p)\psi_{l+n,jmkn}^\pm(r)
    \\ \notag
    & \qquad \qquad\qquad \qquad \times Y_{lm}(\theta,\phi)e^{-i\omega_{mkn}t},
    \\ \label{eqn:harmonicPhi}
    \hat{F}^{\mathrm{ret},ljmkn}_{\phi\pm} &= im\psi_{ljmkn}^\pm(r) Y_{lm}(\theta,\phi)e^{-i\omega_{mkn}t}.
\end{align}
The harmonics are then sampled along a fiducial resonant geodesic by taking $(t,r,\theta,\phi) \rightarrow (\hat{t}_p,\hat{r}_p,\hat{\theta}_p,\hat{\phi}_p)$. When these modes are evaluated along the geodesic worldline, the linear terms in $\hat{t}_p$ and $\hat{\phi}_p$ cancel,
\begin{align}
    m\hat{\phi}_p - \omega_{mkn}\hat{t}_p &= m\Delta \hat{\phi}^{(r)}(q_r) + m\Delta \hat{\phi}^{(\theta)}(q_\theta) 
    \\ \notag
    & \qquad \qquad - \omega_{mkn}\Delta \hat{t}^{(r)}(q_r) - n q_r  
    \\ \notag
    & \qquad \qquad \qquad -  \omega_{mkn}\Delta \hat{t}^{(\theta)}(q_\theta) - k q_\theta,
\end{align}
leaving us with functions that depend on Mino time through the periodic angle variables, $q_r$ or $q_\theta$.

Therefore, we construct $\hat{F}^{\mathrm{ret},ljmkn}_{\alpha\pm}$ as follows:
\begin{enumerate}
    \item Given geodesic data and the mode numbers $(l,j,m,k,n)$, we use the spheroidal harmonic module in Appendix \ref{app:spheroidModule} to calculate the coefficients $b^l_{jmkn}$; the spheroidal eigenvalue $\lambda_{jmkn}$; and the harmonic $S_{jmkn}$, which is evaluated at the discrete polar positions $\Delta\hat{\theta}^{(\theta)}_{j_\theta}$.
    
    \item Next we calculate the homogeneous radial Teukolsky solutions and their derivatives, $R^\pm_{jmkn}$ and $\partial_r R^\pm_{jmkn}$, using the radial Teukolsky module in Appendix \ref{app:teukModule}. The solutions are evaluated at the discrete radial positions $\Delta \hat{r}^{(r)}_{j_r}$.
    
    \item We then supply these discretely sampled functions to our source integration module in Appendix \ref{app:sourceModule}, which calculates the normalization coefficients (or Teukolsky amplitudes) $\hat{C}^\pm_{jmkn}$.
    
    \item The code then evaluates the coupling terms $\beta^{(n)}_{lm}$ and the spherical harmonics $Y_{lm}$ at the discrete polar positions $\Delta\hat{\theta}^{(\theta)}_{j_\theta}$.
    
    \item Given the coefficients $\hat{C}^\pm_{jmkn}$, $b^l_{jmkn}$, and $\beta^{(n)}_{lm}$, along with the discrete samplings of $R_{jmkn}^\pm$, $\partial_r R_{jmkn}^\pm$, and $Y_{lm}$, we construct the (sampled) harmonics $\hat{F}^{\mathrm{ret},ljmkn}_{\alpha\pm,j_r j_\theta}$ using \eqref{eqn:harmonicT}-\eqref{eqn:harmonicPhi}.
\end{enumerate}

\subsection{Spheroidal harmonic module}
\label{app:spheroidModule}

The spheroidal harmonic module calculates the coupling coefficients $b^l_{jmkn}$ and the spheroidal eigenvalue $\lambda_{jmkn}$, and then evaluates the scalar spheroidal harmonics $S_{jmkn}$ with spheroidicity $\gamma_{mkn} = a \omega_{mkn}$ at the points ${\theta}_{j}$ via the following algorithm
\begin{enumerate}
    \item Given harmonic numbers $j$ and $m$ and the spheroidicity $\gamma_{mkn}$, we first solve for the coupling coefficients $b^l_{jmkn}$ and spheroidal eigenvalues $\lambda_{jmkn}$ via an eigenvalue problem \cite{Hugh00,WarbBara10},
    \begin{align}
        \mathbb{K}^N_{mkn} \vec{b}^N_{jmkn} = \tilde{\lambda}^N_{jmkn} \vec{b}^N_{jmkn},
    \end{align}
    where $\mathbb{K}^N_{mkn}$ is a known $N\times N$ matrix. As one increases $N$, the eigenvalue converges to the value of the spheroidal eigenvalue, i.e., $\tilde{\lambda}^N_{jmkn} \approx {\lambda}_{jmkn}$. Additionally, the components of eigenvectors $\vec{b}^N_{jmkn} \doteq (\tilde{b}^{N,|m|}_{jmkn}, \tilde{b}^{N,|m|+1}_{jmkn}, \dots, \tilde{b}^{N,|m|+N-1}_{jmkn})$ converge to the values of the spherical-spheroidal coupling coefficients for $l \leq l_\mathrm{cut}$, i.e., $\tilde{b}^{N,l \leq l_\mathrm{cut}}_{jmkn} \approx b^{l \leq l_\mathrm{cut}}_{jmkn}$. Therefore, we set the cutoff mode $l_\mathrm{cut}$ to be the minimum value of $l>l_\mathrm{max}$ such that
    \begin{align}
        |\tilde{b}^{N,l_\mathrm{cut}}_{jmkn}| < \epsilon_\mathrm{DBL}.
    \end{align}
    
    \item The module begins with an initial value $N=l_\mathrm{max} + 30$, and then increases $N$ in increments of 10 until we meet the convergence criteria
    \begin{align}
        \left\vert 1-\frac{\tilde{b}^{N-10,l}_{jmkn}}{\tilde{b}^{N,l}_{jmkn}}\right\vert < \epsilon_\mathrm{eigen},
    \end{align}
    for all $l \leq l_\mathrm{cutoff}$. In this work we set $\epsilon_\mathrm{eigen} = 10^{-13}$.
    
    Eigenvalues and eigenvectors are calculated using the GSL eigensystems library \cite{GSL09}.
    
    \item Given the coefficients $b^l_{jmkn}$, we then construct $S_{jmkn}$ via the spherical harmonic expansion in \eqref{eqn:swsh}, which is evaluated at the presampled points ${\theta}_{j}$. We calculate the spherical harmonics using the GSL special functions library.
\end{enumerate}

\subsection{Radial Teukolsky module}
\label{app:teukModule}

The radial Teukolsky module calculates the homogeneous solutions $R^\pm_{jmkn}$ and $\partial_r R^\pm_{jmkn}$ using the hyperboloidal transformation proposed by Zengino{\v{g}}lu \cite{Zeng11b},
\begin{align} \label{eqn:hblTransform}
    \Psi^\pm_{jmkn}(r) = r e^{-i(m\varphi\pm \omega_{mkn} r_*)}R^\pm_{jmkn}(r),
\end{align}
where $\varphi(r) = \frac{a}{2M\kappa}\mathrm{ln}\frac{r-r_+}{r-r_-}$ and $\kappa =\sqrt{1-a^2/M^2}$. This transformation removes the leading-order oscillatory behavior of the homogeneous solutions, thereby improving the speed of our numerical solver. Therefore, the module first generates initial data for $\Psi^\pm_{jmkn}$ and $\partial_r \Psi^\pm_{jmkn}$ at the boundary points $r^\pm_\mathrm{init}$ and then numerically integrate the transformed Teukolsky equation \eqref{eqn:hblODE} given in Appendix \ref{app:harm}. Hyperboloidal solutions are stored at user-specified radial points, which are then transformed back to the radial Teukolsky solutions.

We use a combination of methods to generate the initial data. The first method we refer to as the \textit{series algorithm}. For the series algorithm, we first recast the Teukolsky equation into the form of the confluent Heun equation \eqref{eqn:cheODE}. We then use the Frobenius method to expand the confluent Heun solution associated with $\Psi^-_{jmkn}$ around the regular singular point at $r=r_+$ [see \eqref{eqn:R-Expansion}], and we perform an asymptotic expansion about $r=\infty$ for the confluent Heun solution associated with $\Psi^+_{jmkn}$ [see \eqref{eqn:R+Expansion}]. The transformation between the confluent Heun and radial Teukolsky solutions, along with the series expansions about the singular points are provided in Appendix \ref{app:che}. The advantage of transforming to \eqref{eqn:cheODE} is that the coefficients of the confluent Heun expansions satisfy simple three-term recurrence relations, making it straightforward to compute the expansions to very high order. In some cases, however, the series algorithm will suffer catastrophic cancellation when evaluating $\Psi^-_{jmkn}$ away from the horizon, while the asymptotic series may not decay rapidly enough to sufficiently approximate $\Psi^+_{jmkn}$ at finite radii.

Therefore, we also make use of a second method, which we refer to as the \textit{MST algorithm}. For the MST algorithm, we generate initial solutions using the semi-analytic function expansions proposed by Mano-Suzuki-Takasugi (MST) \cite{ManoSuzuTaka96a,ManoSuzuTaka96b} and expanded upon by Fujita and Tagoshi \cite{FujiTago04}. In this MST method, one expresses the homogeneous Teukolsky solutions as series of hypergeometric functions. For these series to converge one must first solve for the eigenvalue $\nu$---known as the renormalized angular momentum---which then determines the coefficients of these series. Typically one solves for $\nu$ using approximate low-frequency expansions \cite{FujiTago04}, root-finding methods \cite{FujiTago05}, or monodromy techniques \cite{CastETC13a,CastETC13b}. The MST series solutions, along with these methods for computing $\nu$, are well-behaved for lower frequencies, but for large values of $j$ and $\omega_{mkn}$ the MST method often suffers from catastrophic cancellations. 

Our old MMA code circumvented this issue by using arbitrary numerical precision, but with CPP we are restricted to the regions of parameter space that do not experience this large precision loss. We implement the MST expansions in CPP using the same methods as MMA, which are described in Sec.~III B of \cite{NasiOsbuEvan19} and Sec.~5.3.2 in \cite{Nasi20}. However, to the best of our knowledge, there is no open-source \texttt{C++} library that computes hypergeometric functions in the complex domain without using arbitrary precision. Therefore, we designed our own special functions library to implement the MST expansions in \texttt{C++} and we keep track of any precision lost during the MST calculation to estimate potential errors in our MST solutions due to catastrophic cancellation.

As a result, given the mode numbers $j$ and $m$, frequency $\omega_{mkn}$, spheroidal eigenvalue $\lambda_{jmkn}$, and an array of $N$ radial sample points $r_i$ where $i=0,1,\dots,N$ and $r_0 < r_1 < \cdots < r_N$, we construct the homogeneous Teukolsky solutions as follows:
\begin{enumerate}
    \item First we define the initial radial points $r^-_\mathrm{init} = r_+ + 0.01 M$ and $r^+_\mathrm{init} = 50 M$. If $r^-_\mathrm{init} > r_0$, then $r^-_\mathrm{init} = 0.9(r_0 - r_+)$, and if $r^+_\mathrm{init} < r_N$, then $r^+_\mathrm{init} = r_N + 20 M$.
    
    \item \label{item:IVP} Next we construct the initial data for $\Psi^\pm_{jmkn}$ and $\partial_r \Psi^\pm_{jmkn}$ at the boundaries $r^\pm_\mathrm{init}$. First we try to calculate the initial data using the series algorithm. If the series algorithm does not meet the convergence criteria described in Appendix \ref{app:che} when evaluated at $r^\pm_\mathrm{init}$ and $\omega_{mkn}r^\pm_\mathrm{init} \leq 0.01$, then we solve for the initial values using the MST method. If the estimated fractional error of an MST solution is $> 10^{-13}$ or $\omega_{mkn}r^\pm_\mathrm{init} > 0.01$, then we move our initial radial point closer to the boundary, and repeat our initial value calculations, beginning with the series algorithm. For the $(+)$ solutions we double the outer radial points, $r^+_\mathrm{init} \rightarrow 2r^+_\mathrm{init}$, and for the $(-)$ solutions we half the distance to the horizon, $(r^-_\mathrm{init}-r_+) \rightarrow (r^-_\mathrm{init}-r_+)/2$. We iterate this process until our initial value solutions converge.
    
    \item After we determine the initial values of $\Psi^\pm_{jmkn}$ and $\partial_r \Psi^\pm_{jmkn}$, we solve \eqref{eqn:hblODE} using an explicit embedded Runge-Kutta Prince-Dormand $(8, 9)$ method and an adaptive stepper from the GSL ODE library. We store solutions at the radial points $r=r_i$, and the stored values of $\Psi^\pm_{jmkn}$ and $\partial_r \Psi^\pm_{jmkn}$ are finally transformed back to the Teukolsky solutions $R^\pm_{jmkn}$ and $\partial_r R^\pm_{jmkn}$ by inverting \eqref{eqn:hblTransform} and its derivative. 
    
\end{enumerate}

\subsection{Source integration module}
\label{app:sourceModule}

Given discrete samplings of $S_{jmkn}$ and $R_{jmkn}^\pm$, along with the output from the geodesic module, we calculate the normalization constants $\hat{C}^\pm_{jmkn}$ using spectral source integration methods. The details of this procedure are discussed in \cite{HoppETC15,NasiOsbuEvan19}.
    
At large frequencies, these integrals become highly oscillatory and experience large numerical cancellations. This is not an issue when constructing the field outside the source region, where our mode-sum over the homogeneous solutions converges rapidly, and, therefore, we do not need to evaluate these highly oscillatory integrals. However, when extended into the source region, the homogeneous solutions are less efficient at capturing the behavior of the perturbing field, as noted by van de Meent and Shah \cite{VandShah15}. In practice, this means that the further we extend our homogeneous solutions, the more extended homogeneous modes we need to include in our mode-sum to accurately calculate the self-force. Therefore, for particularly eccentric orbits of $e\simeq 0.4$, higher-frequency modes can make a dominant contribution to our self-force calculation, and these modes will also have large numerical errors due to the catastrophic cancellations encountered when computing $\hat{C}^\pm_{jmkn}$. If the maximum value encountered in our spectral integral is more than 14 orders of magnitude greater than the computed value of $\hat{C}^\pm_{jmkn}$, then we consider all precision to be lost in our calculation and set $\hat{C}^\pm_{jmkn} = 0$. This is one of the dominant sources of numerical error in our self-force calculations.

\subsection{Secular evolution module}
\label{app:secularEvolutionModule}

Given a background geodesic and a force $f^\mathrm{gen}_{\alpha}$ acting on a scalar charge following that geodesic, the secular evolution module computes the resulting average rate of change of the scalar charge's orbital energy, angular momentum, and Carter constant due to $f^\mathrm{gen}_\alpha$. We denote the secular averages as $\langle \dot{E}\rangle^\mathrm{gen}$, $\langle \dot{L}_z\rangle^\mathrm{gen}$, and $\langle \dot{Q}\rangle^\mathrm{gen}$. They are found by replacing $\mu a^\mu$ with $\mu a^\mu_\mathrm{gen}=(g^{\mu\nu} + u^\mu u^\nu)f_\nu^\mathrm{gen}$ in \eqref{eqn:Jdot}-\eqref{eqn:IQ}.  This module works for any arbitrary force, provided that the input data for $f^\mathrm{gen}_\alpha$ are structured like the output data from the SSF module and $f^\mathrm{gen}_\alpha$ is real. 

We compute $\langle \dot{E}\rangle^\mathrm{gen}$, $\langle \dot{L}_z\rangle^\mathrm{gen}$, and $\langle \dot{Q}\rangle^\mathrm{gen}$ using the following numerical procedures:
\begin{enumerate}
    \item We start with two inputs: (1) worldline data from the geodesic module and (2) force data $f^\mathrm{gen}_{\alpha,j_rj_\theta}$ that is sampled on an $N \times N$ grid in $q_r$ and $q_\theta$. With this data we construct the integrands $\hat{I}^\mathrm{gen}_E$, $\hat{I}^\mathrm{gen}_{L_z}$, and $\hat{I}^\mathrm{gen}_Q$ using \eqref{eqn:IE}-\eqref{eqn:IQ} but replacing $a^\mu$ with $a^\mu_\mathrm{gen}$.
    
    \item Next we evaluate \eqref{eqn:Jdot} using spectral integration methods \cite{HoppETC15}. Because the resonant averages depend on $q_{\theta 0}$, rather than returning a single number, our spectral integrator returns a set of coefficients $\{\tilde{\mathcal{F}}^\mathcal{J}_0, \tilde{\mathcal{F}}^\mathcal{J}_1, \dots, \tilde{\mathcal{F}}^\mathcal{J}_{N_\mathrm{gen}-1}\}$. These coefficients are related to the averages via the discrete Fourier series
    \begin{align} \label{eqn:JdotFourier}
        \langle \dot{\mathcal{J}} \rangle^\mathrm{gen} &\approx \frac{q^2}{\mu}\mathrm{Re}\Bigg[\tilde{\mathcal{F}}^\mathcal{J}_0 + \sum_{k = 1}^{N_\mathrm{gen}-1} \tilde{\mathcal{F}}^\mathcal{J}_k e^{-ikq_{\theta 0}} \Bigg],
    \end{align}
    where $\mathrm{Re}[X]$ is the real part of $X$ and again we use $\mathcal{J}$ to represent $E$, $L_z$, or $Q$. Details of the integration can be found in Appendix \ref{app:fluxAvgDFT}.
    
    \item To test the accuracy of \eqref{eqn:JdotFourier}, we down-sample the force data $f^\mathrm{gen}_{\alpha,j_rj_\theta}$ to a $N/2 \times N/2$ grid and repeat our calculations. We then estimate the error in our Fourier coefficients by taking the fractional difference between our original Fourier coefficients $\tilde{\mathcal{F}}_{\mathcal{J}}^k$, and those produced by down-sampling the input data. 
    
    \item We then output the coefficients $\tilde{\mathcal{F}}^{E}_k$, $\tilde{\mathcal{F}}^{L_z}_k$, and $\tilde{\mathcal{F}}^{Q}_k$ and their estimated errors for $0 \leq k \leq N_\mathrm{gen}-1$. From this output, we can evaluate $\langle \dot{E}\rangle^\mathrm{gen}$, $\langle \dot{L}_z\rangle^\mathrm{gen}$, and $\langle \dot{Q}\rangle^\mathrm{gen}$ for any initial phase $q_{\theta0}$ using \eqref{eqn:JdotFourier}.
\end{enumerate}

\subsection{Regularization module}
\label{app:regModule}

The first part of the regularization module computes the SSF regularization parameters $A_\alpha$, $B_\alpha$, and $D^{(2)}_\alpha$ [see \eqref{eqn:regExpansion}]: 
\begin{enumerate}
    \item Given data from the geodesic module and a sample number $N_\mathrm{SSF}$, we evaluate $A_\alpha$, $B_\alpha$, and $D^{(2)}_\alpha$ on an $N_\mathrm{SSF} \times N_\mathrm{SSF}$ grid spanned by $q_r$ and $q_\theta$. This results in the discretely sampled parameters $A_{\alpha,j_rj_\theta}$, $B_{\alpha,j_rj_\theta}$, and $D^{(2)}_{\alpha,j_rj_\theta}$. This sampling is analogous to the construction of $\hat{F}^{\mathrm{ret},lm}_{\alpha\pm,j_rj_\theta}$ in \eqref{eqn:discreteSSF}.
\end{enumerate}
While we can regularize the SSF (or some quantity that depends on the SSF) with only these three parameters, the resulting mode-sum over $l$  [see \eqref{eqn:modeSumReg}] will only decay like $l^{-4}$. Consequently, truncating the sum at some $l_\mathrm{max}$ gives us a truncation error that scales like $l_\mathrm{max}^{-3}$. If we use values of $l_\mathrm{max} \sim O(20)$, the standard mode-sum regularization will only produce regular SSF data that is known to about two digits of precision. However, we can reduce the error in our regularized data by fitting for the higher-order regularization parameters. 

Therefore the second part of our regularization module fits for the regular component $H^\mathrm{R}$ of a formally divergent self-force quantity $H$ using the known higher-order structure of the singular component $H^\mathrm{S}$ given in \eqref{eqn:regExpansion}. We perform this fit by modifying the procedure proposed by van de Meent and Shah \cite{VandShah15}. First we decompose $H$ into a series of spherical harmonic $l$-modes, $H^l$, from which we define the finite, but unregularized quantity,
\begin{align}
    H_{l_\mathrm{cut}} &= \sum_{l=0}^{l_\mathrm{cut}} H^l = \sum_{l=0}^{l_\mathrm{cut}} \left(H^{\mathrm{R},l} + H^{\mathrm{S},l} \right)
    \\
    & = H^{\mathrm{R}}_{l_\mathrm{cut}} + H^{\mathrm{S}}_{l_\mathrm{cut}},
\end{align}
where $H = H_{l_\mathrm{cut}}$ as $l_\mathrm{cut}\rightarrow \infty$, while $H^{\mathrm{R}}_{l_\mathrm{cut}}$ and $ H^{\mathrm{S}}_{l_\mathrm{cut}}$ refer to the regular and singular pieces of $H_{l_\mathrm{cut}}$, respectively.

We then assume, based on the singular expansion in \eqref{eqn:regExpansion}, that $H^{\mathrm{S}}_{l_\mathrm{cut}}$ takes the form
\begin{align}
    H^{\mathrm{S}}_{l_\mathrm{cut}} &= \frac{1}{2}H^{{A}}(l_\mathrm{cut}+1)^2 + H^{{B}}(l_\mathrm{cut}+1)
    \\ \notag
    & \qquad + \sum_{l=0}^{l_\mathrm{cut}} \sum_{n = 1}^{\infty}
    \prod_{k=1}^n\frac{H^{D(2n)}_\alpha}{(2l+2k+1)(2l-2k+1)},
\end{align}
where $H^{{A}}$, $H^{{B}}$, and $H^{D(2n)}$ are the $l$-independent regularization parameters for the divergent quantity $H$. Furthermore, we approximate that $H^{\mathrm{R}}_{l_\mathrm{cut}} \approx H^{\mathrm{R}}$ for a sufficiently large choice of $l_\mathrm{cut}$, since the regular multipole contributions $H^{\mathrm{R},l}$ should decay exponentially with $l$. This gives us the approximate model
\begin{align} \label{eqn:fitModel}
    H_{l_\mathrm{cut}} &\approx {H}^{\mathrm{R}} + \frac{1}{2}{H}^{{A}}(l_\mathrm{cut}+1)^2 + {H}^{{B}}(l_\mathrm{cut}+1)
    \\ \notag
    & \qquad + \sum_{l=0}^{l_\mathrm{cut}} \sum_{n = 1}^{n_\mathrm{cut}}
    \prod_{k=1}^n\frac{{H}^{D(2n)}_\alpha}{(2l+2k+1)(2l-2k+1)}.
\end{align}
Using known values of $H_{l_\mathrm{cut}}$, we fit for the set of unknown parameters $\{{H}^{\mathrm{R}}, {H}^{{A}}, {H}^{{B}}, \dots, {H}^{{D(2n_\mathrm{cut})}}\}$, where we truncate this set of free parameters at $n_\mathrm{cut}$. From our fits we estimate $H^{\mathrm{R}}$. If any of the regularization parameters are known, then we modify our model by moving the known parameters to the left-hand side of \eqref{eqn:fitModel} and fitting for the remaining unknowns.

Therefore, we estimate $H^\mathrm{R}$ using the following fitting procedure:
\begin{enumerate}
    \item Given $H^l$ in the range $0 \leq l \leq l_\mathrm{max}$ and some set of known regularization parameters $\{H^A, H^B, \dots, H^{D(2n')}\}$, we construct $H_{l_\mathrm{cut}}$ for $12 \leq l_\mathrm{cut} \leq l_\mathrm{max}$.
    
    \item Next we define the tuning parameters $l_\mathrm{cut,min}$ and $l_\mathrm{cut,max}$, which select the values of $H_{l_\mathrm{cut}}$ that are used in our fit. As noted in \cite{VandShah15}, if we choose a value of $l_\mathrm{cut}$ that is too low, then the approximation in \eqref{eqn:fitModel} becomes less accurate and will not produce a good estimate of ${H}^{\mathrm{R}}$. On the other hand, if we choose a value of  $l_\mathrm{cut}$ that is too high, then we may include $l$-modes that are dominated by numerical error, and this error will bias our fits. Because we do not know \textit{a priori} which values of $l_\mathrm{cut}$ are too low or too high, we allow the tuning parameters to vary in the ranges $12 \leq l_\mathrm{cut,min} < l_\mathrm{max}$ and $l_\mathrm{cut,min} < l_\mathrm{cut,max} \leq l_\mathrm{max}$.
    
    \item \label{item:fit1} Given values for $l_\mathrm{cut,min}$, $l_\mathrm{cut,max}$, and $n_\mathrm{cut}$, we perform a least squares fit of \eqref{eqn:fitModel} using the set of known values $\{H_{l_\mathrm{cut,min}}, H_{l_\mathrm{cut,min}+1}, \dots, H_{l_\mathrm{cut,max}} \}$ and the set of known parameters $\{H^A, H^B, \dots, H^{D(2n')}\}$. This gives us a set of fit values for $\{H^\mathrm{R}, H^{{D}(2n'+2)}, \dots, H^{{D}(2n_\mathrm{cut})}\}$. We refer to this set of fits as $\tilde{h}_1 = \{\tilde{h}_1^\mathrm{R}, \tilde{h}_1^{{D}(2n'+2)}, \dots, \tilde{h}_1^{{D}(2n_\mathrm{cut})}\}$.
    
    \item \label{item:fitRepeat} We repeat step \ref{item:fit1} for all acceptable values of $l_\mathrm{cut,min}$, $l_\mathrm{cut,max}$, and $n_\mathrm{cut,max}$ (i.e., we require the number of known parameters to be greater than the number of free parameters), giving us a several different $\tilde{h}_1$ sets. Note that we restrict $n_\mathrm{cut}$ to the range $3 \leq n_\mathrm{cut} \leq 6$.
    
    \item We then repeat steps \ref{item:fit1}--\ref{item:fitRepeat}, but this time we leave the highest-order known parameter $H^{D(2n')}$ as an unknown variable. This gives us fits for the set $\{H^\mathrm{R}, H^{{D}(2n')}, \dots, H^{{D}(2n_\mathrm{cut})}\}$, which we denote as $\tilde{h}_2 = \{\tilde{h}_2^\mathrm{R}, \tilde{h}_2^{{D}(2n')}, \dots, \tilde{h}_2^{{D}(2n_\mathrm{cut})}\}$. 
    
    \item Next, we take the ten sets of $\tilde{h}_1$ that best fit the data $H_{l_\mathrm{cut}}$ and the ten sets of $\tilde{h}_2$ that best minimize $|\tilde{h}_2^{{D}(2n')}-H^{D(2n')}|$. Combining these twenty sets, we construct a set of twenty fits for $H^\mathrm{R}$ (i.e., ten values of $\tilde{h}_1^\mathrm{R}$ and ten of $\tilde{h}_2^\mathrm{R}$). 
    
    \item Finally we produce two estimates for the uncertainty in our fitted value of $H^\mathrm{R}$. First we estimate its uncertainty $\sigma^\mathrm{R}$ by taking the median absolute deviation of the combined set of $\tilde{h}_1^\mathrm{R}$ and $\tilde{h}_2^\mathrm{R}$ values. Second, we estimate its uncertainty to be the standard deviation of this same set. If the standard deviation is greater than the median absolute deviation, then we take this to be the final uncertainty in $H^\mathrm{R}$, and the value of $H^\mathrm{R}$ is given by the mean of our combined set of $\tilde{h}_1^\mathrm{R}$ and $\tilde{h}_2^\mathrm{R}$ values. If the median absolute deviation is greater than the standard deviation, then we take this to be the final uncertainty in $H^\mathrm{R}$, and the value of $H^\mathrm{R}$ is given by the median of our combined set.
\end{enumerate}
The above fitting procedure gives a simple estimate for $H^\mathrm{R}$, but neglects how errors in the input data $H^l$ may impact the fitting procedure. A more sophisticated algorithm could incorporate these errors, but we leave that for future work. We find that our method gives consistent estimates even when varying $l_\mathrm{max}$ or the number of known regularization parameters. Therefore, the robustness of the current algorithm is sufficient for this work.

\section{Hyperboloidal transformations of the radial Teukolsky equation}
\label{app:harm}

As mentioned in Appendix \ref{app:harmonicModule}, we make use of the hyperboloidal transformation \eqref{eqn:hblTransform} to put the radial Teukolsky equation into a form that is more amenable for numerical integration. The transformed homogeneous equations then take the forms
\begin{align} \label{eqn:hblODE}
    \left[\frac{d^2}{dr^2} - G^\pm_{mkn}(r) \frac{d}{dr} -  U^\pm_{jmkn}(r) \right]\Psi^\pm_{jmkn}(r)= 0,
\end{align}
where the potentials are given by
\begin{align} \notag
    G^\pm_{mkn}(r) &= \frac{2\left[a^2 + rM - ir(ma\pm (r^2+a^2)\omega_{mkn})\right]}{r\Delta}, 
    \\
    U^\pm_{jmkn}(r) &= \frac{2ia(m\pm a\omega_{mkn})}{r\Delta}
    \\ \notag
    & \qquad + \frac{\lambda_{jmkn} r^2+2Mr+2a^2}{r^2\Delta}
    \\ \notag
    & \qquad \qquad + \frac{2ma\omega_{mkn}(r^2+a^2)(1\pm 1)}{\Delta^2}.
\end{align}

\section{Confluent Heun equation and expansions about its singular points}
\label{app:che}

The confluent Heun equation takes the general form
\begin{equation} \label{eqn:cheODE}
    \frac{d^2w}{dz^2} + \left(\frac{\gamma_\mathrm{CH}}{z} + \frac{\delta_\mathrm{CH}}{z-1} + \varepsilon_\mathrm{CH} \right)
    \frac{dw}{dz} + \frac{\alpha_\mathrm{CH} z - q_\mathrm{CH}}{z(z-1)} w = 0,
\end{equation}
where $\alpha_\mathrm{CH}$, $\gamma_\mathrm{CH}$, $\delta_\mathrm{CH}$, $\varepsilon_\mathrm{CH}$, and $q_\mathrm{CH}$ are free parameters. It has regular singularities at $z=0$ and $z=1$, and an irregular singularity of Poincar{\'e}
rank 1 at $z = \infty$. This matches the singular structure of the radial Teukolsky equation \eqref{eqn:teukODE}. Therefore, we can reexpress \eqref{eqn:teukODE} in terms of \eqref{eqn:cheODE} via the following transformations,
\begin{align}
    \epsilon_\mathrm{CH} \kappa (z - 1) &= \omega( r- r_+),
    \\
    R(z) &= z^{a_\mathrm{CH}} (z-1)^{b_\mathrm{CH}} e^{\pm i\epsilon_\mathrm{CH} \kappa z} w(z),
\end{align}
where $\epsilon_\mathrm{CH} = 2M \omega$, $\kappa = \sqrt{1 - \chi^2}$, $\chi = a/M$,
$a_\mathrm{CH}=\pm i\epsilon_-$ and $b_\mathrm{CH}=\pm i\epsilon_-$, $\epsilon_\pm = (\epsilon_\mathrm{CH} \pm \tau_\mathrm{CH})/2$, and $\tau_\mathrm{CH} = (\epsilon_\mathrm{CH} - m \chi)/\kappa $. Note that we have dropped the mode subscripts, i.e., $R_{jmkn}\rightarrow R$ and $\omega_{mkn}\rightarrow \omega$, to simplify notation.
The Teukolsky variables are then related to the confluent Heun parameters via
\begin{align} \notag
	\gamma_\mathrm{CH} &= 1+2a_\mathrm{CH},
	\\ \notag
	\delta_\mathrm{CH} &= 1+2b_\mathrm{CH}, 
	\\
	\varepsilon_\mathrm{CH} &= \pm 2i\epsilon_\mathrm{CH}\kappa,
	\\ \notag
	\alpha_\mathrm{CH} &= \pm 2i\epsilon_\mathrm{CH}\kappa\left(1 \mp i\epsilon_\mathrm{CH}+a_\mathrm{CH}+b_\mathrm{CH} \right),
	\\ \notag
	q_\mathrm{CH} &= \lambda - \frac{1}{2}(\epsilon_\mathrm{CH}^2-\tau_\mathrm{CH}^2) + \epsilon_\mathrm{CH} m\chi
	-\epsilon_\mathrm{CH}^2(1-\kappa)
	\\ \notag
	&\qquad
	-b_\mathrm{CH}-a_\mathrm{CH}- 2a_\mathrm{CH}b_\mathrm{CH} \pm i\epsilon_\mathrm{CH}\kappa(1+2a_\mathrm{CH}),
\end{align}
where $\lambda$ is the spheroidal eigenvalue. For simplicity we take $b_\mathrm{CH} = i\epsilon_-$ and $a_\mathrm{CH} = -i\epsilon_-$ from here on.

After recasting the Teukolsky equation in this confluent Heun form, we use the Frobenius method to generate series expansions of the confluent Heun solutions about $z=1$, which corresponds to the regular singular point $r = r_+$ of the Teukolsky equation. We then connect the confluent Heun expansion to $R^-$, giving 
\begin{align} \label{eqn:R-Expansion}
    R^-(r) = f^-(r)
    \sum_{j=0}^\infty a^-_{2,j} \left(\frac{r-r_+}{2M\kappa} \right)^{j},
\end{align}
where
\begin{align} \label{eqn:prefactorR-}
    f^-(r) = \tilde{B}^\mathrm{trans}\left(\frac{r-r_-}{2M\kappa} \right)^{-i\epsilon_-}
    \left({r-r_+} \right)^{-i\epsilon_+} e^{-i\omega(r-r_+)},
\end{align}
and the series  coefficients satisfy a three-term recurrence relation,
\begin{equation}
    A^-_{2,n} a^-_{2,n - 1} + B^-_{2,n} a^-_{2,n} + C^-_{2,n} a^-_{2,n + 1} = 0,
\end{equation}
with
\begin{align} \notag
    A^-_{2,n} &= \alpha_\mathrm{CH} +\varepsilon_\mathrm{CH} (n + \lambda_2 - 1),
    \\ \notag
    B^-_{2,n} &= n^2 + n(\gamma_\mathrm{CH} + \delta_\mathrm{CH} - \varepsilon_\mathrm{CH} + 2 \lambda_2 - 1)
    \\ \notag
    & \qquad + \lambda_2(\gamma_\mathrm{CH} + \delta_\mathrm{CH} - \epsilon_\mathrm{CH} + \lambda_2 - 1) - q_\mathrm{CH},
    \\
    C^-_{2,n} &= -(n + 1 + \lambda_2)(n + \gamma_\mathrm{CH}  + \lambda_2),
\end{align}
and the monodromy eigenvalue $\lambda_2 = 1 - \gamma_\mathrm{CH}$. Series coefficients are generated recursively using the initial conditions $a^-_{2,-1} = 0$ and $a^-_{2,0} = 1$. In this work we take the amplitude in \eqref{eqn:prefactorR-} to be
\begin{align}
    \tilde{B}^\mathrm{trans} = 2^{-i\epsilon_+} \kappa^{i\epsilon_-} e^{i\epsilon_+},
\end{align}
so that \eqref{eqn:R-Expansion} matches the normalization of $R^-(r)$ in our MST series calculations.

Similarly, we generate asymptotic expansions of the confluent Heun solutions about $z=\infty$. Connecting these series expansions to $R^+$, we find that
\begin{align} \label{eqn:R+Expansion}
    R^+(r) \sim f^+(r)
    \sum_{j=0}^\infty a^+_{1,j} \left(\frac{r-r_-}{2M\kappa}\right)^{-j},
\end{align}
where the prefactor is given by
\begin{align} \label{eqn:prefactorR+}
    f^+(r) = \tilde{C}^\mathrm{trans} 
    \left(\frac{r-r_+}{r-r_-}\right)^{i\epsilon_+}
    \left(r-r_-\right)^{-(1 - i\epsilon_\mathrm{CH})}e^{i\omega r},
\end{align}
and the series coefficients satisfy the three-term recurrence relation,
\begin{equation}
    A^+_{1,n} a^+_{1,n - 1} + B^+_{1,n} a^+_{1,n} + C^+_{1,n} a^+_{1,n + 1} = 0,
\end{equation}
where
\begin{align} \notag
    A^+_{1,n} &= -(\alpha_\mathrm{CH} +  \varepsilon_\mathrm{CH}(n-1))(\alpha_\mathrm{CH} +  \varepsilon_\mathrm{CH}(n-\gamma_\mathrm{CH})),
    \\ \notag
    B^+_{1,n} &= \alpha_\mathrm{CH} ^2+
    \alpha_\mathrm{CH} \varepsilon_\mathrm{CH}  \big((1-\beta_\mathrm{CH} +2 n) +\varepsilon_\mathrm{CH}\big)
    \\ \notag
    & \qquad \qquad +\varepsilon_\mathrm{CH} ^2 (n (1-\beta_\mathrm{CH}+\varepsilon_\mathrm{CH}+n)-q_\mathrm{CH}),
    \\ 
    C^+_{1,n} &= -(n+1) \varepsilon_\mathrm{CH} ^3,
\end{align}
with $a^+_{1,0} = 1$, $a^+_{1,-1} = 0$, and $\beta_\mathrm{CH} = \gamma_\mathrm{CH} + \delta_\mathrm{CH}$. In this work we take the amplitude in \eqref{eqn:prefactorR+} to be
\begin{align}
    \tilde{C}^\mathrm{trans} = 2^{i\epsilon_\mathrm{CH}},
\end{align}
so that \eqref{eqn:R+Expansion} matches the normalization of $R^+(r)$ in our MST series calculations. 

We can connect \eqref{eqn:R-Expansion} and \eqref{eqn:R+Expansion} to the hyperboloidal functions $\Psi^\pm$ and their derivatives via \eqref{eqn:hblTransform}, allowing us to generate initial values for our numerical solvers. \eqref{eqn:R-Expansion} converges for $(r-r+)/(2M\kappa) < 1$, and therefore is suitable for generating initial values for the radial Teukolsky solutions at the initial radial point $r^-_\mathrm{init} \leq r_+ + 0.01 M$, provided $\kappa > 0.005$ or $\chi^2 < 0.999975$. This condition is met in this work. However, because the coefficients are complex and can alternate sign, it is possible for \eqref{eqn:R-Expansion} to experience catastrophic cancellation before converging to some required precision goal. Therefore, when computing initial data with this expansion, we track any potential precision loss. If more than 4 digits of precision are lost before the series converges to a precision $< 10^{-13}$, then we consider the Frobenius expansion method to have failed.

On the other hand, \eqref{eqn:R+Expansion} does not formally converge, just as we expect for an asymptotic series, with the coefficients growing like $a_{1,n}^+ \sim n a^+_{1,n-1}/\varepsilon_\mathrm{CH}$ as $n\rightarrow \infty$. However, for small values of $(2M\kappa)/(r-r_-)$ the series will initially decay. Therefore, when we truncate the sum at some finite value of $j$, \eqref{eqn:R+Expansion} provides a sufficiently accurate approximation of $R^+$. We evaluate \eqref{eqn:R+Expansion} at $r^+_\mathrm{init} \geq 50 M$ and perform the sum until the last two terms meet the convergence criteria,
\begin{align}
    \left\vert \left({\sum_{k=0}^{j} \frac{a^+_{1,k}}{a^+_{1,j}} \left(\frac{r^+_\mathrm{init}-r_-}{2M\kappa}\right)^{-k+j}}\right) \right\vert^{-1} < \epsilon_\mathrm{DBL}.
\end{align}
If this convergence criteria is not met before the series begins to diverge, then we consider the asymptotic series method to have failed.

\section{Spectral integration of the secular averages}
\label{app:fluxAvgDFT}

Consider some real-valued function $\mathcal{J}$ that depends on the worldline of the scalar charge as it follows a resonant geodesic. Then its evolution in Mino time $\lambda$ can be parametrized in terms of the angle variables $q_r = \Upsilon_r \lambda$ and $q_\theta = \Upsilon_\theta\lambda$, e.g., $\mathcal{J}(q_r, q_\theta)$. Now we average $\mathcal{J}$ over one resonant Mino period $\Lambda$ via the integral,
\begin{align} \label{eqn:resAvgX}
    \langle \mathcal{J} \rangle = \frac{1}{\Lambda} \int_0^\Lambda \mathcal{J}(q_r = \Upsilon_r \lambda , q_\theta = \Upsilon_\theta \lambda + q_{\theta 0})\, d\lambda,
\end{align}
where $q_{\theta 0}$ sets the initial phase of the charge's resonant motion at $\lambda = 0$.

To numerically evaluate \eqref{eqn:resAvgX}, first we take the two-dimensional discrete Fourier transform (DFT) of $\mathcal{J}$,
\begin{align}
    \tilde{f}^\mathcal{J}_{kn} = \frac{1}{N^2} \sum_{a=0}^{N-1} \sum_{b=0}^{N-1} \mathcal{J}\left(\frac{2\pi b}{N}, \frac{2\pi a}{N}\right) e^{2\pi i\left(\frac{ak + bn}{N}\right)},
\end{align}
and then we approximate the integrand as a truncated Fourier series
\begin{align} \label{eqn:IEFourier}
    \mathbf{F}[\mathcal{J}] &= \tilde{f}^\mathcal{J}_{00} 
    +2 \mathrm{Re}\sum_{k=1}^{N/2}\sum_{n=1}^{N/2} \tilde{f}^\mathcal{J}_{kn}e^{-i(kq_\theta+nq_r)}
    \\ \notag
    & \qquad \;\; + 2 \mathrm{Re}\sum_{k=1}^{N/2}\sum_{n=1}^{N/2} \tilde{f}^\mathcal{J}_{k,N-n}e^{-i(kq_\theta-nq_r)}
    \\ \notag
    & \qquad \qquad + 2 \mathrm{Re}\sum_{k=1}^{N/2}\tilde{f}^\mathcal{J}_{k0}e^{-ikq_\theta} + 2 \mathrm{Re}\sum_{n=1}^{N/2}\tilde{f}^\mathcal{J}_{0n}e^{-inq_r},
\end{align}
where $\mathrm{Re}[f]$ refers to the real part of $f$. In \eqref{eqn:IEFourier} we have taken advantage of the fact that, because $\mathcal{J}$ is real, $\tilde{f}^\mathcal{J}_{kn} = \overline{\tilde{f}^\mathcal{J}_{-k,-n}}$.
For sufficiently large values of $N$, the series representation $\mathcal{F}[\mathcal{J}]$ faithfully approximates $\mathcal{J}$ to machine precision \cite{HoppETC15}, i.e.,
\begin{align}
    \left\vert 1 - \frac{\mathbf{F}[\mathcal{J}]}{\mathcal{J}} \right\vert \lesssim \epsilon_\mathrm{DBL}.
\end{align}
In general, we do not know the minimum value of $N$ that meets this criteria before we construct $\mathbf{F}[\mathcal{J}]$.
    
Finally, we evaluate \eqref{eqn:resAvgX} by replacing $\mathcal{J}$ with $\mathbf{F}[\mathcal{J}]$. Only the zero-frequency modes will contribute to the integral. Because $q_r = \Upsilon_r/\beta_r = \Upsilon_\theta/ \beta_\theta = \Upsilon$ (see Sec.~\ref{sec:geo}), the integral reduces to a sum over the $(k,n)$-modes that satisfy $\beta_\theta k - \beta_r n = 0$ with $k>0$ and $n>0$. This gives a Fourier series representation of $\langle \mathcal{J} \rangle$,
\begin{align} \label{eqn:JdotFourier2}
    \langle \mathcal{J} \rangle &\approx \tilde{\mathcal{F}}^\mathcal{J}_0 + 2\mathrm{Re}\sum_{k = 1}^{N'} \tilde{\mathcal{F}}^\mathcal{J}_k e^{-ikq_{\theta 0}},
\end{align}
where $\tilde{\mathcal{F}}^\mathcal{J}_0 = \tilde{f}^\mathcal{J}_{00}$,
\begin{align}
    \tilde{\mathcal{F}}^\mathcal{J}_k &= \frac{1}{\Upsilon_t}\tilde{f}^\mathcal{J}_{k,N - \beta_\theta k/\beta_r}, 
    & 
    N' &= \mathrm{floor}\left[\frac{\beta_rN}{2\beta_\theta}\right],
\end{align}
and $\mathrm{floor}[X]$ refers to the largest integer that is less than $X$.

\section{Fourier representation of radiative averages}
\label{app:fourierFluxes}

Because the radiative averages $\langle \dot{\mathcal{J}}\rangle^{\mathcal{H}/\infty}_\mathrm{rad}$ in \eqref{eqn:XdotRad} will vary with respect to $q_{\theta 0}$, we can express each average as a Fourier series,
\begin{equation} \label{eqn:fourierFlux}
    \langle \dot{\mathcal{J}}\rangle^{\mathcal{H}/\infty}_\mathrm{rad} = \tilde{\mathcal{F}}^{\mathcal{J},\mathcal{H}/\infty}_{0} + 2 \mathrm{Re} \sum_{b = 1}^\infty \tilde{\mathcal{F}}^{\mathcal{J},\mathcal{H}/\infty}_{b} e^{-ibq_{\theta 0}},
\end{equation}
where we have taken advantage of the fact that $\tilde{\mathcal{F}}^{\mathcal{J},\mathcal{H}/\infty}_{b} = \overline{\tilde{\mathcal{F}}^{\mathcal{J},\mathcal{H}/\infty}_{-b}}$ since $\langle \dot{\mathcal{J}}\rangle^{\mathcal{H}/\infty}_\mathrm{rad}$ is real-valued. The advantage of this Fourier representation is that the sum in \eqref{eqn:fourierFlux} is rapidly convergent. Thus, in a numerical calculation we can truncate the series after summing over just the first few terms. This means we only need to calculate the first few coefficients (e.g., $b \lesssim 8$) in order to accurately approximate $\langle \dot{\mathcal{J}}\rangle^{\mathcal{H}/\infty}_\mathrm{rad}$ for any value of $q_{\theta 0}$.
As an example, we outline an efficient method for calculating $\tilde{\mathcal{F}}^{\mathcal{J},\infty}_{b}$ using the fiducial normalization coefficients $\hat{C}^+_{jmkn}$, though these methods generalize to $\tilde{\mathcal{F}}^{\mathcal{J},\mathcal{H}}_{b}$ as well.

Using \eqref{eqn:fiducalCjmkn} and \eqref{eqn:XdotRad}, the coefficients take the form
\begin{align}
    \tilde{\mathcal{F}}^{\mathcal{J},\infty}_b &= \frac{1}{4\pi} \sum_{jmN}\sum_{(k,n)_N}\sum_{(k',n')_N} \mathcal{A}^{\mathcal{J}}_{mkn}\omega_{mk'n'}
    \\ \notag
    & \qquad \qquad
    \times \hat{C}^+_{jmkn}\overline{\hat{C}^+_{jmk'n'}} \int_0^{2\pi} \frac{d q_{\theta 0}}{2\pi}
    e^{i(b+k-k')q_{\theta 0}},
    \\ \label{eqn:fluxCoeffFullSum}
    &= \frac{1}{4\pi} \sum_{j=0}^\infty \sum_{k=-\infty}^\infty \tilde{\mathcal{F}}^{\mathcal{J},\infty}_{b,jk},
    \\ \label{eqn:fluxCoeffFullSubSum}
    \tilde{\mathcal{F}}^{\mathcal{J},\infty}_{b,jk} &= \sum_{m=-j}^j \sum_{N=-\infty}^\infty \mathcal{A}^{\mathcal{J}}_{mk(N- \beta_\theta{k})/\beta_r}\,\omega_{mN}
    \\ \notag
    & \qquad \times
    \hat{C}^+_{jmk(N-\beta_\theta{k})/\beta_r}\overline{\hat{{C}}^+_{jm(b+k)(N-\beta_\theta b-\beta_\theta {k})/\beta_r}},
\end{align}
where $\omega_{mN} = m\Omega_\phi + N \Omega$.
First we find that when $b = 0$ our expressions reduce to our nonresonant mode-sum expressions for our averaged quantities,
\begin{align} \notag
    \tilde{\mathcal{F}}^{\mathcal{J},\infty}_0 &= \frac{1}{4\pi} \sum_{jmkN}\mathcal{A}^{\mathcal{J}}_{mk(N- \beta_\theta{k})/\beta_r}\,\omega_{mN} \left\vert \hat{C}^+_{jmk(N-\tilde{k})/\beta_r} \right\vert^2,
    \\
    &= \frac{1}{4\pi}  \sum_{jmkn}\mathcal{A}^{\mathcal{J}}_{mkn}\omega_{mkn} \vert \hat{C}^+_{jmkn} \vert^2,
\end{align}
just as we expect.

While we can calculate $\tilde{\mathcal{F}}^\infty_{\mathcal{J},b}$ directly from \eqref{eqn:fluxCoeffFullSum}, in practice we use the symmetries of the amplitudes to simplify this expression and only sum over positive frequency modes $\omega_{mN}>0$. Separating the $\omega_{mN}>0$ and $\omega_{mN}<0$ terms in \eqref{eqn:fluxCoeffFullSubSum}, we find that $\tilde{\mathcal{F}}^{\mathcal{J},\infty}_{b,jk} = \overline{\tilde{\mathcal{F}}^{\mathcal{J},\infty}_{-b,j-k}}$. Combining this with $\tilde{\mathcal{F}}^{\mathcal{J},\mathcal{H}/\infty}_{b} = \overline{\tilde{\mathcal{F}}^{\mathcal{J},\mathcal{H}/\infty}_{-b}}$ tells us that the Fourier coefficients must be purely real. Taking advantage of these properties, we can arrange our sums so that we only need to consider $m$ and $N$ values such that $\omega_{mN}>0$,
\begin{align}
    \tilde{\mathcal{F}}^{\mathcal{J},\infty}_b
    &= \frac{1}{4\pi} \sum_{jk} \sum_{\omega_{mN}>0} \omega_{mN}\mathcal{B}^{\mathcal{J}}_{mkbN}
    \\ \notag
    & \qquad \times
    \hat{C}^+_{jmk(N-\beta_\theta{k})/\beta_r}\overline{\hat{{C}}^+_{jm(b+k)(N-\beta_\theta b-\beta_\theta {k})/\beta_r}},
    \\ \notag
    \mathcal{B}^{\mathcal{J}}_{mkbN} &= \mathcal{A}^{\mathcal{J}}_{mk(N- \beta_\theta{k})/\beta_r} + \mathcal{A}^{\mathcal{J}}_{m(k+b)(N- \beta_\theta b - \beta_\theta k)/\beta_r}.
\end{align}
Furthermore, because the mode amplitudes $\hat{C}^\pm_{jmkn}$ vanish if $j + m + k = \mathrm{odd}$, $\tilde{\mathcal{F}}^{\mathcal{J},\infty}_b = 0$ if $b = \mathrm{odd}$. So we only need to calculate $\tilde{\mathcal{F}}^{\mathcal{J},\infty}_{2b'}$ for $b' > 0$. This is consistent with what we see in full numerical calculations of $\langle \dot{\mathcal{J}}\rangle^{\mathcal{H}/\infty}_\mathrm{rad}$.




\bibliography{parent}

\end{document}